\documentclass[10pt,pra,aps,amsmath,amssymb,floatfix,showpacs,longbibliography,notitlepage]{revtex4-1}
\usepackage{shortCut}

\graphicspath{{FIG/EPS/}}

\begin{document}
\title{
The effect of helicity on the correlation time of large scale turbulent flows
}
\author{Alexandre \textsc{Cameron}}
\email[]{alexandre.cameron@ens.fr}
\author{Alexandros \textsc{Alexakis}} 
\email[]{alexakis@lps.ens.fr}
\author{Marc-{\'E}tienne~\textsc{Brachet}}
\email[]{brachet@physique.ens.fr}
\affiliation{Laboratoire de Physique Statistique, 
  \'{E}cole Normale Sup\'{e}rieure, PSL Research University; 
  Universit\'{e} Paris Diderot Sorbonne Paris-Cit\'{e}; 
  Sorbonne Universit\'{e}s UPMC Univ Paris 06; CNRS; 
  24 rue Lhomond, 75005 Paris, France}
\date{\today}
\begin{abstract}
Solutions of the forced Navier-Stokes equation have been conjectured to 
thermalize at scales larger than the forcing scale, similar to an absolute 
equilibrium obtained for the spectrally-truncated Euler equation. 
Using direct numeric simulations of Taylor-Green flows and 
general-periodic helical flows, we present results on the probability 
density function, energy spectrum, auto-correlation function and 
correlation time that compare the two systems.
In the case of highly helical flows, we derive an analytic expression describing 
the correlation time for the absolute equilibrium of helical flows that is different 
from the $E^{-1/2}k^{-1}$-scaling law of weakly helical flows. This model predicts a new 
helicity-based scaling law for the correlation time as $\tau(k)\sim H^{-1/2}k^{-1/2}$. 
This scaling law is verified in simulations of the truncated Euler equation.
In simulations of the Navier-Stokes equations the large scale modes of forced 
Taylor-Green symmetric flows (with zero total helicity and large separation of scales) 
follow the same properties as absolute equilibrium including a 
$\tau(k) \sim E^{-1/2}k^{-1}$ scaling for the correlation time. General-periodic helical 
flows also show similarities between the two systems, however the largest scales 
of the forced flows deviate from the absolute equilibrium solutions. 

\end{abstract}

\maketitle

\section{Introduction}                                           
\label{sec:intro}                                                     

Experiments and numerical studies
\cite{champagne_fine-scale_1978,kaneda_energy_2003} are known to reproduce,
up to intermittency corrections \cite{frisch_turbulence:_1996},
the energy spectrum power law predicted by Kolmogorov's 1941 theory of turbulence
\cite{kolmogorov_local_1941}:
$E(k) \propto \epsilon^{2/3}k^{-5/3}$ where $k$ and $\epsilon$ denote the 
wavenumber and the energy dissipation respectively. 
This power law is valid in the inertial range, {\it i.e.} for wavenumbers satisfying 
$\kf{} < k < k_{d}$ where $\kf$ and $k_{d} \sim \kf{} Re^{3/4}$ denote the forcing wavenumber 
and the viscous dissipation wavenumber respectively. The Reynolds number, $Re$,
is defined as $Re \equiv U /\visco{} k_f$ where $U$ stands for the root mean square velocity
and $\visco{}$ denotes the viscosity.
In order to maximize the size of this inertial range, most experiments and direct 
numeric simulations (DNS) forced the flow at the largest scale of the system.
At these scales the dynamics of the system can be described in terms of the 
Richardson cascade \cite{frisch_turbulence:_1996} where large scales transfer their 
energy to smaller scales with large eddies breaking into smaller eddies. This simple 
description cannot be used to describe the dynamics at scales larger than the forcing 
scale (from now on referred to as the {\it large scales}). Indeed, contrary to the inertial 
range, large scales do not have a direct flux of energy coming from the forcing scale.
Some theoretical arguments 
\cite{forster_large-distance_1977,kraichnan_is_1989,frisch_turbulence:_1996}
indicate that they can be described by the dynamics of absolute equilibrium solutions 
of the spectrally truncated Euler equation (\TEE{}) given by
\begin{align}
	\partial_t \uvec + \mathbb{P}_{\km}[\uvec \cdot \grad \uvec + \grad P ]=0
	\quad \text{and} \quad
	\diV \uvec =0 \,,
	\label{eq:TE}
\end{align}
where $\uvec$, $P$ and $\km{}$ denote the velocity field, the pressure field and 
the truncation wavenumber respectively. The operator $\mathbb{P}_{\km}$ enforces a 
spherical truncation in the \TEE{}. It acts in Fourier space as a small scale filter. The modes 
whose wavenumbers satisfy $|{\kvec}|\le \km{}$, are unaffected by the projection whereas
the amplitudes of the other modes are all set to zero. Despite keeping the
amplitude of the truncated modes to zero, the projection conserves the total
energy and helicity of the flows. After a transient state, 
the system reaches an absolute equilibrium state which has been conjecture to follow 
a Boltzmann-Gibbs distribution depending only on the initial energy and helicity of the flow.
Under these assumptions, Kraichnan \cite{kraichnan_helical_1973} predicted the form 
of the energy and helicity spectrum. DNS of the \TEE{} performed in 
\cite{cichowlas_effective_2005,krstulovic_cascades_2009}
verified the absolute equilibrium energy and helicity spectra predicted in 
\cite{kraichnan_helical_1973}.

The \TEE{} ~\eqref{eq:TE} needs to be contrasted with the forced Navier-Stokes equation (\NSE{}) 
that governs the evolution of viscous flows and is given by
\begin{align}
	\partial_t \uvec + (\uvec \cdot \grad) \uvec = - \grad P 
	+ \visco{} \Laplace \uvec + \Fvec 
	\quad \text{and} \quad
	\diV \uvec = 0,	
	\label{eq:NS}
\end{align}
where $\uvec$, $P$  and \visco{} denote the velocity field, the pressure field and 
the viscosity respectively. $\Fvec$ denotes the forcing field that is acting at a 
particular wavenumber $k_f$. Note that unlike the \TEE{}, in the \NSE{}, energy needs 
to be supplied by the forcing term, in order to compensate viscous dissipation. As result of
the presence of viscosity, the solutions of the equation are dependent on the Reynolds number.
Furthermore, the energy and helicity of the system at steady state are not determined 
by the initial conditions as in the \TEE{}, but depend on the forcing, the domain size and 
the viscosity. Despite these differences, the large scales could, in principle, be modeled as 
in the state of absolute equilibrium predicted in \cite{kraichnan_helical_1973}
due to the absence of mean flux of energy . 
Recently, Dallas {\it et al.} \cite{dallas_statistical_2015} performed turbulent DNS with 
enough resolution to model flows with significant scale separation between the forcing scale 
$\ell_f\propto1/\kf$ and the domain scale $L$. In their study, it was 
reported  that large scale spectra are in agreement with the absolute equilibrium
theory, although some deviations at scales the order of the domain size were reported. 
Among other reasons, these deviations could be attributed to large scale 
instabilities \cite{frisch1987large,cameron_large-scale_2016}.

Agreement in the spectra however does not guarantee the presence of an absolute equilibrium, 
since the temporal dynamics can be different. The differences (or the equivalence) between the 
temporal dynamic of the large scale modes of solution of the \NSE{} and the absolute
equilibrium solutions of the \TEE{} have not yet been investigated. Note that the equivalence of 
the temporal dynamic of the two systems is a much more stringent condition than
the simple equivalence of their spatial spectra.
The correlation time of the system is a good measurement to asses the temporal 
dynamic of large scale modes. In pseudo-spectral DNS, the Eulerian correlation time 
of the modes of the velocity field is the best suited to describe the temporal dynamic of the system.
Numerical solutions have already been used to analyze the temporal evolution of statistical 
equilibrium in the statistically stationary regime without helicity in \cite{cichowlas_effective_2005} and 
in the transitory regime with helicity in \cite{krstulovic_cascades_2009}. But the characterization of 
the correlation time and comparison with the large scales of the \NSE{} has not yet been performed.


The aim of the present paper is to compare the statistics of individual modes
and the temporal properties of the \TEE{} solutions and 
those of the large scales of the \NSE{}. 
In the next section \ref{sec:TT} we review absolute equilibrium theory and
derive analytic results for the correlation time of incompressible flows solutions of the \TEE{}
for arbitrary helicity. 
In section  \ref{sec:TE} we validate these predictions with numerical solutions of the 
\TEE{} while 
in section \ref{sec:FNS} we test if they also apply to the large scales of solutions of the \NSE{}
for different forcing functions.
Our conclusions are drawn in the last section.

\section{Absolute equilibrium and Thermalization theory}     
\label{sec:TT}                                                                               
\subsection{Energy and helicity spectra}                                 
The derivation of absolute equilibrium statistics for helical flows was carried out by Kraichnan 
\cite{lee_statistical_1952,kraichnan_helical_1973} in an analogy to micro-canonical 
ensembles in statistical thermodynamics \cite{landau_statistical_1980}. Similarly to the 
micro-canonical ensemble, the \TEE{} conserves the total energy, $E$. In addition, 
the \TEE{} also conserves the total helicity, $H$, which is another global quadratic 
quantity in velocity
\begin{eqnarray}
	E=\frac{1}{2} \frac{1}{L^3} \int \vert\uvec\vert^{2} d \rvec 
		\quad \text{and} \quad
	H=\frac{1}{2} \frac{1}{L^3} \int \uvec \cdot \omvec d \rvec ,
		\label{eq:nrghel}
\end{eqnarray}
where $L^3 = \int  d \rvec$ and $\omvec = \roT \uvec$. 
In analogy with the thermodynamic canonical ensemble, in a statistically 
steady state, absolute equilibrium solutions of the \TEE{} will correspond to a flow in a state 
$\uvec$ with probability $\mathcal{P}({\uvec})$ that follows the Boltzmann-Gibbs distribution
\begin{align}
   \mathcal{P}({\uvec}) = \frac{1}{Z}e^{-\mathcal{C}(\mathbf{u})}.
	\label{eq:bltzm}
\end{align}
The functional $\mathcal{C}(\mathbf{u})$ is a linear combination of the energy 
$E$ and the helicity $H$ of the flow
\begin{align}
	\mathcal{C}(\mathbf{u}) = \alpha E + \beta H = \alpha \left( E  - \Kr{} \frac{H}{\km{}} \right)
	\quad \text{and} \quad
	\Kr{} = -\frac{\beta \km{}}{\alpha} 
	\label{eq:aebh}
\end{align}
where $\alpha$ and $\beta$ are two parameters introduced by Kraichnan. These parameters 
unequivocally define a class of absolute equilibrium solutions of the \TEE{} with a fixed energy 
$E$ and helicity $H$. What we presently call the Kraichnan number: \Kr{} is a dimensionless 
combination of $\alpha$, $\beta$ and \km{}. The Kraichnan number indicates 
the degree of helicity of the flow. When \Kr{}=0, the flow does not have helicity,
whereas when $\vert \Kr{} \vert= 1$ the flow is maximally helical. As of consequence of the 
definition of $\mathcal{C}(\mathbf{u}) $, velocity modes are independent Gaussian variables. 
The partition function $Z$ used as normalization in eq.~\eqref{eq:bltzm} is defined by 
\begin{align}
	Z = \int \mathcal{D} \uvec e^{ -\mathcal{C} ( \uvec)  } \, .
	\label{eq:partf}
\end{align}

Similarly to statistical thermodynamics, Boltzmann-Gibbs weights can be used to compute
statistical averages over the space of incompressible flows. The average of a generic
observable $f(\uvec)$ is then given by
\begin{align}
	\langle f(\uvec) \rangle = 
	\frac{1}{Z}\int \mathcal{D}\uvec f(\uvec) e^{ -\mathcal{C} (\uvec)}
	\label{eq:avr}
\end{align}
In the case of the \TEE{}, the truncation in wavenumber implies that the functional integral can 
be done over a finite number of Fourier modes $\widetilde{\uvec}_{\kvec}$ 
that satisfy the incompressibility condition ${\kvec}\cdot \widetilde{\uvec}_{\kvec}=0$.
This last condition can be simplified by using the Craya-Herring 
\cite{craya_contribution_1957,herring_approach_1974} helical decomposition.
Each Fourier mode $\widetilde{\uvec}_{\kvec}$ is written as the sum of two modes of opposite 
helicity: $\widetilde{\uvec}_{\kvec} = \widetilde{\uvec}_{\kvec}^+ +\widetilde{\uvec}_{\kvec}^-$, where 
\begin{align}
	\widetilde{\uvec}^{\pm}_{\kvec} = \frac{1}{2} \left[ 
	\widetilde{\uvec}_{\kvec} \pm \frac{\widetilde{\omvec}_{\kvec}}{k} \right ]
\end{align}
with $\widetilde{\omvec}_{\kvec}=i\kvec\times\widetilde{\uvec}_{\kvec}$. This leads to 
two independent complex amplitudes $\widetilde{\uvec}^{\pm}_{\kvec}$ for each Fourier mode 
of an incompressible flow.
Using this statistical average of eq.~\eqref{eq:avr}, the average energy and helicity of 
the modes of the flow can be derived analytically. Since the PDF of every mode 
of the velocity field follows a Gaussian distribution, 
the average energy 
$\langle e_{\kvec} \rangle = \frac{1}{2} \langle \left\vert \widetilde{\uvec}_{\kvec} \right\vert^2 \rangle$
and average helicity 
$\langle h_{\kvec} \rangle= \frac{1}{2} \langle \widetilde{\uvec}_{-\kvec} 
\cdot \widetilde{\omvec}_{\kvec} \rangle$ 
of each wavevector are given by
\begin{align}
	\langle e_{\kvec} \rangle 
	= \frac{ \alpha^{-1} }{ 1 - \left( \Kr{} \frac{k}{\km{}} \right)^{2} }
	\quad \text{and} \quad
	\langle h_{\kvec} \rangle
	= \frac{\beta}{\alpha} \frac{ \alpha^{-1}k^2 }{ 1 - \left( \Kr{} \frac{k}{\km{}} \right)^{2} }
	= \frac{\beta}{\alpha} k^2 \langle e_{\kvec} \rangle \, .
	\label{eq:KrSt}
\end{align}
When $|\Kr{}| \to 1$, the energy is confined in the modes in the smallest scales 
of wavenumber near $\km{}$. 

The absolute equilibrium distribution can be expressed using the Craya-Herring 
\cite{craya_contribution_1957,herring_approach_1974} helical decomposition 
for the energies 
$\langle e_{\kvec}^\pm \rangle = \frac{1}{2} 
\langle \left\vert \tilde{\uvec}_{\kvec}^\pm \right\vert^2 \rangle$
and helicities $\langle h_{\kvec}^\pm \rangle =  \pm k
\langle e_{\kvec}^\pm \rangle$
\begin{align}
	\langle e^{\pm}_{\kvec} \rangle
	= \frac{1}{2} \frac{ \alpha^{-1} }{ 1 - (\pm) \Kr{} \frac{k}{\km{}} }
	\quad \text{and} \quad
	\langle h^{\pm}_{\kvec} \rangle
	= \frac{1}{2} \frac{ \pm \alpha^{-1}k }{ 1 - (\pm) \Kr{} \frac{k}{\km{}} }
	\, .
	\label{eq:CrHe}
\end{align}
When $\Kr{} \to 1$, the energy is confined in the small scale modes and
more precisely in their positive helical component.
The energy spectra are then obtained by sphericaly averaging over all modes of the same modulus
\begin{align}
	E^\pm(k) \equiv \sum_{|{\bf k}|=k} \langle e^{\pm}_{\kvec} \rangle 
	\simeq \frac{ 2 \pi k^2 \alpha^{-1} }{ 1 - (\pm) \Kr{} \frac{k}{\km{}} }
	\quad \mathrm{and} \quad E(k)=E^+(k)+E^-(k)
	\simeq \frac{ 4 \pi k^2 \alpha^{-1} }{ 1 - \left ( \Kr{} \frac{k}{\km{}} \right)^2}
\end{align}
and similarly for the helicity spectra
\begin{align}
	H^\pm(k) \equiv \sum_{|{\bf k}|=k} \langle h^{\pm}_{\kvec} \rangle 
	\simeq \frac{ \pm 2 \pi k^3 \alpha^{-1} }{ 1 - (\pm) \Kr{} \frac{k}{\km{}} }
	\quad \mathrm{and} \quad H(k)=H^+(k)+H^-(k)
	\simeq \frac{ 4 \pi k^4 \alpha^{-2} \beta }{ 1 - \left ( \Kr{} \frac{k}{\km{}} \right)^2}
	\,.
\end{align}

\subsection{Correlation times of absolute equilibrium solutions} 
\label{sec:abseq} 
The derivations so far predict the ensemble average of 
the energy and helicity per mode. However they do not characterize the temporal 
properties of the system. The computation of the spectra only requires 
the knowledge of the conserved quantities of the \TEE{} and not the equation 
itself. To describe with more depth the properties of the solutions of the \TEE{} or the 
\NSE{} in the thermalization domain, the temporal properties of the flows must be analyzed.

In \cite{cichowlas_effective_2005}, Cichowlas {\it et al.} studied the correlation time of 
absolute equilibrium solutions of the \TEE{} without helicity. It was shown 
that the correlation time $\tau_{\kvec}^{E}$ (defined more precisely below: eq.~\eqref{eq:gam:T2}) 
depends on the energy $E$ of the flow and is inversely proportional to the wavenumber $\kvec$
\begin{align}
	\tau_{\kvec}^{E} \propto k^{-1} E^{-1/2 } \, .
	\label{eq:tau:sclE}
\end{align}

In this present work, we extend this result for flows with an arbitrary helicity and 
show that a new power law emerges when the flow is strongly helical. This
new power law depends on the helicity $H$ of the flow rather than the energy and reads
\begin{align}
	\tau_{\kvec}^{H} \propto k^{-1/2} H^{-1/2} \, .
	\label{eq:tau:sclH}
\end{align}
This new power law is valid for highly helical flows and wavenumbers 
in the range $k_c\ll k \ll \km$ where $k_c\propto \km (1-|\Kr|) \ln (1-|\Kr|) $,
while the non-helical scaling law of eq.~\eqref{eq:tau:sclE} is valid for $k\ll k_c$. In what 
follows, we give a sketch of the derivation of eq.~\eqref{eq:tau:sclE} and \eqref{eq:tau:sclH} 
while the full derivation is presented in appx.~\ref{sec:apx:ct}.

The correlation time will be built using the short time approximation 
of the correlation function. The temporal correlation function $\Gamma_{\kvec}(t)$ of a 
mode is defined as
\begin{align}
	\Gamma_{\kvec}(t) & = \frac{\overline{ \uvec_{\kvec}^{*}(s)  \uvec_{\kvec}(s+t)} }
	{\overline{ \vert \uvec_{\kvec} (s) \vert ^2 }}
	\quad \text{ where } \quad
	\overline{ f(s) } = \lim_{T\to \infty} \frac{1}{2T} \int_{-T}^{T} f(s) \d s \,.
	\label{eq:gam}
\end{align}
It satisfies the relations $\Gamma_{\kvec}(0)=1$ and 
$\Gamma_{\kvec}(t) = \Gamma_{\kvec}(-t)$. If the system looses memory as time
elapses, the correlation function also satisfies $\Gamma_{\kvec}(\infty)=0$. 
Thus, the correlation function assesses how fast a mode 
de-correlates from its initial value. Using the Taylor expansion of the correlation 
function near $t=0$, the correlation function can be written as 
\begin{align}
	\Gamma_{\kvec}(t)  =  1 - \frac{1}{2}t^2 \tau_{\kvec}^{-2} + \dots 
		\quad \text{with} \quad
	{\tau_{\kvec}^{-2}} = - \partial_t^2 \Gamma_{\kvec}|_{t=0}
	= - \frac{\overline{ \uvec_{\kvec}^{*}(s) \partial^{2}_{t} \uvec_{\kvec}(s+t) }|_{t=0} }
	{\overline{ \vert \uvec_{\kvec} (s) \vert ^2 }} \, ,
	\label{eq:gam:T1}
\end{align}	 
where $\tau_{\kvec} $ will be referred to as the parabolic correlation time. The term 
$- \overline{ \uvec_{\kvec}^{*}(s) \partial^{2}_{t} \uvec_{\kvec}(s+t) }|_{t=0}$
can be rewritten as $\overline{ \vert \partial_{t} \uvec_{\kvec}(s+t) \vert^2}_{t=0}$
using an integration by parts.
Assuming that the \TEE{} system is ergodic, the averages over time can be replaced by 
the statistical averages defined in eq.~\eqref{eq:avr}. The correlation time can 
then be expressed as
\begin{align}
	\tau_{\kvec} = \sqrt{\frac{ \langle \vert \uvec_{\kvec}\vert^2 \rangle 
	}{ \langle \vert \partial_{t} \uvec_{\kvec} \vert^2 \rangle }} \,.
	\label{eq:gam:T2}
\end{align}

The expression of $\langle \vert \uvec_{\kvec}\vert^2 \rangle $ is given by the 
absolute equilibrium statistics. On the other hand, the expression of 
$\langle \vert \partial_{t} \uvec_{\kvec} \vert^2 \rangle$
can be computed using the temporal evolution equation of the mode given 
by eq.~\eqref{eq:TE} in the case of the \TEE{}. 
The property that the modes of the velocity field are independent Gaussian 
variables is used to compute the averages.

In the limit where $\Kr{} \to 1$ and $ k / \km{} \to 0 $, it is possible to compute an 
asymptotic expression of the correlation time with the Craya-Herring helical 
decomposition \cite{craya_contribution_1957,herring_approach_1974} (see 
appx.~\ref{sec:apx:ct}). In this limit, most of the energy is concentrated in the 
positive helical components of the modes near \km{}. The interactions of these 
modes are the dominant terms in the temporal evolution equation and give a 
theoretical prediction for the correlation time
\begin{align}
  \tau_{\kvec} \simeq
	\sqrt{\frac{\langle \vert \uvec^{+}_{\kvec} \vert^2 \rangle }{
	\langle \vert \partial_t \uvec^{+}_{\kvec}\vert^2\rangle}} 
	\underset{\substack{\Kr{} \to 1 \\ k / \km{} \to 0}}{=} 
	\sqrt{\frac{ \frac{15}{8} (1-\Kr{} ) - \frac{k }{ \km{} \ln(1-\Kr{} )} 
	}{ 4 \pi \alpha^{-1} k^2 (1 - s_{\kvec} \Kr{} \frac{k }{ \km{} } ) }} 
	\quad \text{with} \quad
	\alpha = \frac{\tanh^{-1}(\Kr{})- \Kr{}}{ 2 E_{tot} \Kr{}^{3} } \, .
	\label{eq:tau:kr}
\end{align}

\begin{figure}[!ht]
  \centering
  	\includegraphics[width=5.9cm, trim= 0 0 0 0, clip=true]{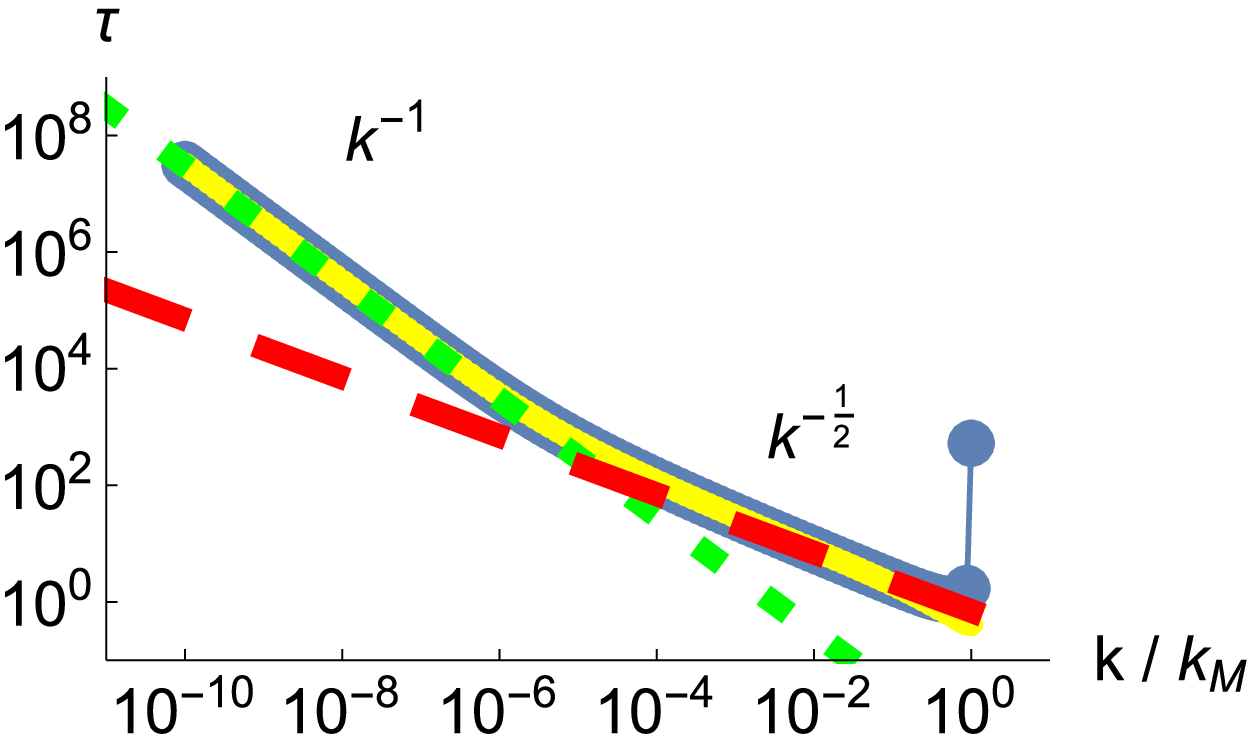}
  	\includegraphics[width=5.9cm, trim= 0 0 0 0, clip=true]{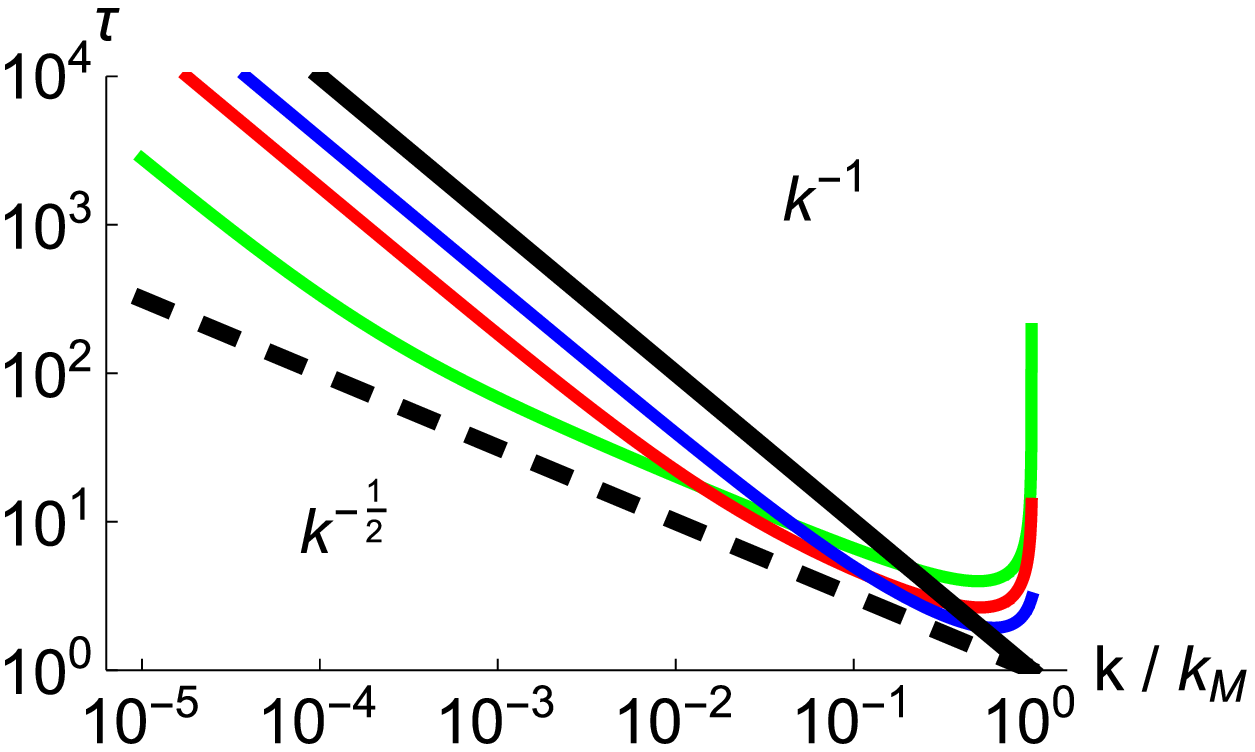}
  	\includegraphics[width=5.9cm, trim= 0 0 0 0, clip=true]{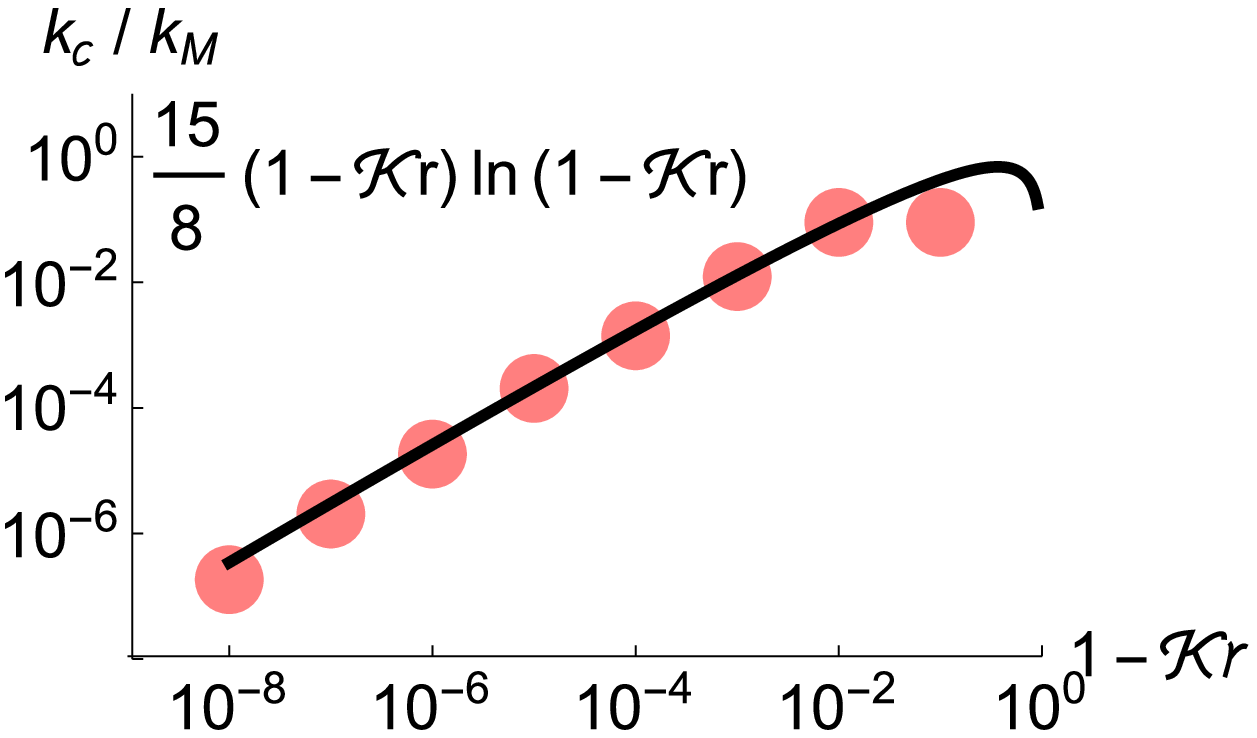}
  	\caption[Thermalization theory: correlation time]
  	{Left) Correlation time as a function of the wavenumber for $1-\Kr{} =10^{-6} $.
  	The results are represented with the full line with dark dots for the
  	positive helical components and in bright dots for the negative helical components.
  	The $k^{-1/2}$-scaling law is represented with the dashed line.
  	The $k^{-1}$-scaling law is represented with the dotted line.
	Center) Dependence of the correlation time for the positive helical component. 
	The Kraichnan number is increased at fixed energy: 
	$\Kr{} \in \lbrace 1-10^{-1} ; 1-10^{-2} ; 1-10^{-4}  \rbrace $,
	represented by blue, red and green lines respectively. The
	dotted line represents the $k^{-\frac{1}{2}}$-scaling law and the full line represents 
	the $k^{-1}$-scaling law.
	Right)
	Transition wavenumber as a function of $1-\Kr{}$, the semi-analytic
	prediction is represented with dots and the $\frac{15}{8} (1-\Kr{}) \ln(1-\Kr{})$ scaling law
	is represented with the full line.
	}
  \label{fig:fig1}
\end{figure}

The left panel of fig.~\ref{fig:fig1} represents the correlation time $\tau_k$
as a function of $k$ for a Kraichnan number near one: $\Kr{} = 1- 10^{-6} $.
Both power laws $k^{-1}$ and $k^{-1/2}$ are represented. The $k^{-1}$-power law
is valid in the largest wavenumbers: $k\ll k_c$, while the $k^{-1/2}$-power law 
is valid for an intermediate range of wavenumber: $k_c \ll k\ll \km$.
The center panel of fig.~\ref{fig:fig1} represents the dependence of the correlation time at
different Kraichnan numbers for fixed total energy $E_{tot}$. As $\Kr{} \to 1$, the 
correlation time in the small scales increases and the correlation time in the large 
scales decreases. For absolute equilibrium solutions of the \TEE{}, increasing the helicity 
slows down the dynamic of the small scales and makes the dynamic of the large scales 
more rapid.
The right panel of fig.~\ref{fig:fig1} represents the dependence of the transition 
wavenumber $k_c$ with $1-\Kr$. The value of the transition wavenumber was 
estimated using the intersection of the two power laws. The dark curve on the graph 
indicates that the critical wavenumber follows closely the $A(1-\Kr{})\ln(1-\Kr{})$ prediction.
Even though the helicity does not appear explicitly in eq.~\eqref{eq:tau:kr}, the correlation
time has a $k^{-\frac{1}{2}}$-scaling for intermediate wavenumbers when 
$ 1- \Kr{} \ll k/\km{} \ll 1 $. This scaling is similar to the helicity-based correlation time 
(see eq.~\eqref{eq:tau:sclH}) and appears for a range of Kraichnan numbers corresponding 
to highly helical flows.

\section{Truncated Euler DNS}                            
\label{sec:TE}                                                        

In the previous section, we discussed some predictions on the properties of the PDF, 
the standard derivation and the correlation time of solutions of the \TEE{}. 
We will now check their validity in periodic flows with and
without \TG{} symmetries \cite{brachet_small-scale_1983}. The DNS with 
\TG{} symmetries are performed using the pseudo-spectral code TYGRES 
\cite{pouquet_dynamics_2010} and those without \TG{} symmetries were
performed using the pseudo-spectral code GHOST 
\cite{mininni_nonlocal_2008,mininni_hybrid_2011}.
The major advantage of studying flows with \TG{} symmetries is that the 
symmetries can be used to gain a factor $32$ both in storage and execution time.
However, flows with \TG{} symmetries have a total helicity equal to zero and
consequently always have a Kraichnan number equal to zero. In order to study helical 
flows, DNS have to be performed in the general-periodic domain without \TG{} 
symmetry. This last configuration will be referred to as general-periodic flow in 
opposition to \TG{} symmetric flows.

\subsection{No helicity: inviscid Taylor-Green flows}      
\label{sec:TE:sub:NoH}                                                        
The first flows used to probe the statistical properties of the \TEE{} have \TG{}
symmetries that impose the total helicity to be equal to zero. As a consequence 
of \TG{} symmetries \cite{nore_decaying_1997}, the Fourier expansion of 
the flow can be expressed with the following simplified expression
\begin{align}
	\begin{bmatrix}
		u^x_{\rvec} \\ u^y_{\rvec} \\ u^z_{\rvec}
	\end{bmatrix} 
	= \sum^{\infty}_{k_x=0} \sum^{\infty}_{k_y=0} \sum^{\infty}_{k_z=0} 
	\begin{bmatrix}
		u^x_{\kvec} \times \sin k_x x \; \cos k_y y \; \cos k_z z \\
		u^y_{\kvec} \times \cos k_x x \; \sin k_y y \; \cos k_z z \\
		u^z_{\kvec} \times \cos k_x x \; \cos k_y y \; \sin k_z z
	\end{bmatrix} 
		\; ,
	\label{eq:TG:def}
\end{align}
where $\uvec_{\kvec} \in \mathbb{R}^{3}$ if $k_x$, $k_y$, $k_z$ are all odd or all 
even integers and $\uvec_{\kvec} =0$ otherwise. All the properties of the Fourier 
coefficients related to \TG{} symmetries can be found in the appendix of 
\cite{nore_decaying_1997}. Specific properties useful to understand the number of 
independent variables of \TG{} symmetric flows are presented in appx.\ref{sec:apx:TGP}. 

Incompressible random flows with \TG{} symmetries and energy equipartition were used 
to initialize the simulations. Since \TG{} symmetric flow do not have helicity 
($H=0$ and $\Kr{}=0$), the thermalization theory of Sect. \ref{sec:abseq} predicts that 
they should follow
\begin{align}
	\langle e_{\kvec} \rangle 
	= \alpha^{-1}
	\quad \text{thus} \quad
	E(k) = \frac{4 \pi k^2}{ \km{}^3} \alpha^{-1}
		\quad \text{and} \quad
	\tau_{\kvec} \underset{\substack{k / \km{} \to 0}}{=} \sqrt{\frac{45 \alpha}{112}} \frac{1}{k}
	\, .
	\label{eq:tau:nH}
\end{align}

\begin{figure}[!ht]
  \centering
  \includegraphics[height=\fheight, trim= 0 0 0 0, clip=true]{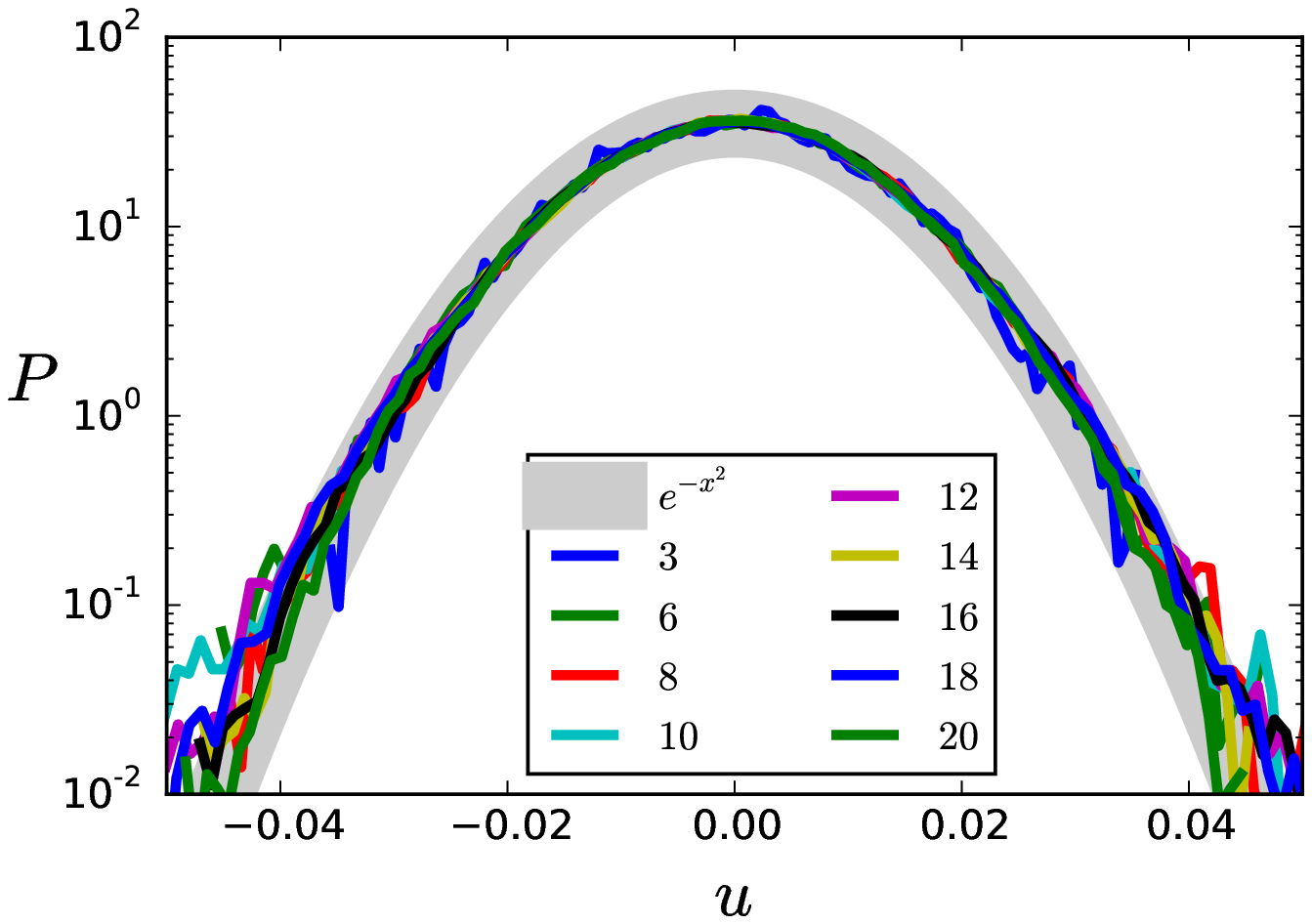}
  \includegraphics[height=\fheight, trim= 0 0 0 0, clip=true]{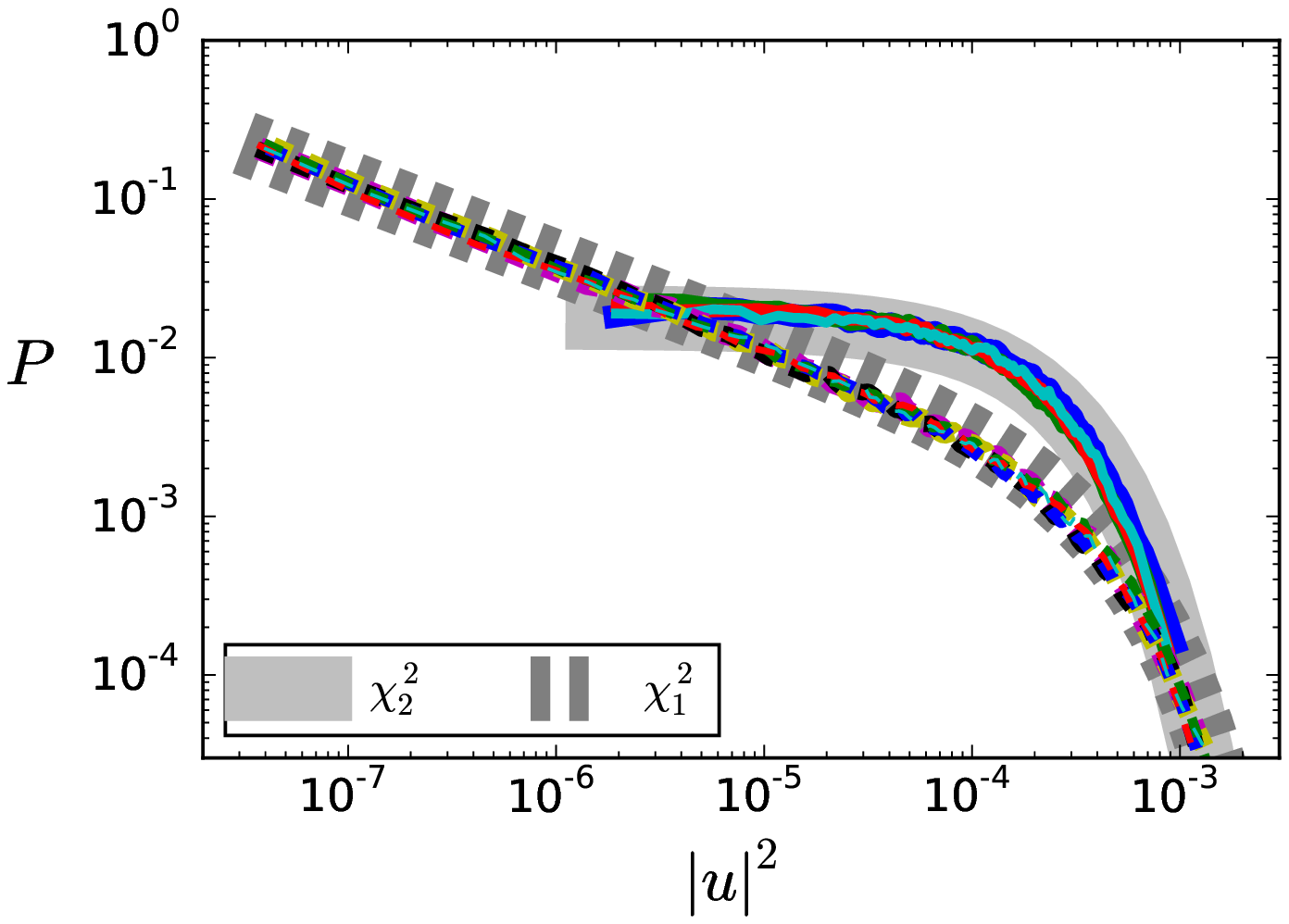}
  \caption[Taylor-Green truncated Euler: PDF]
  {Left) PDF of the amplitude of the velocity of modes with one degree of 
  freedom of Taylor-Green symmetric DNS in semi-logarithmic scale.
  Right) PDF of energy of the modes of Taylor-Green symmetric DNS of the 
  truncated Euler equation in logarithmic scale: (thick) theoretical predictions,
  (thin dashed) one degree of freedom, (thin full) two degrees of freedom. 
  }
  \label{fig:fig2}
\end{figure}

The first hypothesis in the absolute equilibrium theory is that the components of velocity
are independent Gaussian variables. As detailed in appx.\ref{sec:apx:TGP}, in \TG{} flows,
the even modes with one of their components equal to zero -- {\it i.e.} of the form 
$(0,k_y,k_z)$, $(k_x,0,k_z)$ or $(k_x,k_y,0)$ -- and the $xy$-diagonal modes -- {\it i.e.} 
of the form $(k_{\perp},k_{\perp},k_{\parallel})$ -- both only have one degree of freedom 
corresponding to their real amplitude. All other \TG{} modes have two degrees of freedom.
To test this assumption, temporal series of the modes are recorded and analyzed to 
extract the PDF. PDFs of modes with one degree of freedom are presented in the left panel 
of the fig.~\ref{fig:fig2} at different wavenumbers. On the semi-logarithmic scale, the distributions 
of the modes follow the parabolic trend characteristic of Gaussian distribution. 
The right panel of fig.~\ref{fig:fig2} represents the distribution of energy of the different
velocity modes. By definition, the energy is the sum of the square of the velocity components.
If at a particular wave number $\bf k$ there are $g$ modes that are independent and Gaussian 
then the distribution of the energy $e_{\kvec}$ that lies at that wavenumber must follow 
a $\chi^2_{g}$-distribution \cite{weisstein_chi-squared_2017} 
\begin{align}
	\chi^{2}_{g}(e_{\kvec}) = \frac{1}{2^{\frac{g}{2}}\Gamma(\frac{g}{2})} 
	(e_{\kvec})^{\frac{g}{2}-1} \exp\left[{-\frac{e_{\kvec}}{2} }\right]
	\,.
	\label{eq:chi2}
\end{align}
where $g$ denotes the number of independent Gaussian variables (see appx.\ref{sec:apx:chi2}). 
Thus, the power-law behavior at small values of $e_{\kvec}$ reveals the degrees of 
freedom involved. Due to the symmetries of the \TG{} flows different wavenumbers have 
different degrees of freedom. The modes -- $(0,k_y,k_z)$, $(k_x,0,k_z)$, $(k_x,k_y,0)$ and 
$(k_{\perp},k_{\perp},k_{\parallel})$ -- have one degree of freedom and should have an 
energy distribution following a $\chi^2_{1}$-law. All other modes have two degrees of freedom
and have an energy distribution following a $\chi^2_{2}$-law. 
If the energy follows a $\chi^2$-distribution, it does not necessarily imply that the velocity 
has a Gaussian statistics, it is only a characteristic of the sum of Gaussian distributions. However, 
we will only check the Gaussian statistics of modes with one degree of freedom 
and look at the energy distribution of the other modes.

\begin{figure}[!ht]
  \centering
  \includegraphics[height=6.cm, trim= 0 0 0 0, clip=true]{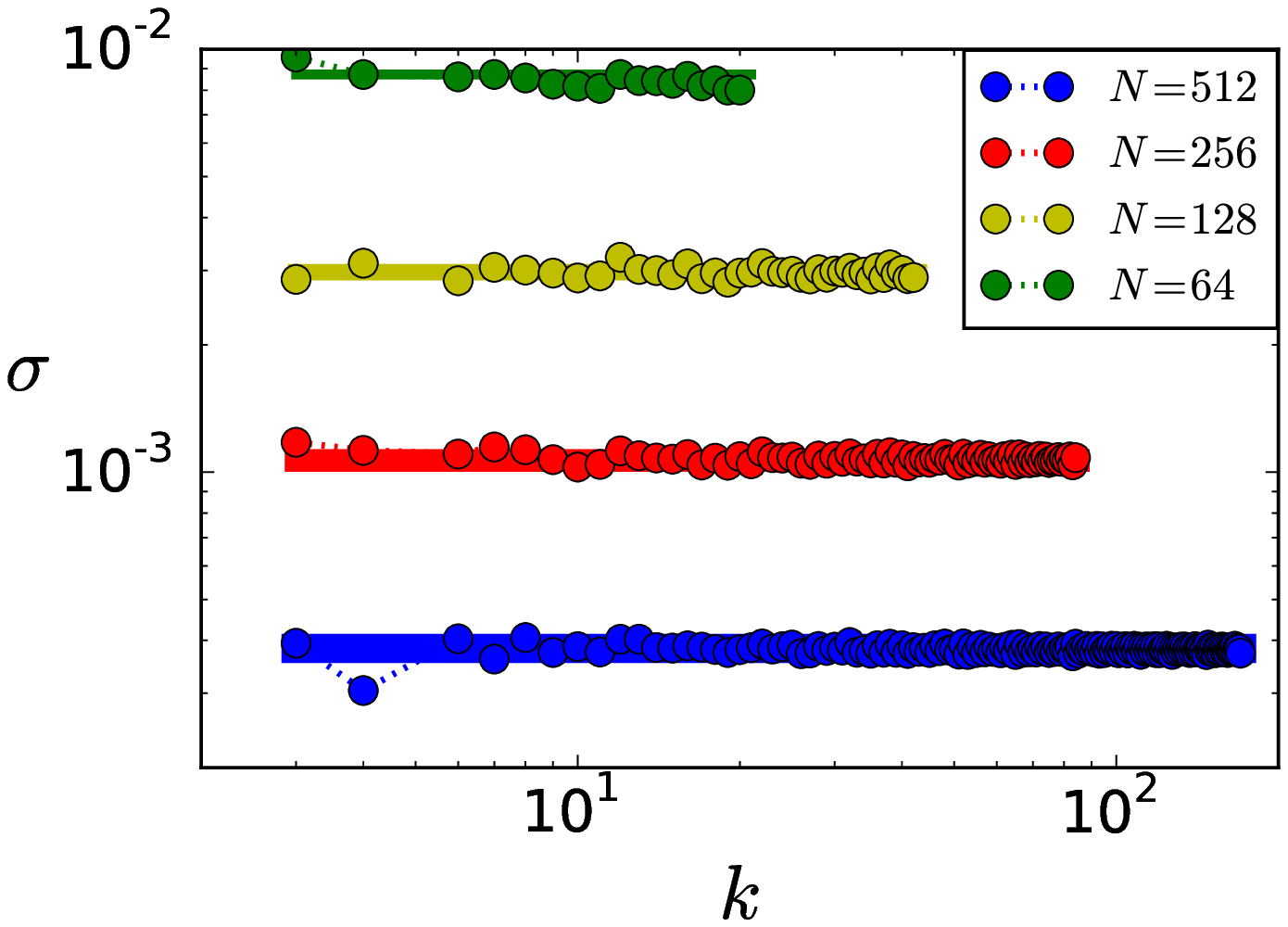}
  \includegraphics[height=6.cm, trim= 0 0 0 0, clip=true]{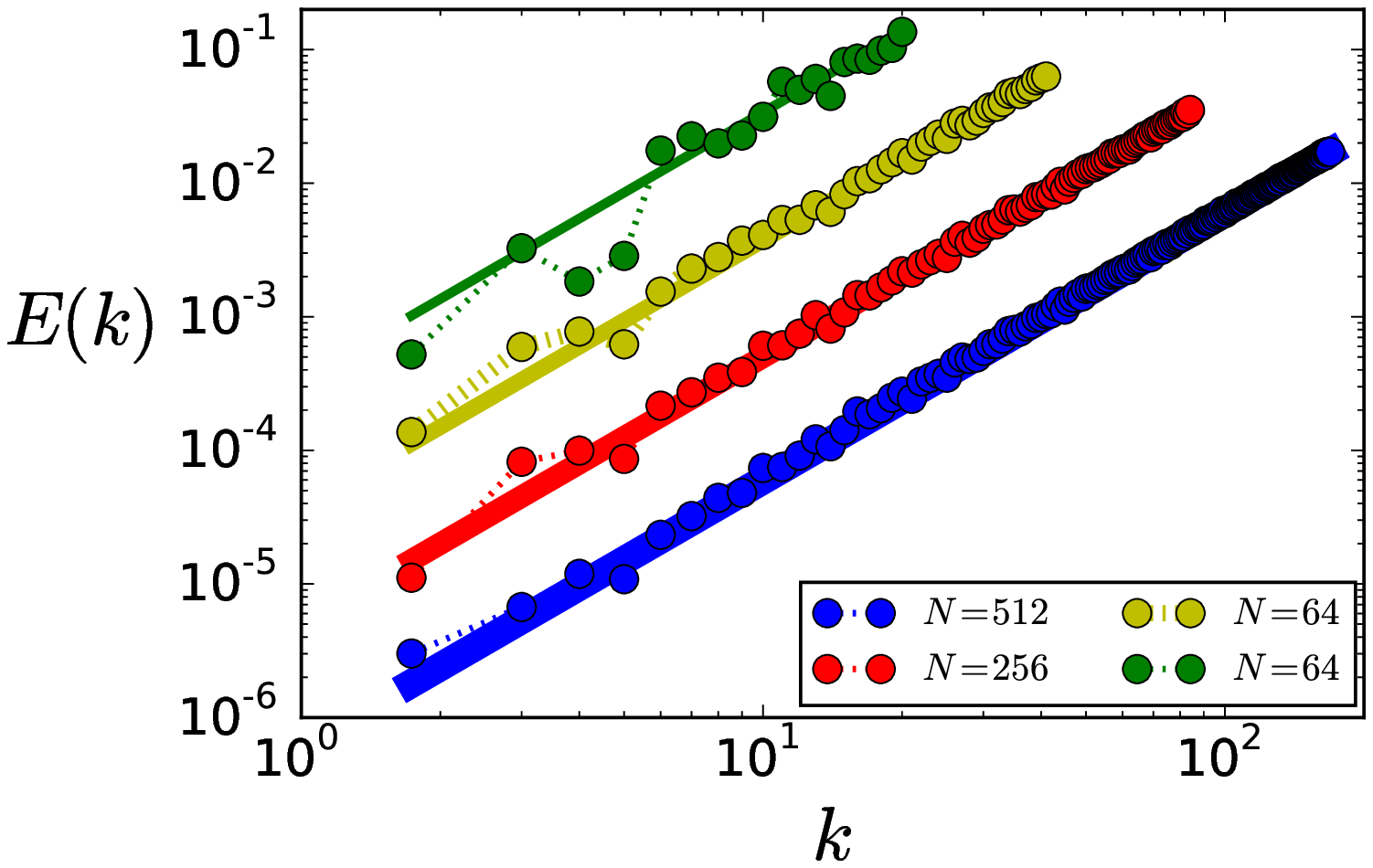}
  \caption[Taylor-Green truncated Euler: standard deviation]
  {Left) Average standard deviation of the velocity
  of Taylor-Green symmetric DNS of the truncated Euler equation
  for different resolutions.
  Right) Energy spectrum of Taylor-Green symmetric DNS of the 
  truncated Euler equation at different resolutions $N=\lbrace 64, 128, 256, 512\rbrace$
  for a fixed energy.
  }
  \label{fig:fig3}
\end{figure}

To check the equipartition of the energy, the standard deviation of modes can be
computed with their temporal series or the energy of all the modes within a shell of 
constant wavenumbers can be summed. The first method can be performed with 
the temporal series used to plot the PDF of the modes in fig.~\ref{fig:fig2}. 
Using this time average, the standard deviation, $\sigma^2 =\langle e_{k}\rangle$, 
is represented in the left panel of fig.~\ref{fig:fig3}. The second method, with 
the shell-summed energy at a fixed time, is presented in the right panel of fig.~\ref{fig:fig3}.
If the system satisfies energy equipartition and is ergodic, the amount of energy per shell 
should be proportional to the surface of the shell, $4\pi k^2$. The $k^2$-power law followed
by the energy spectrum in the right panel of fig.~\ref{fig:fig3} is therefore consistent with 
the equipartition of energy and ergodicity.
	
\subsubsection{Correlation time computation}
\label{sec:TE:sub:tau}
\begin{figure}[!ht]
  \centering
  \includegraphics[height=6.25cm, trim= 0 0 0 0, clip=true]{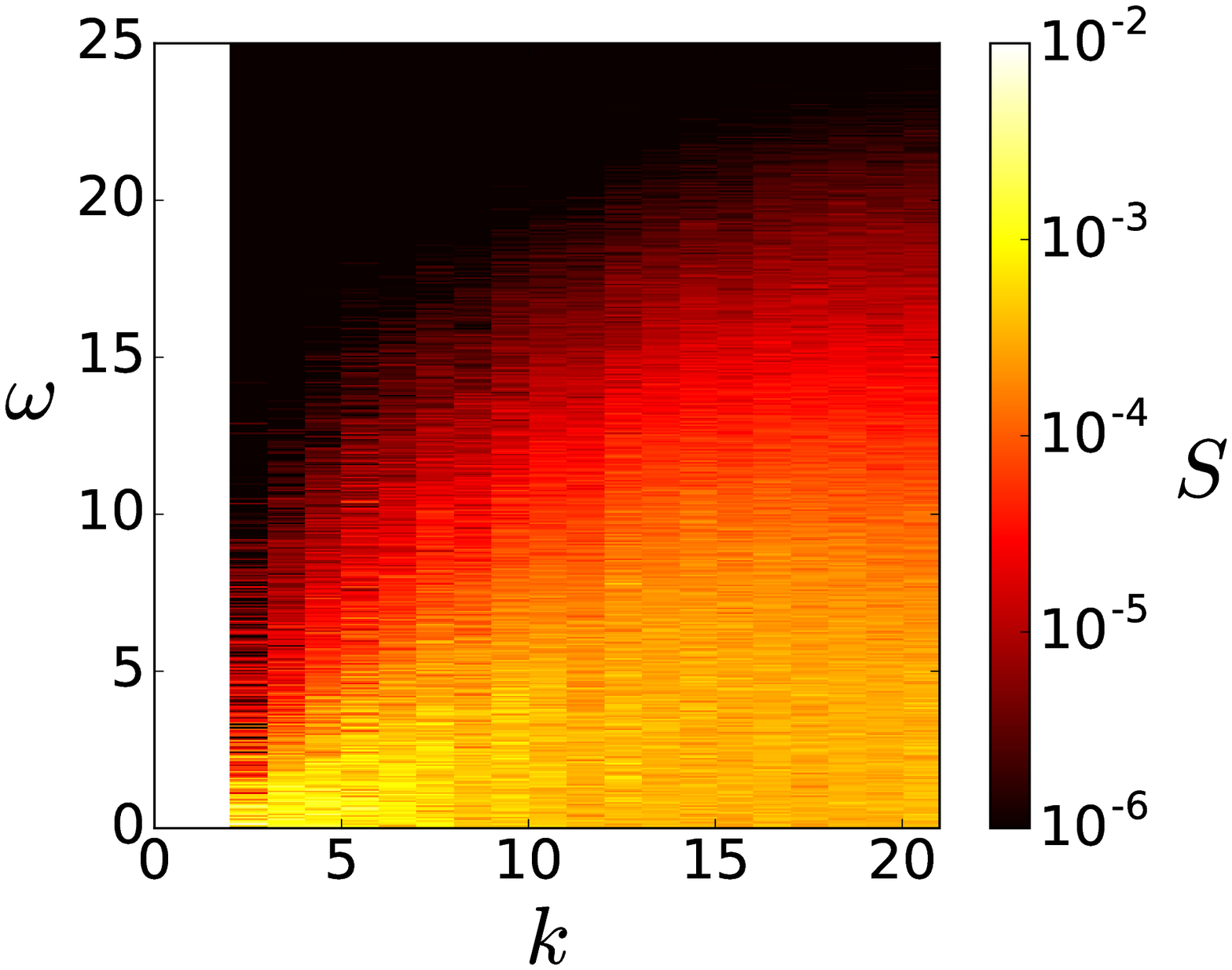}
  \includegraphics[height=6.25cm, trim= 0 0 0 0, clip=true]{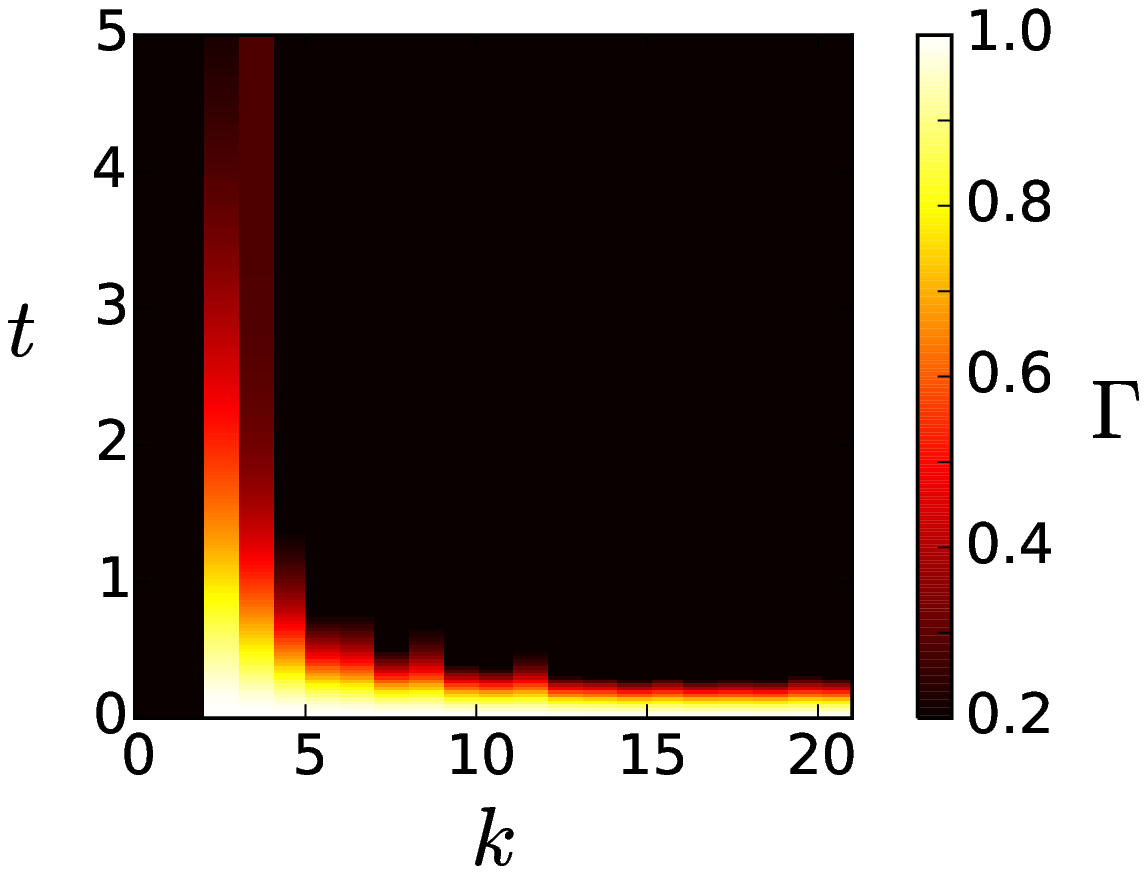}
  \caption[Taylor-Green truncated Euler: spatio-temporal spectrum]
  {Spatio-temporal spectrum of Taylor-Green symmetric DNS of
  the truncated Euler equation. 
  The numeric data is represented with dots and the theoretical prediction with full lines.
  Left) Power spectrum $S(k, \omega)$.
  Right) Correlation function $\Gamma(k, t)$.}
  \label{fig:fig4}
\end{figure}

The temporal statistics of the modes depend on the evolution equation and not 
only on the conserved quantities. Since the correlation function and consequently the
correlation time are specific properties of the evolution equation, their measurement
characterizes the temporal evolution of the system. In order to compute the correlation 
function and assess the correlation time, we developed a method similar to that used to 
produce spatio-temporal spectrum in wave experiments 
\cite{cobelli_global_2009,miquel_role_2014,leoni_spatio-temporal_2015},
where the spatio-spectral spectrum $S(k,\omega)$ is first calculated 
and the correlation function $\Gamma(k, t)$ is obtained by a Fourier transform using the 
Wiener-Khinchin theorem \cite{weisstein_wiener-khinchin_2017}. 
This method is explained in appx.~\ref{sec:apx:cort}.
Fig.~\ref{fig:fig4} presents in the left panel the spatio-temporal color-plot of 
the power spectrum $S(k,\omega)$ and in the right panel the correlation function 
$\Gamma(k, t)$.  

\begin{figure}[!!ht]
  \centering
  \includegraphics[height=\fheight, trim= 0 0 0 0, clip=true]{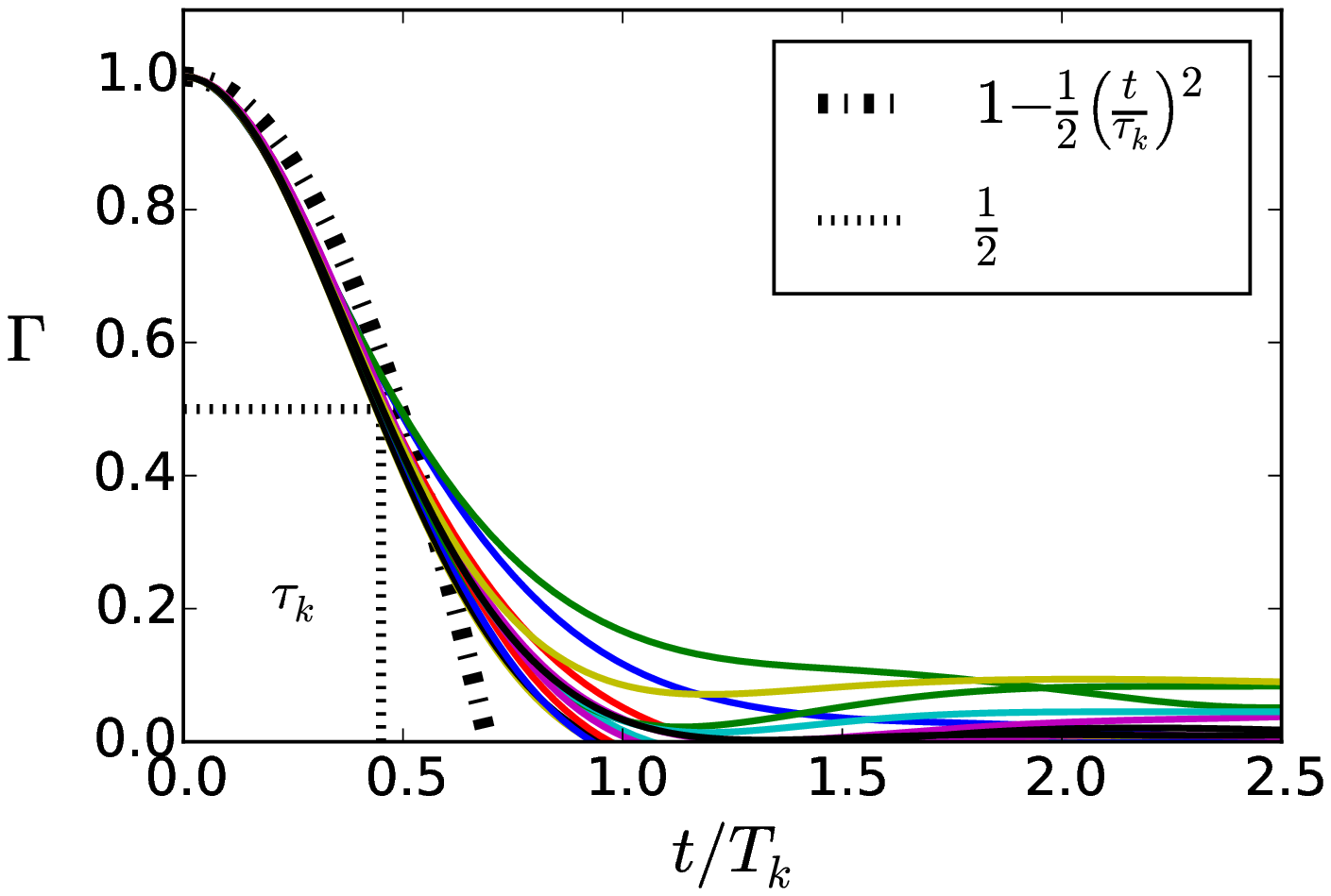}
  \includegraphics[height=\fheight, trim= 0 0 0 0, clip=true]{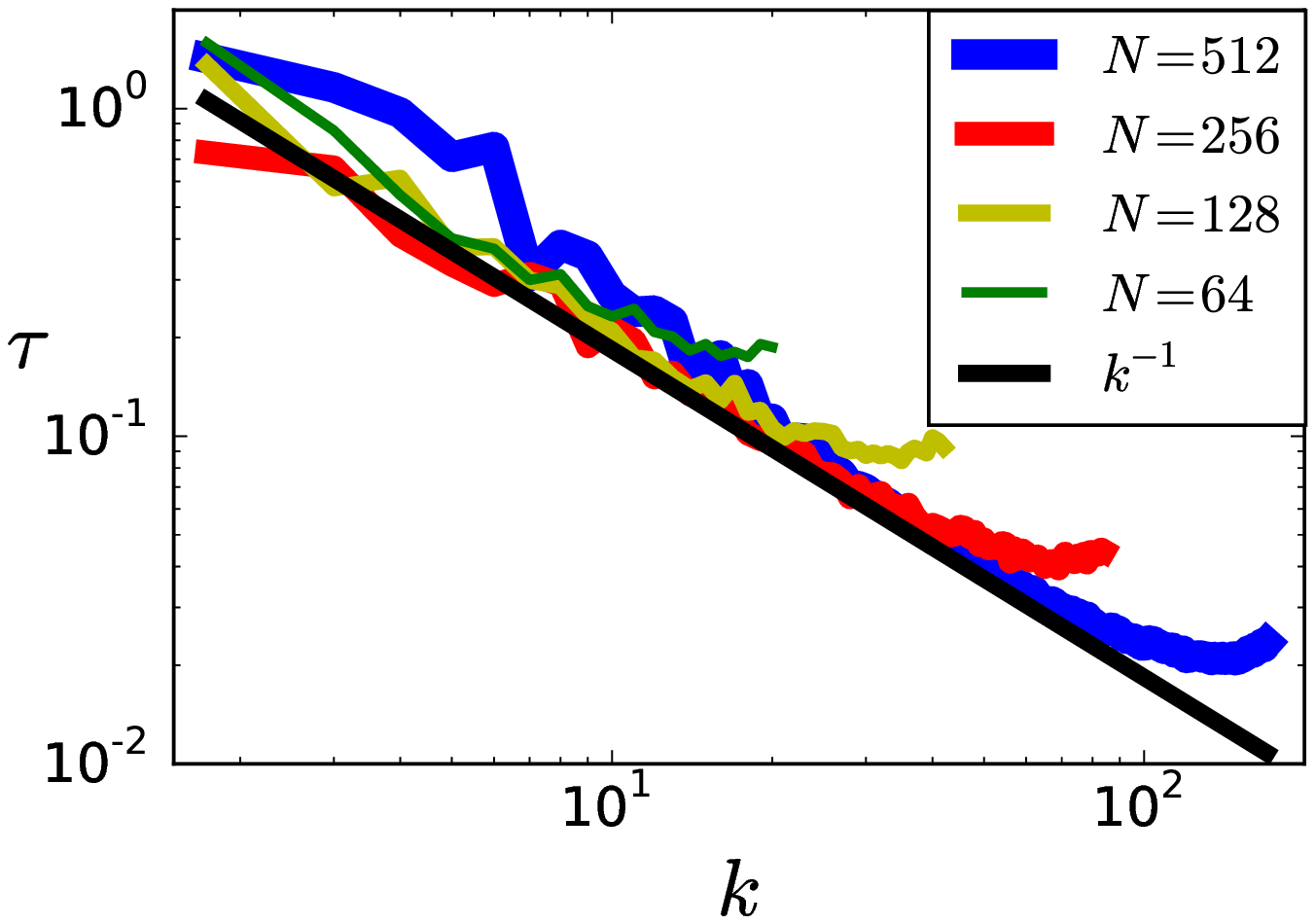}
  \caption[Taylor-Green truncated Euler: time-correlations]
  {Temporal correlation properties of Taylor-Green symmetric DNS of
  the truncated Euler equation. 
  Left) Correlation function as a function of time.
  Right) Correlation time as a function of wavenumber. The $k^{-1}$-power
  law is plotted with the normalization deriving from eq.~\eqref{eq:tau:nH}.
	}
  \label{fig:fig5}
\end{figure}

The correlation functions for different velocity modes are shown in 
the left panel of fig.~\ref{fig:fig5}. The time of each correlation function has 
been re-scaled by the correlation time measured. As shown in the left panel of 
fig.~\ref{fig:fig5}, the correlation functions collapse on the same curve for times 
in the range $0 \leq t \leq \tau_{k}$. Since long time-series are required 
for the correlation function to converge at long times, the curves only collapse for 
$t < \tau_{k} $. 
This collapse at small 
time confirms that the parabolic assumption made in eq.~\eqref{eq:gam:T2} is 
valid for absolute equilibrium solutions of the \TEE{}. Furthermore, the good agreement of 
the curve until $\tau_{k}$ also confirms that the half-height time is a good
proxy to measure the correlation time. The right panel of fig.~\ref{fig:fig5} 
represents the correlation time of an absolute equilibrium solution of the \TEE{}. It 
shows that the correlation time follows a $k^{-1}$-scaling law characteristic of 
an energy-based correlation time. The thermodynamic model developed in the 
previous section is therefore in excellent agreement with the measurements 
carried out with \TG{} symmetric DNS.

\subsection{Inviscid flows with helicity}                              
\label{sec:TE:sub:wtH}                                                         

The next set of DNS of the \TEE{} were carried out using the pseudo-spectral code 
GHOST \cite{mininni_nonlocal_2008,mininni_hybrid_2011}. In these DNS, for every 
wavenumber four real degrees of freedom exist. Indeed, the Craya-Herring 
helical decomposition \cite{craya_contribution_1957,herring_approach_1974}
states that every Fourier mode of velocity can be separated into a positive and 
a negative helical component as expressed in eq.~\eqref{eq:CrHe}. The two helical 
components are modulated by their complex amplitude. Since there is no additional 
restriction on the amplitude of the mode, the velocity modes have four real degrees of 
freedom.

All flows are initialized with the same energy, but different helicity which determines 
the Kraichan number $\Kr{} = - \km{} \beta / \alpha$. This parameter is not present for the \TG{} flows 
and leads to major differences in the global aspect of the PDF of the modes. 
For the same wavenumber, the helical components of helical flows have different energies.

\begin{figure}[!ht]
  \centering
  \includegraphics[width=5.5cm, trim= 0 0 0 0, clip=true]{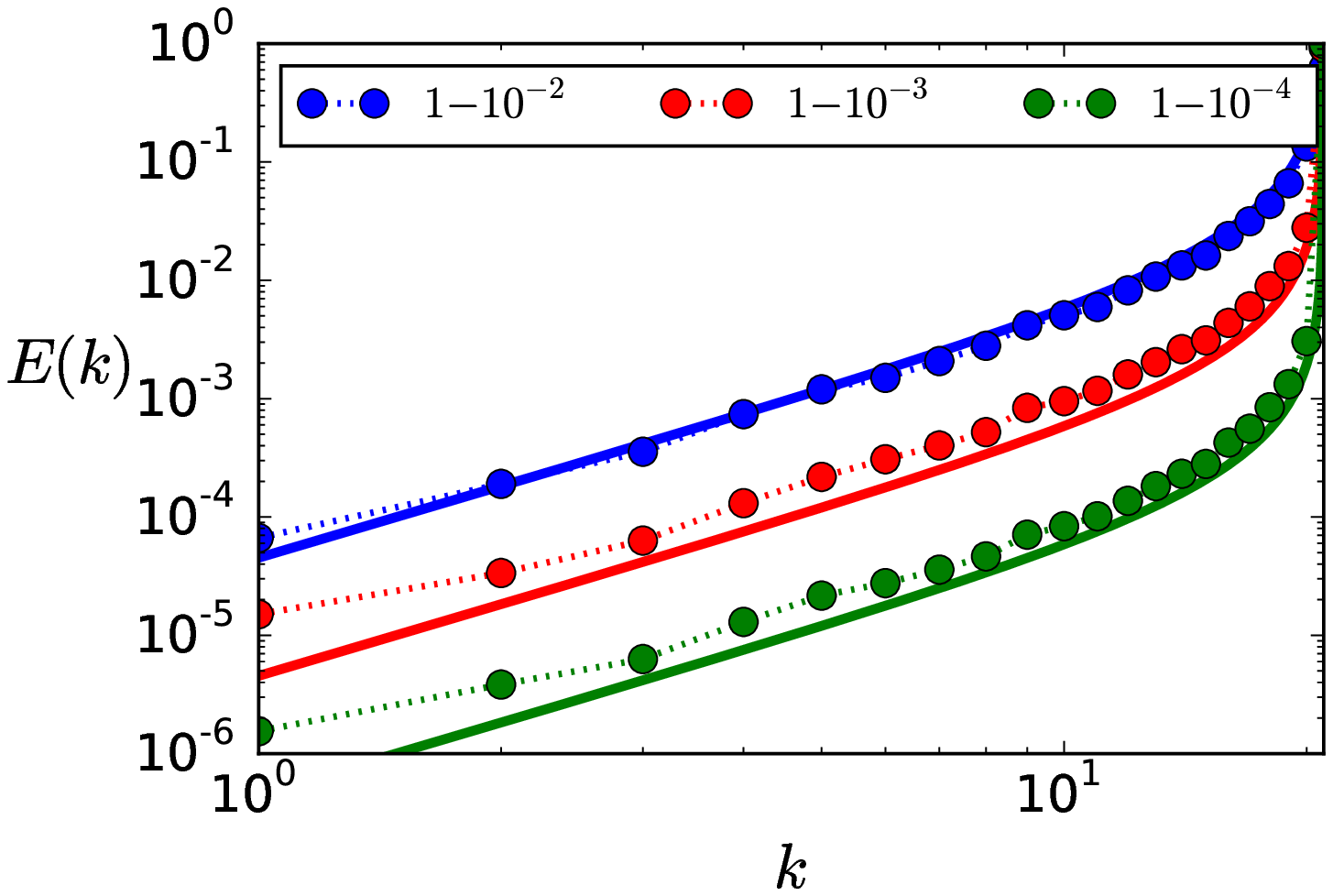}
  \includegraphics[width=5.5cm, trim= 0 0 0 0, clip=true]{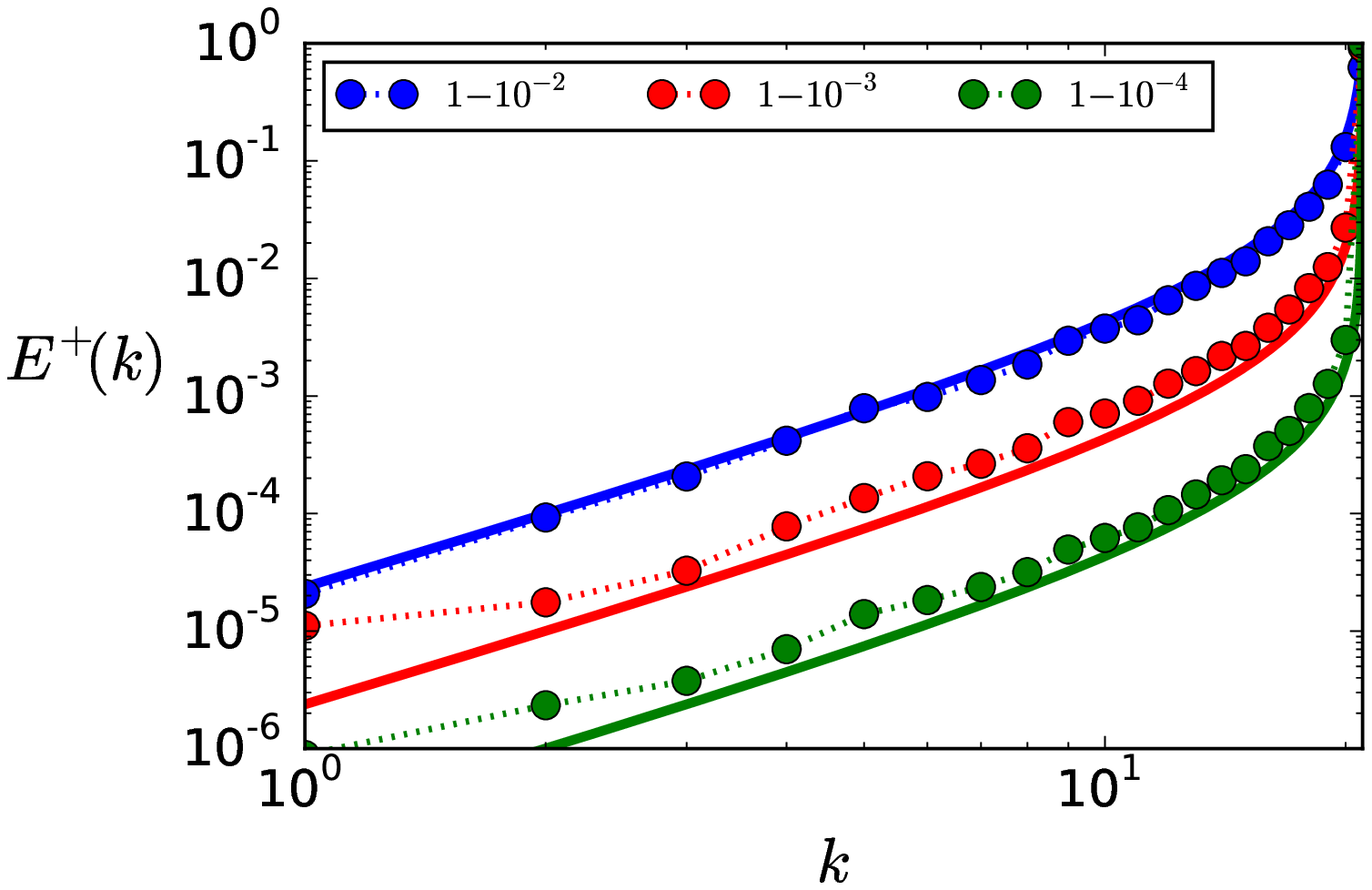}
  \includegraphics[width=5.5cm, trim= 0 0 0 0, clip=true]{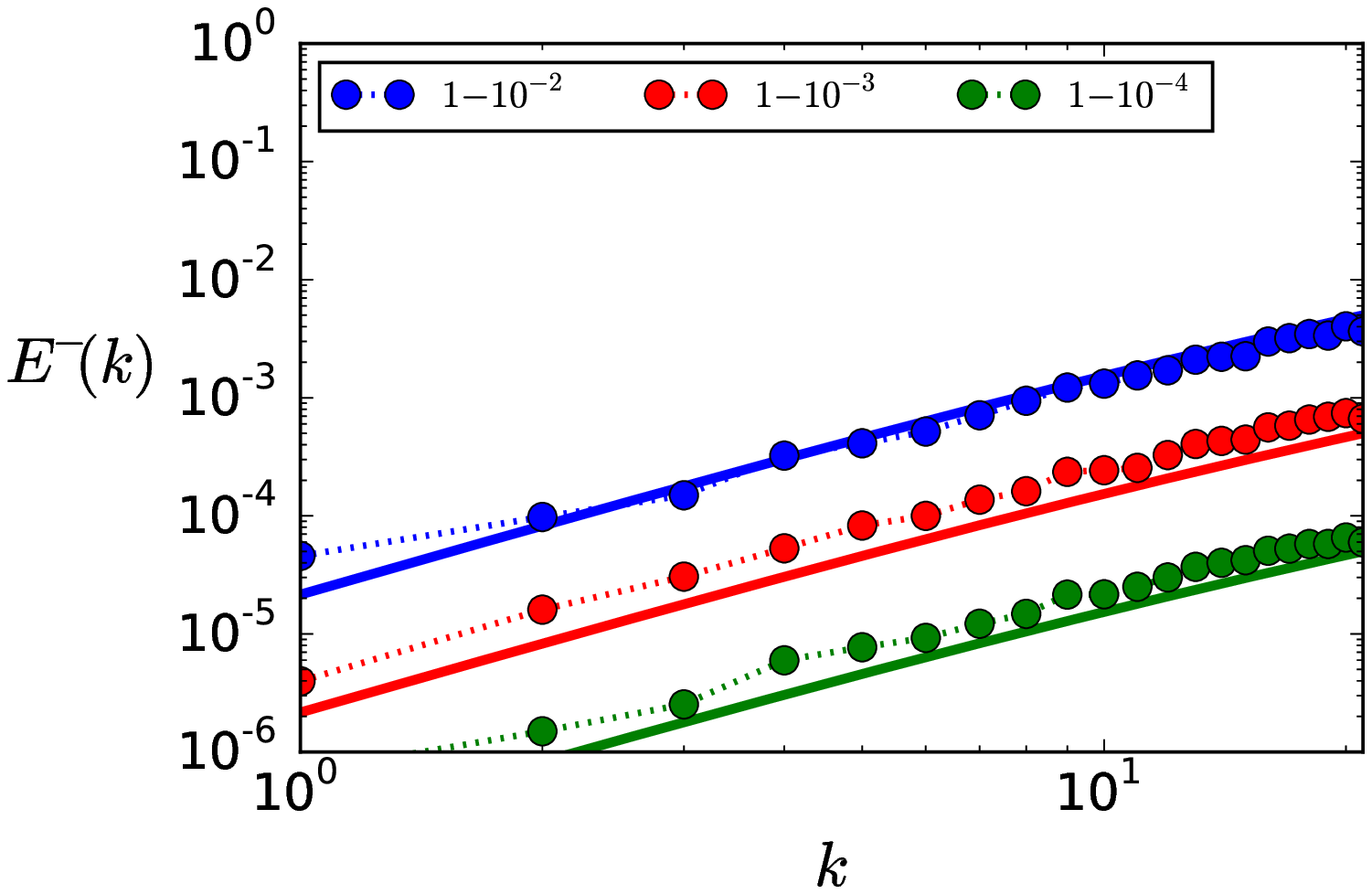}
  \caption[Truncated Euler: energy spectrum v. Kraichnan number]
  {Energy spectrum for different Kraichnan numbers 
  $\Kr{} \in \lbrace 1-10^{-2} ; 1-10^{-3} ; 1-10^{-4} \rbrace$ for 
  general-periodic solutions of the truncated Euler equation. 
  The numeric data is represented with dots and the theoretical prediction with full lines.
	Left) Energy spectrum associated to the total velocity.
	Center) Energy spectrum associated to the positive helical component of the velocity.
	Right) Energy spectrum associated to the negative helical component of the velocity.
  On both panels, the full lines are associated with the energy of the total velocity;
  the dotted lines are associated with the energy of the positive helical component
  of the velocity; and the dashed lines are associated with the energy of the negative 
  helical component of velocity.
	}
  \label{fig:fig6}
\end{figure}

Fig.~\ref{fig:fig6} represents the energy spectrum of the highly helical flows
$\Kr{} = \lbrace 1-10^{-2} ; 1-10^{-3} ; 1-10^{-4} \rbrace$.
The results in the left panel come from DNS and those in the right panel are 
from the absolute equilibrium theory \cite{kraichnan_helical_1973}. Theoretical 
and numerical results match, which confirms the validity of absolute equilibrium 
theory.

\begin{figure}[!ht]
  \centering
  \includegraphics[width=\fwidth, trim= 0 0 0 0, clip=true]{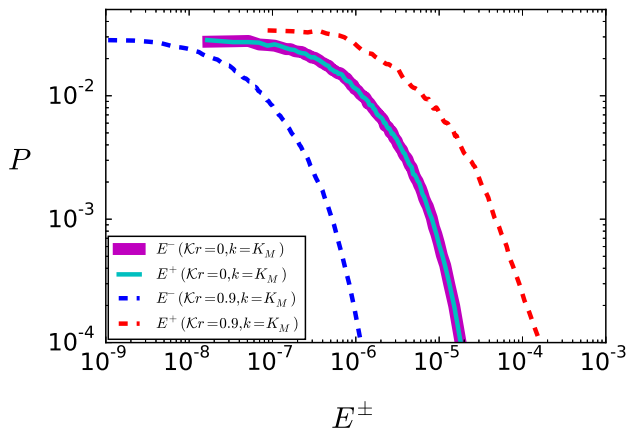}
  \caption[Truncated Euler: impact of helicity on the PDF]
  {Separation of the PDFs due to helicity in solutions of the truncated Euler equation.
  The PDFs have been rescaled to plateau at the same level near zero
  in order to compare the width of the tail of the distribution at large values.  
  At $k=\km{}$, the distributions 
  of probability have an off-set for $\Kr{}=0.9$ because helicity greatly affects the 
  distribution of energy in the small scales.
  }	
  \label{fig:fig7}
\end{figure}

Fig.~\ref{fig:fig7} presents the PDF of the helical components of the velocity of two 
general-periodic flows with different Kraichnan numbers. The first flow has a Kraichnan 
number of zero and consequently does not have any helicity. The second flow has a 
Kraichnan number of $0.9$ and is thus highly helical. For every wavenumber and 
every Kraichnan number, the energy of the positive and negative helical components of 
the velocity follow a $\chi^2_{2}$-distribution. The $\chi^2_{2}$-distribution is characteristic 
of the sum of the square of two independent Gaussian variables. The hypothesis of 
Gaussian-distributed velocity modes is therefore in agreement with the DNS. 
The PDFs of the positive and 
negative helical modes at $k=\km{}$ collapse for the non-helical flow at $\Kr{}=0$ 
but do not collapse for the highly helical flow at $\Kr{}=0.9$. The separation of the 
PDFs at $\Kr{}=0.9$ is consistent with the statistics presented in eq.~\eqref{eq:CrHe}.
All the characteristics of the energy distribution of the DNS are in good agreement 
with the properties predicted by the absolute equilibrium theory.

\begin{figure}[!ht]
  \centering
  \includegraphics[width=5.85cm, trim= 0 0 0 0, clip=true]{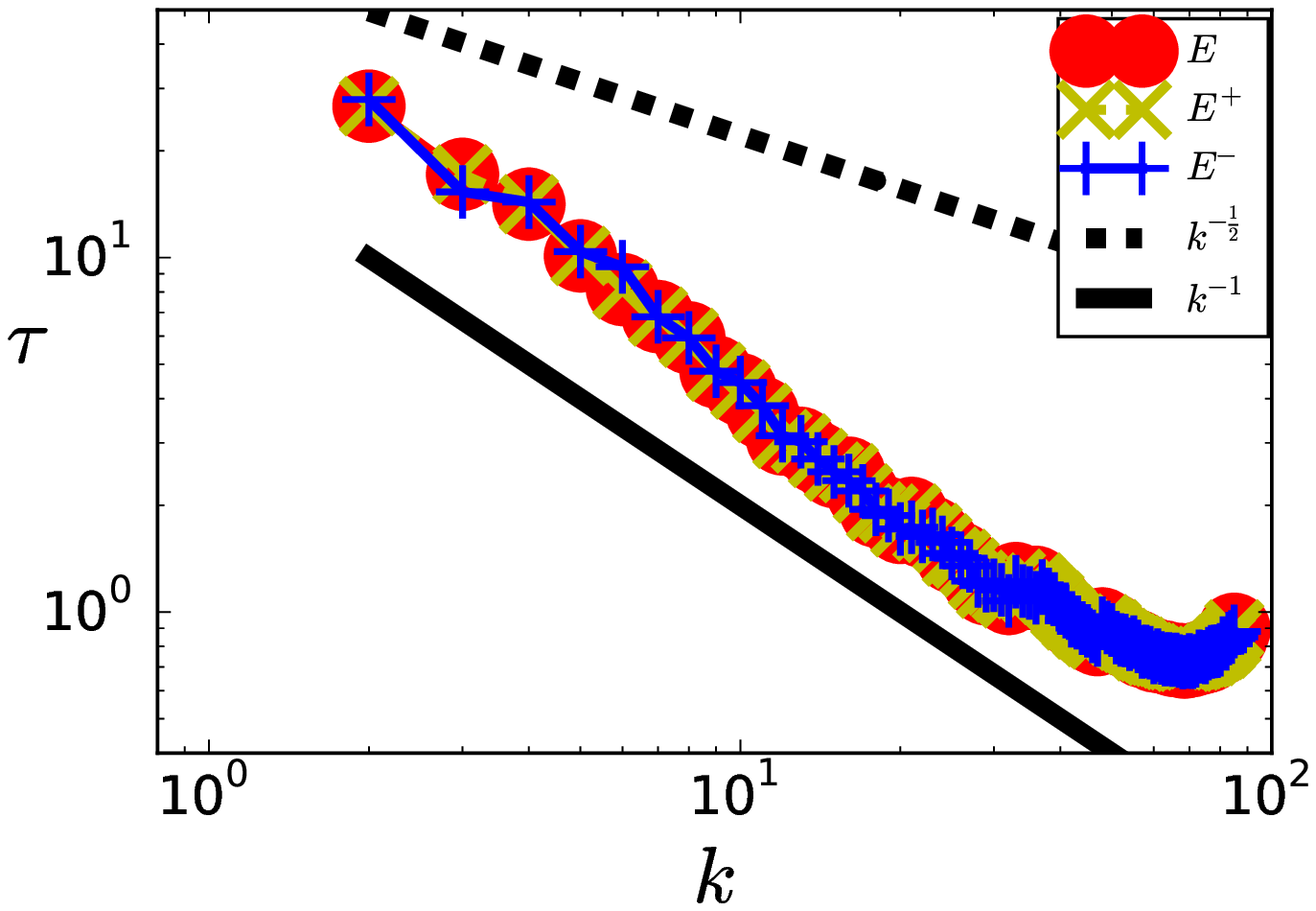}
  \includegraphics[width=5.85cm, trim= 0 0 0 0, clip=true]{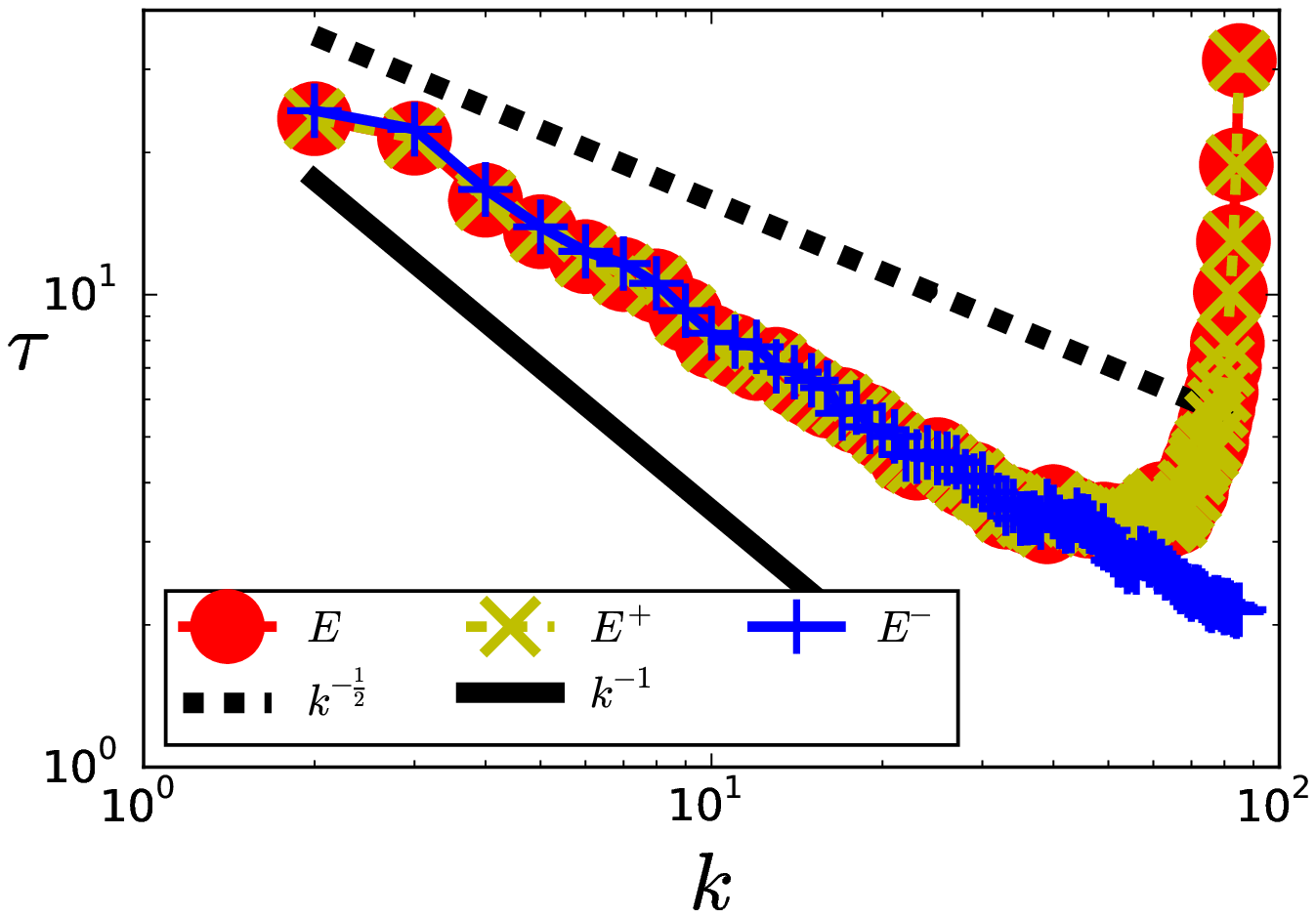}
  \includegraphics[width=5.85cm, trim= 0 0 0 0, clip=true]{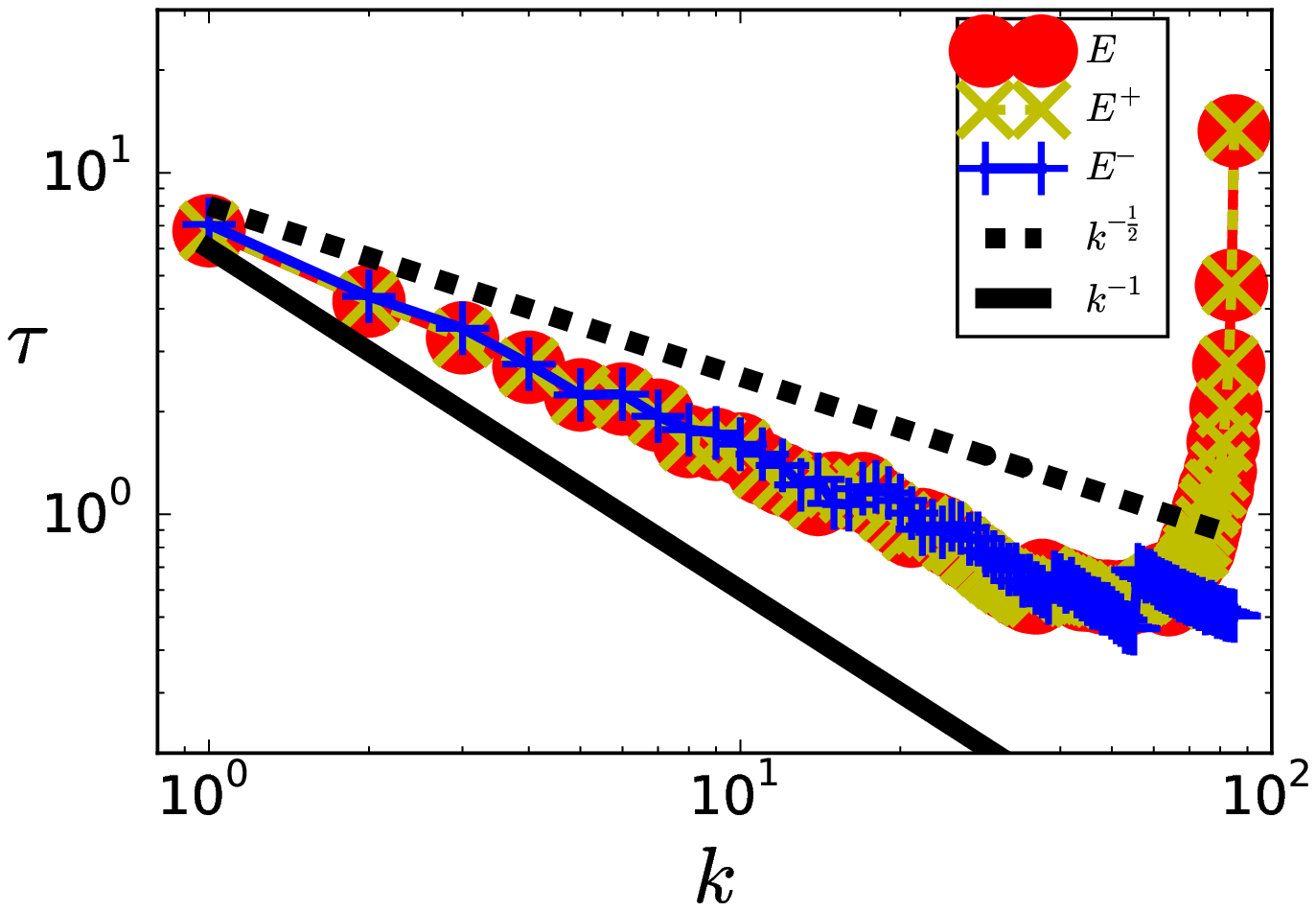}
  \caption[Truncated Euler: correlation time]
  {Correlation time of DNS of general-periodic solutions of 
  the truncated Euler equation. 
  Left) Non-helical flow $\Kr{} = 0$.
  Center) Slightly helical flow $\Kr{} = 0.85$.
  Right) Highly helical flow $\Kr{} = 0.9$.
   }
  \label{fig:fig8}
\end{figure}

The last property that will be examined concerns the evolution of the correlation time as 
the flow becomes highly helical ($\Kr{} \to 1$). The thermalization theory, presented in 
sec.~\ref{sec:TT}, predicts that, as the system becomes highly helical, the energy-based 
correlation time ($\tau^{E}_{k}$ in eq.~\eqref{eq:tau:sclE}) should transition to a 
helicity-based correlation time ($\tau^{H}_{k}$ in eq.~\eqref{eq:tau:sclH}) in intermediate 
scales $ (1-\Kr{})\ll k/\km \ll 1 \,$. With the current computational power, it is difficult to show 
the same number of orders of magnitude as in fig.~\ref{fig:fig1}. Consequently, the results 
presented in fig.~\ref{fig:fig8} are not able to show both scaling laws for a given \Kr{}. The left 
panel of fig.~\ref{fig:fig8} presents the correlation time of a non-helical flow at $\Kr{} = 0$ 
while the right panel of fig.~\ref{fig:fig8} presents the correlation time of a highly helical flow 
at $\Kr{} = 0.9$. The correlation time of the non-helical flow exhibits a $k^{-1}$-scaling law
characteristic of an energy-based correlation time, while the correlation time of the highly 
helical flow is closer to a $k^{-\frac{1}{2}}$-scaling law characteristic of a helicity-based. 
Additionally, in the right panel of fig.~\ref{fig:fig8}, the correlation time peaks in the small scales,
which is characteristic of the localization of the energy near \km{} of helical flows. In the 
highly helical case, the difference between the positive and negative helical components 
of the velocity can also be observed in the small scales. While both helical components collapse 
in the non-helical case, they are different in the small scales in the highly helical case. All
these observations are in agreement with the correlation time predicted by the thermalization theory. 
DNS were also performed at higher Kraichnan numbers and showed a persistence of the 
helicity-based correlation time.

Even though the transition of the scaling law is hard to observe in the DNS carried out, 
results indicate that the transition of correlation time regime occurs for Kraichnan numbers 
in the range $0.8 \leq \Kr{} \leq 0.9$. The center panel of fig.~\ref{fig:fig8} shows that for 
slightly less helical flows with $\Kr{}=0.85$, the correlation follows a power law with an exponent
between $-1$ and $-\frac{1}{2}$. A clear visualization of the transition is demanding in
computational power, since it occurs on nearly one order of magnitude in the left panel of 
fig.~\ref{fig:fig1}.

\section{Navier-Stokes DNS}                               
\label{sec:FNS}                                                      

Because DNS of the \NSE{} must have a converged spectrum at wavenumbers larger 
than the forcing wavenumber, their properties cannot be 
assessed with scale separations as large as those of DNS of the \TEE{}. 
Using the same codes as in the previous section, we now turn to the study of how 
the PDF, the standard deviation and the correlation time behave in the 
large scales for DNS of the \NSE{}. We will also compare
these results with the observation made on DNS of the \TEE{}.

\subsection{No helicity: Taylor-Green flows}                     
\label{sec:FNS:sub:NoH}                                                      

\TG{} flows are studied with different forcing wavenumbers 
$\kf{} \in \lbrace 11\sqrt{3} ;  35\sqrt{3} ; 59\sqrt{3}\rbrace$. The forcing was imposed on 
the flow by fixing to $0.125$ the amplitude of the odd modes 
$[ \kf{} , \kf{} , \kf{} ]/\sqrt{3}$. The other parameter of the system 
is the viscosity \visco{} which was also adjusted to reach 
a turbulent regime. To compare the properties of flows at different
forcing wavenumbers, the Reynolds number $Re = U / (\visco{} \kf{})$
was set to $6.56$. 

\begin{figure}[!ht]
  \centering
  \includegraphics[width=8cm, trim= 0 0 0 0, clip=true]{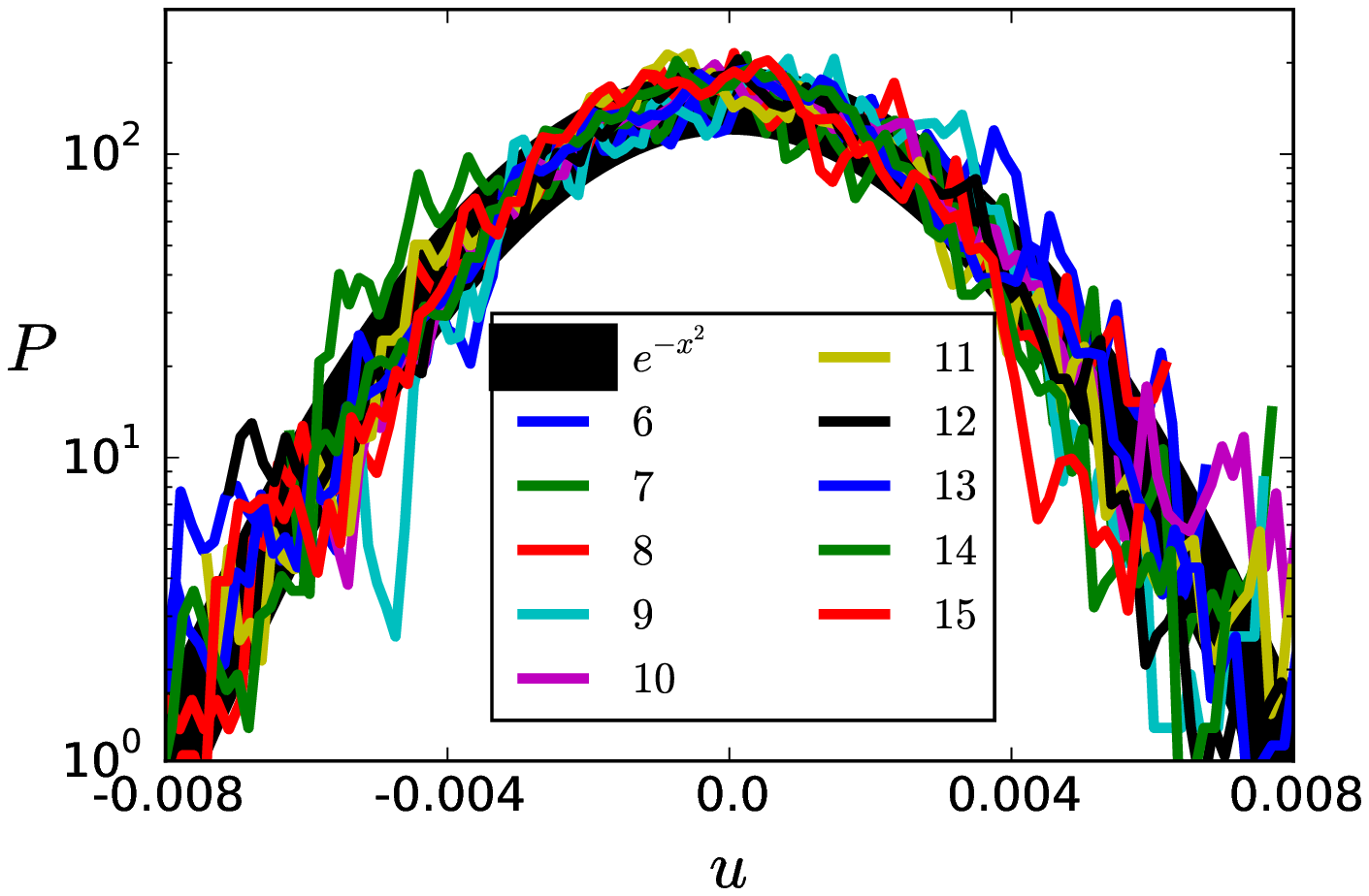}
  \includegraphics[width=8cm, trim= 0 0 0 0, clip=true]{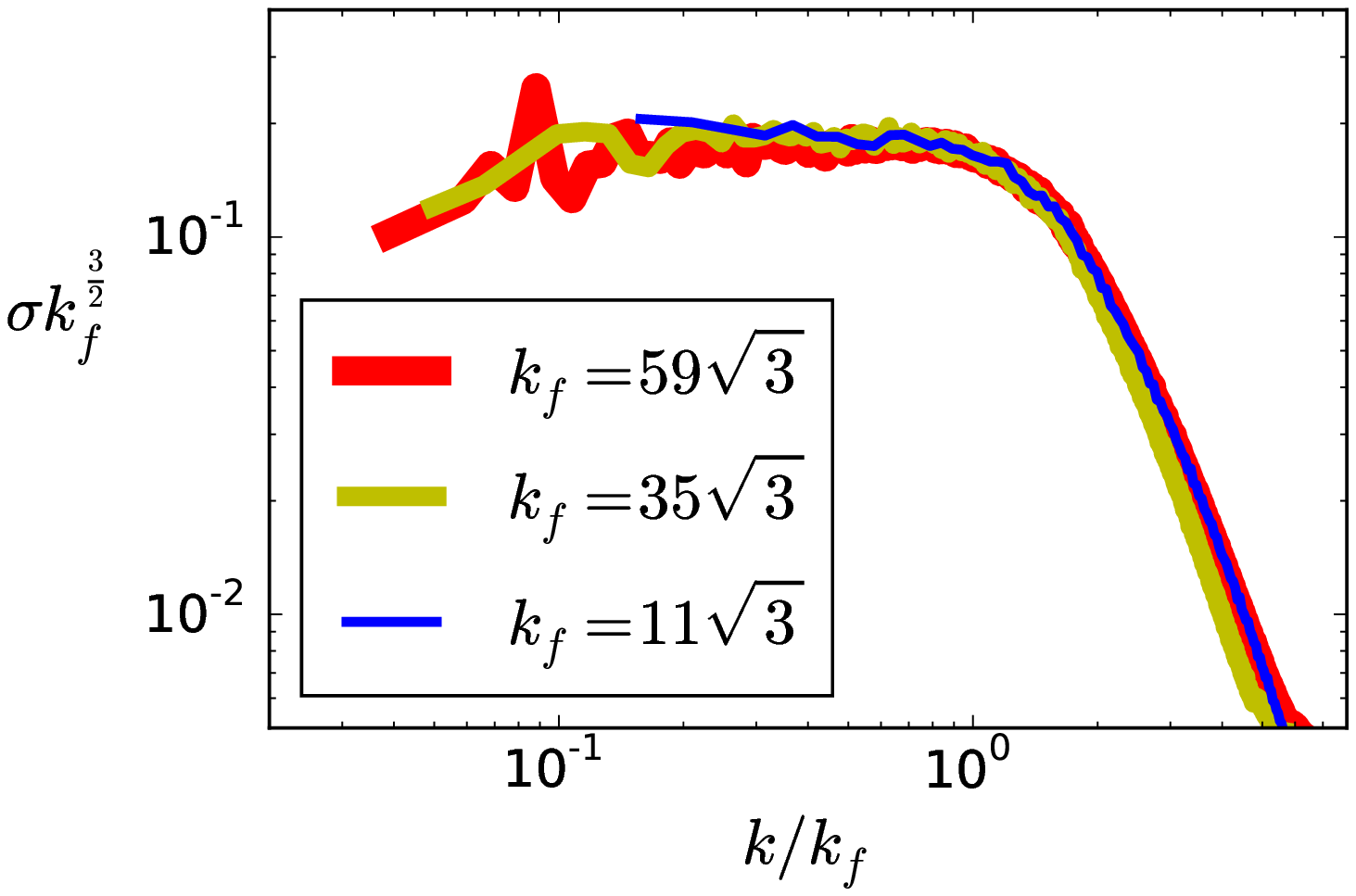}
  \caption[Taylor-Green Navier-Stokes: PDF]
  {PDF of modes of Taylor-Green symmetric DNS of the forced 
  Navier-Stokes equation with a forcing wavenumber at $\kf{}=11\sqrt{3}$.
  Left) PDF of even modes with one degree of freedom at scales larger than the forcing scale.
  Right) Standard deviation of the velocity modes. 
}
  \label{fig:fig9}
\end{figure}

Fig.~\ref{fig:fig9} represents some properties of the PDF of velocity modes with one 
degree of freedom in the six planes $k_x \in  \lbrace 0 ; 1 \rbrace$,  
$k_y \in  \lbrace 0 ; 1 \rbrace$ or $k_z\in  \lbrace 0 ; 1 \rbrace$. These PDFs have 
a clear Gaussian behavior as highlighted by their parabolic shape on the semi-logarithmic 
plot in the left panel of fig.~\ref{fig:fig9}. The datasets in the left panel, 
representing modes with $k<k_f$, collapse on the same curve, which indicates that 
their standard deviation is identical. 
These two trends are also observed in the right panel of fig.~\ref{fig:fig9}, 
where the standard deviation of the PDFs are represented for different forcing scales. 
In the right panel of fig.~\ref{fig:fig9}, the standard deviations, $\sigma$, are compensated 
by a factor $\kf^{3/2}$ to take into account that the total energy is spread out on more 
modes as the forcing wavenumber increases. The wavenumber is also rescaled by the
forcing wavenumber in order to compare the results. Using this scaling, the datasets at 
different resolutions collapse on the same curve. At large scales, the compensated 
standard deviation plateaus, indicating an analogue of the equipartition in energy of 
absolute equilibrium without helicity $\Kr{}=0$. Below the forcing scale, the standard 
deviation rapidly decreases because of the forward cascade and viscosity.

\begin{figure}[!ht]
  \centering
  \includegraphics[width=\fwidth, trim= 0 0 0 0, clip=true]{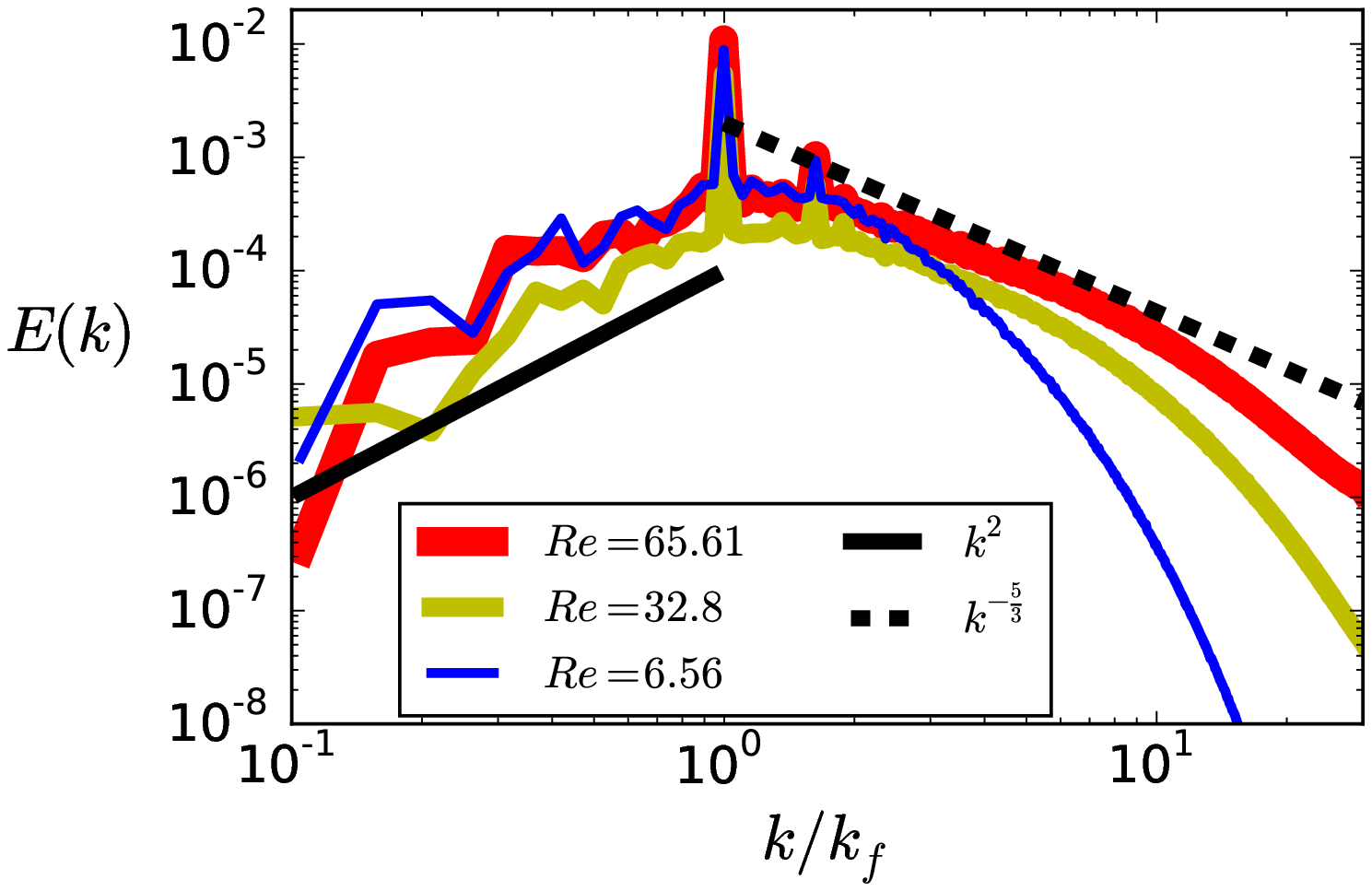}
  \includegraphics[width=\fwidth, trim= 0 0 0 0, clip=true]{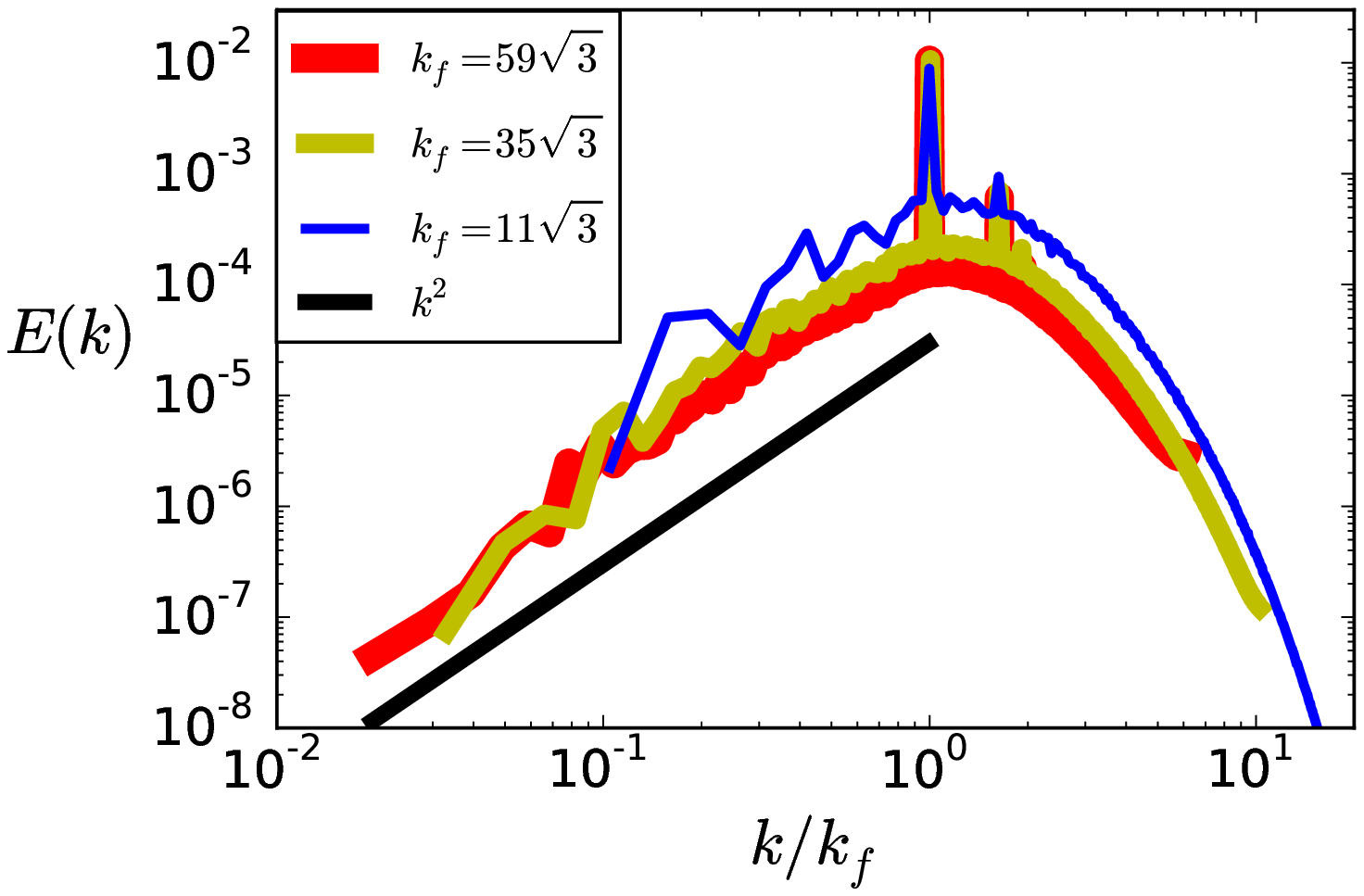}
  \caption[Taylor-Green Navier-Stokes: energy spectrum]
  {Energy spectrum of Taylor-Green symmetric flows solutions of
  the forced Navier-Stokes equation.
  Left) Fixed forcing mode $\kf{}=11\sqrt{3}$ at different Reynolds numbers 
  $\Rey{} \in \lbrace 11\sqrt{3} ; 35\sqrt{3} ; 59\sqrt{3} \rbrace$.
  Right) Fixed Reynolds number $Re=6.56 $ at different 
  forcing modes $\kf{} \in \lbrace 11\sqrt{3} ; 35\sqrt{3} ; 59\sqrt{3} \rbrace$. }
  \label{fig:fig10}
\end{figure}

Energy equipartition in the large scale modes can directly be observed in the right panel of 
fig.~\ref{fig:fig9}. It can also indirectly be observed in the energy spectrum presented 
in fig.~\ref{fig:fig10}. Since energy shells contain a number of modes proportional to 
their surface, $ 4 \pi k^2$, systems satisfying equipartition in energy should have an 
energy spectrum proportional to $k^2$. The spectra presented on both panels of 
fig.~\ref{fig:fig10} are thus consistent with equipartition in energy. 
In left panel of fig.~\ref{fig:fig10}, the energy spectrum is presented at different 
Reynolds numbers for a fixed forcing wavenumber $\kf{}=11\sqrt{3}$. At the largest Reynolds 
number, the energy spectrum reaches the Kolmogorov's $k^{-5/3}$-scaling in the 
inertial range and has also the equipartition $k^2$-scaling in the thermalization 
domain. Even though Kolmogorov's scaling is not present in the other curves with 
smaller Reynolds numbers, the equipartition scaling law is still observable in the 
large scales. The smallest Reynolds number was then used to compute the DNS of 
the right panel of fig.\ref{fig:fig10} where the wavenumber varies at fixed
Reynolds number. The equipartition scaling of the energy spectrum can be observed 
on the three forcing wavenumbers used. The curve with the largest forcing scale, 
$\kf{} = 59\sqrt{3}$, shows that the energy spectrum follows a $k^2$-scaling for nearly two 
decades.

\begin{figure}[!ht]
  \centering
  \includegraphics[height=\fheight, trim= 0 0 0 0, clip=true]{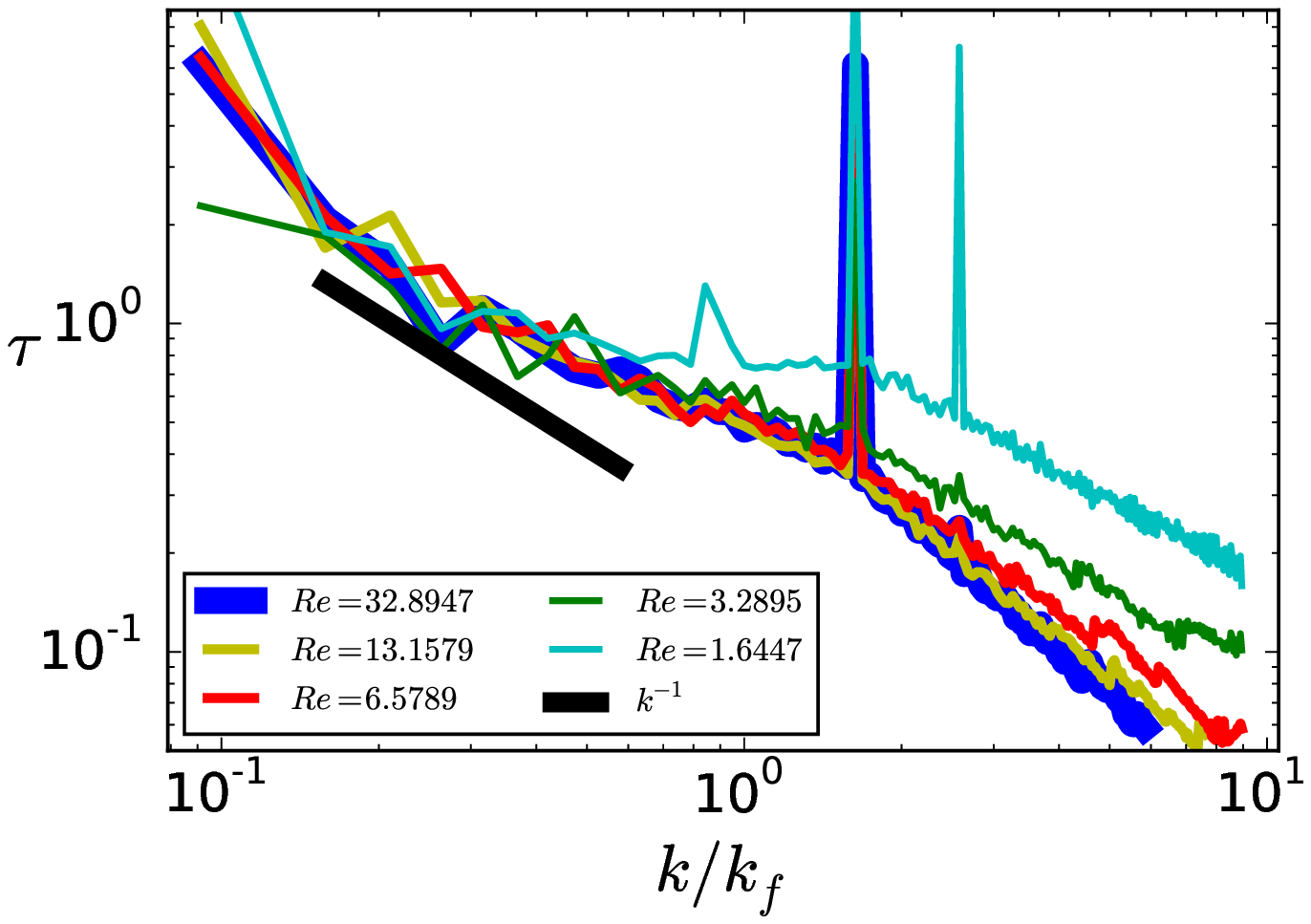}
  \includegraphics[height=6.15cm, trim= 0 0 0 0, clip=true]{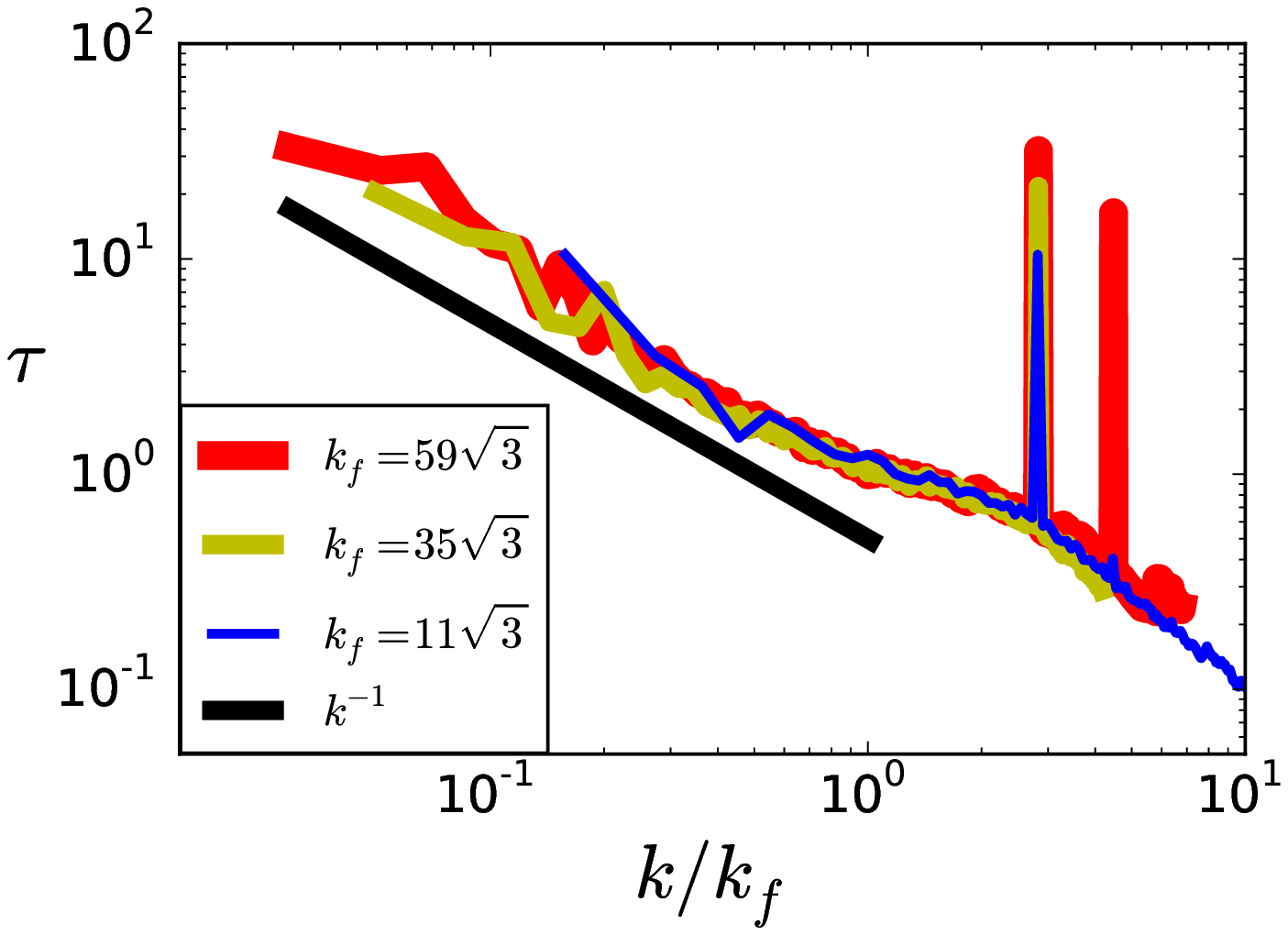}
  \caption[Taylor-Green Navier-Stokes: correlation time]
  {Correlation time of Taylor-Green symmetric DNS of the 
  forced Navier-Stokes equation.
  Left) Different Reynolds numbers at fixed forcing wavenumber $\kf{} =11\sqrt{3}$.
  Right) Different forcing wavenumber $\kf{} \in \lbrace 11\sqrt{3} ; 35\sqrt{3} ; 59\sqrt{3} \rbrace$
  at fixed Reynolds number.
}
  \label{fig:fig11}
\end{figure}

We now turn to the temporal correlation of the flows presented in 
fig.~\ref{fig:fig11}. The left panel shows the dependence on viscosity of the correlation 
time for flows forced at $\kf=11\sqrt{3}$. The correlations are computed using the algorithm 
presented in subsec.~\ref{sec:TE:sub:tau}. In the small scales, the correlation time 
slowly decreases as viscosity decreases whereas the correlation time rapidly stabilizes 
on a $k^{-1}$-power law in the large scales. The major peak observed in the flow does 
not correspond to the forcing wavevector which is not located on the planes used to 
compute the correlation time. This peak corresponds to a harmonic of the forcing 
located in one of the planes used in the correlation time procedure. 
At high Reynolds number, in the large scales, the correlation 
time aligns on a curve, which is consistent with the energy-based correlation time 
$k^{-1}$-scaling. 

The $k^{-1}$-scaling law of the correlation time can be observed on the right panel 
of fig.~\ref{fig:fig11} which shows the correlation time for three scale separations 
$\kf{} = \lbrace 11\sqrt{3} ; 35\sqrt{3} ; 59\sqrt{3} \rbrace$. The Reynolds number used 
in these DNS is based on the smallest viscosity used in the left panel of fig.~\ref{fig:fig11}. The 
correlation times for the three scale separations collapse on the $k^{-1}$-power law. 
The data at the largest forcing wavenumber $\kf{}=59\sqrt{3}$ shows a trend which strongly 
agrees with the energy-based correlation time $k^{-1}$-scaling.

\subsection{Helical flows: \ABC{} flows}                           
\label{sec:FNS:sub:wH}                                                       

In order to study the impact of helicity on the velocity modes in large scales, 
DNS were carried out on general-periodic flows solutions of the \NSE{} given
at eq.~\eqref{eq:NS} with an $\ABC{}$ forcing \cite{dombre_chaotic_1986}
\begin{align}
	F^{\ABC{}}_x = F_{0} ( C \sin \kf{} z + B \cos \kf{} y )
	\; , \;
	F^{\ABC{}}_y = F_{0} ( A \sin \kf{} x + C \cos \kf{} z )
	\; , \;
	F^{\ABC{}}_z = F_{0} ( B \sin \kf{} y + A \cos \kf{} y ) \, .
	\label{eq:ABC}
\end{align}
where $F_{0}$ is the intensity of the forcing. The three dimensionless parameters 
$A$, $B$ and $C$ were set to one. The main characteristic 
of $\ABC{}$ flows is their Beltrami property: $\roT F^{\ABC{}} = \kf{} F^{\ABC{}}$. This 
property makes them exact (but in general unstable) solutions of the \TEE{}. All the \ABC{} DNS presented are 
done at $\kf{}=20$.
A non-helical variant of the \ABC{} forcing, that we will be referred to as the \CBA{} forcing, can 
be built by switching the sine components of the \ABC{} forcing to cosine components. 
\begin{align}
	F^{\CBA{}}_x = F_{0} ( C \cos \kf{} z + B \cos \kf{} y )
	\; , \;
	F^{\CBA{}}_y = F_{0} ( A \cos \kf{} x + C \cos \kf{} z )
	\; , \;
	F^{\CBA{}}_z = F_{0} ( B \cos \kf{} y + A \cos \kf{} y ) \, .
	\label{eq:CBA}
\end{align}
The \CBA{} forcing has already been used as a non-helical reference of the \ABC{} flow in 
\cite{cameron_fate_2016}. At fixed coefficients, the \CBA{} forcing has the same energy as 
the \ABC{} forcing.

\begin{figure}[!ht]
  \centering
  \includegraphics[height=4.cm, trim= 0 0 0 0, clip=true]{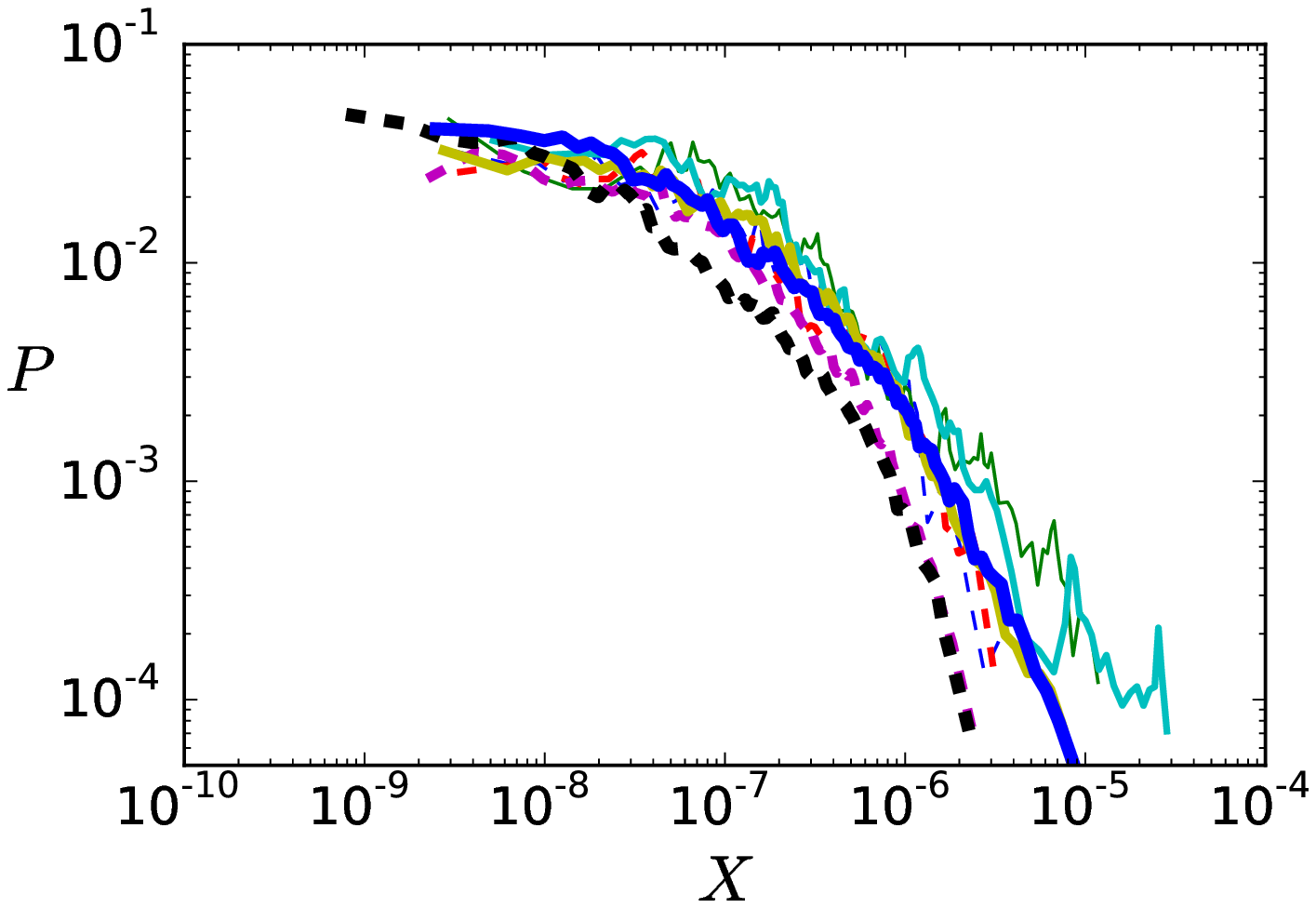}
  \includegraphics[height=4.cm, trim= 0 0 0 0, clip=true]{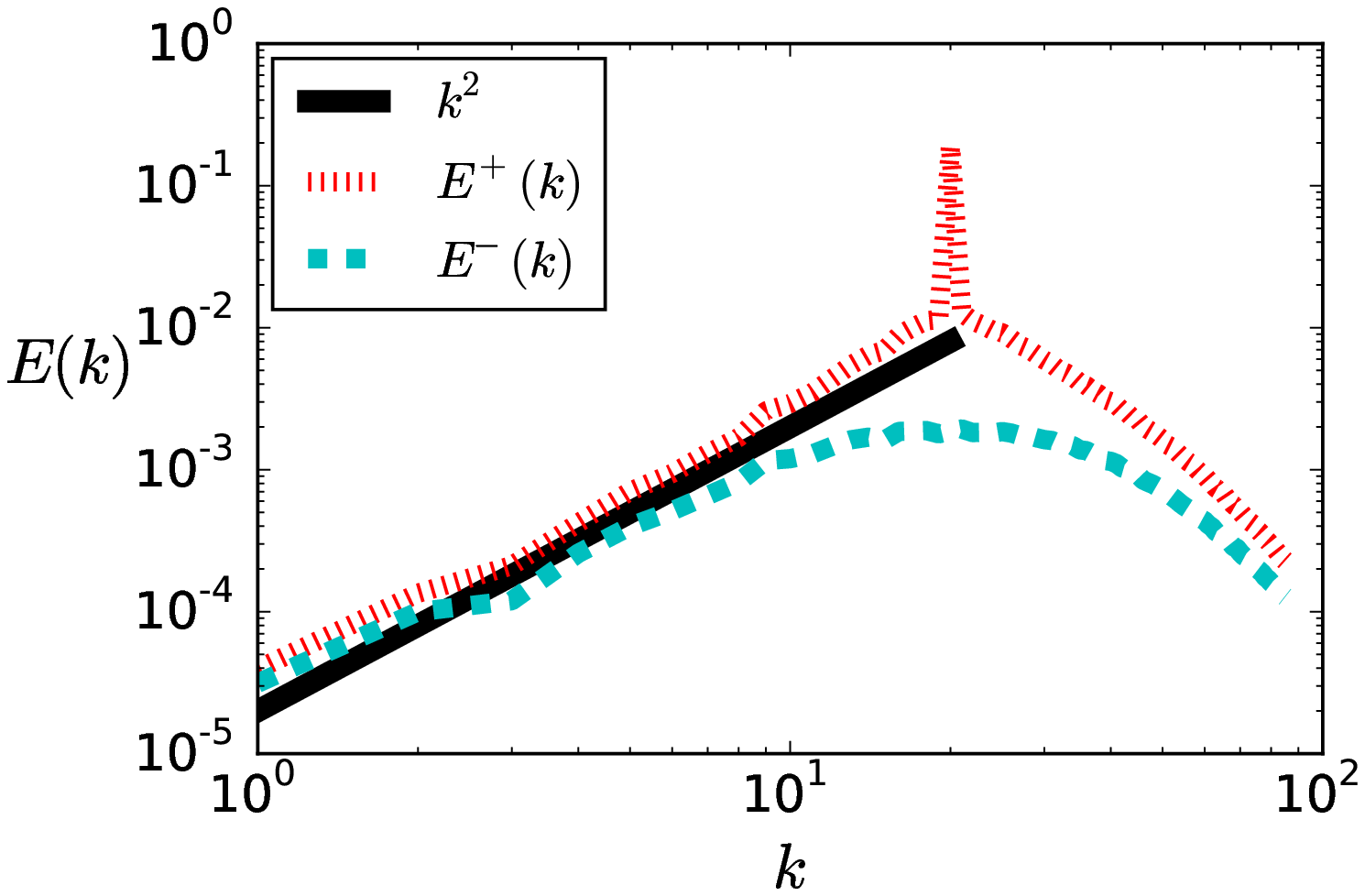}
  \includegraphics[height=4.cm, trim= 0 0 0 0, clip=true]{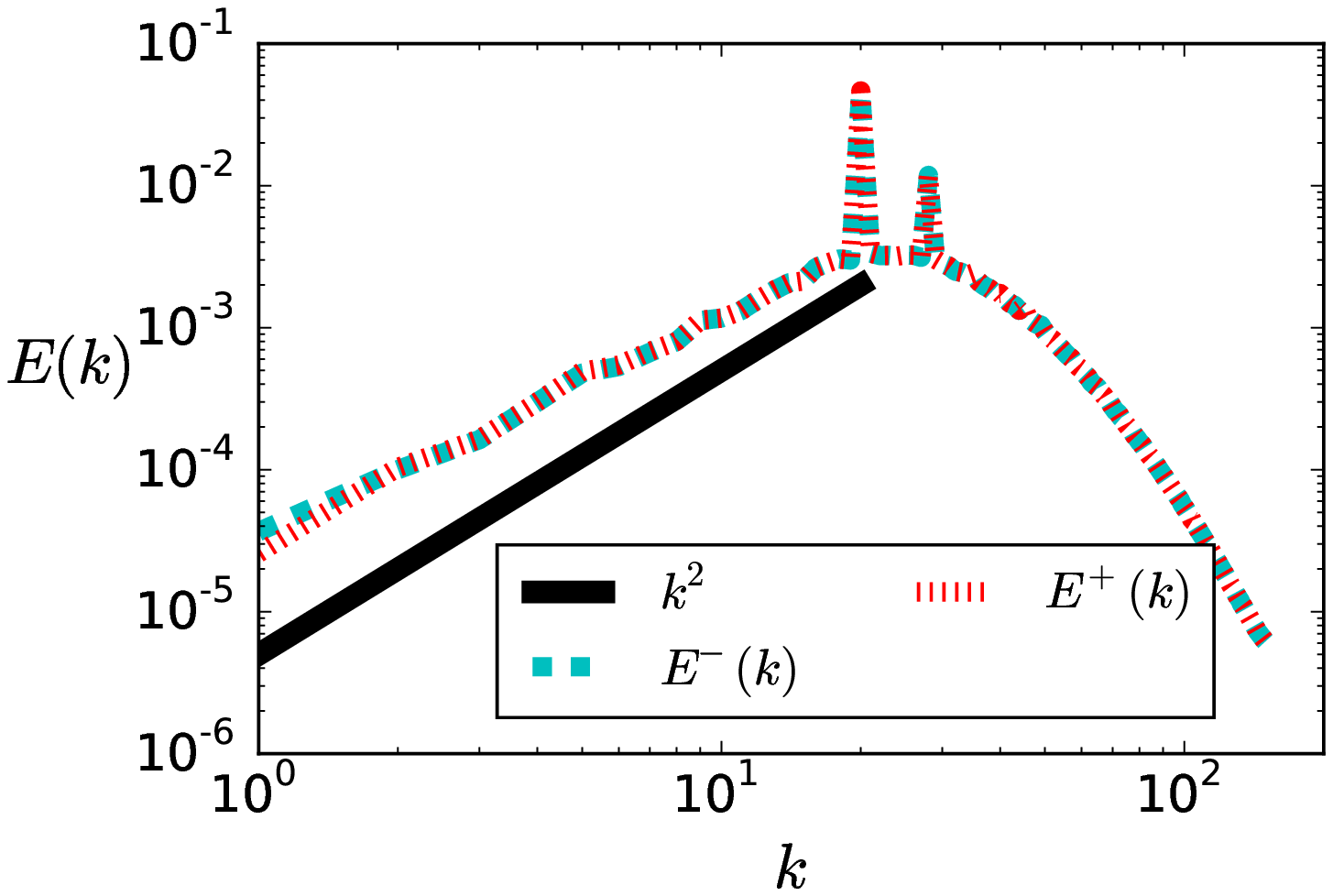}
  \caption[Navier-Stokes: PDF and energy spectrum]
  {Statistical characteristics of general-periodic DNS of the 
  Navier-Stokes equation.
  Left) \ABC{} forcing, PDF of modes before the forcing scale. The full lines 
  represent the positive helical components of the velocity and the dashed lines 
  represent the negative helical components of the velocity.
  Center) \ABC{} forcing, energy spectrum of positive and negative helical 
  components of the velocity.
  Right) \CBA{} forcing, energy spectrum of positive and negative helical 
  components of the velocity.
}
  \label{fig:fig12}
\end{figure}

The PDFs of the positive and negative helical components of the modes in the large scales are 
presented in the left panel of fig.~\ref{fig:fig12} for DNS of the \NSE{} with an \ABC{} forcing. The 
positive helical component of the velocity is represented with full lines and the negative helical 
component of velocity is represented with dashed lines. All PDFs plateau near zero and have an 
fast decay at high values. Because general-periodic DNS 
do not reach the same scale separation and Reynolds number as \TG{} symmetric DNS, the 
comparison with the $\chi^2_2$-distribution is not as clear as in fig.~\ref{fig:fig9}.
The PDFs of the two helical components do not have the same exponential tail as observed in 
absolute equilibrium solutions of the \TEE{}. Since solutions of the \NSE{} are not as helical 
as the solution of the \TEE{} considered in the previous section, the separation of the tail of 
the distributions is not as wide as in the case of absolute equilibrium solutions of the \TEE{}.
The center panel of fig.~\ref{fig:fig12} represents the energy spectrum of the two helical 
components of DNS of the \NSE{} with an \ABC{} forcing. The positive helical component 
of the velocity has more energy than its negative counterpart, which is consistent with the 
separation of the tail of the distribution presented in the left panel. Both components follow 
a $k^2$-scaling consistent with equipartition. In the large scales, the general features of the 
modes match the properties of absolute equilibrium solutions of the \TEE{}. 
The right panel of fig.~\ref{fig:fig12} represents the energy spectrum of the two helical 
components of DNS of the \NSE{} with a \CBA{} forcing. The energy spectrum is not as close 
to the $k^2$-scaling as the energy spectrum resulting from an \ABC{} forcing. We note that the 
energy spectrum of the non-helical \TG{} symmetric flows, presented in the right panel of 
fig.~\ref{fig:fig10}, deviates from the $k^2$-scaling at the smallest scale separations. The 
convergence study carried out in the right panel of fig.~\ref{fig:fig4} 
indicates that such a deviation can happen at small scale separations.

\begin{figure}[!ht]
  \centering
  \includegraphics[height=5.8cm, trim= 0 0 0 0, clip=true]{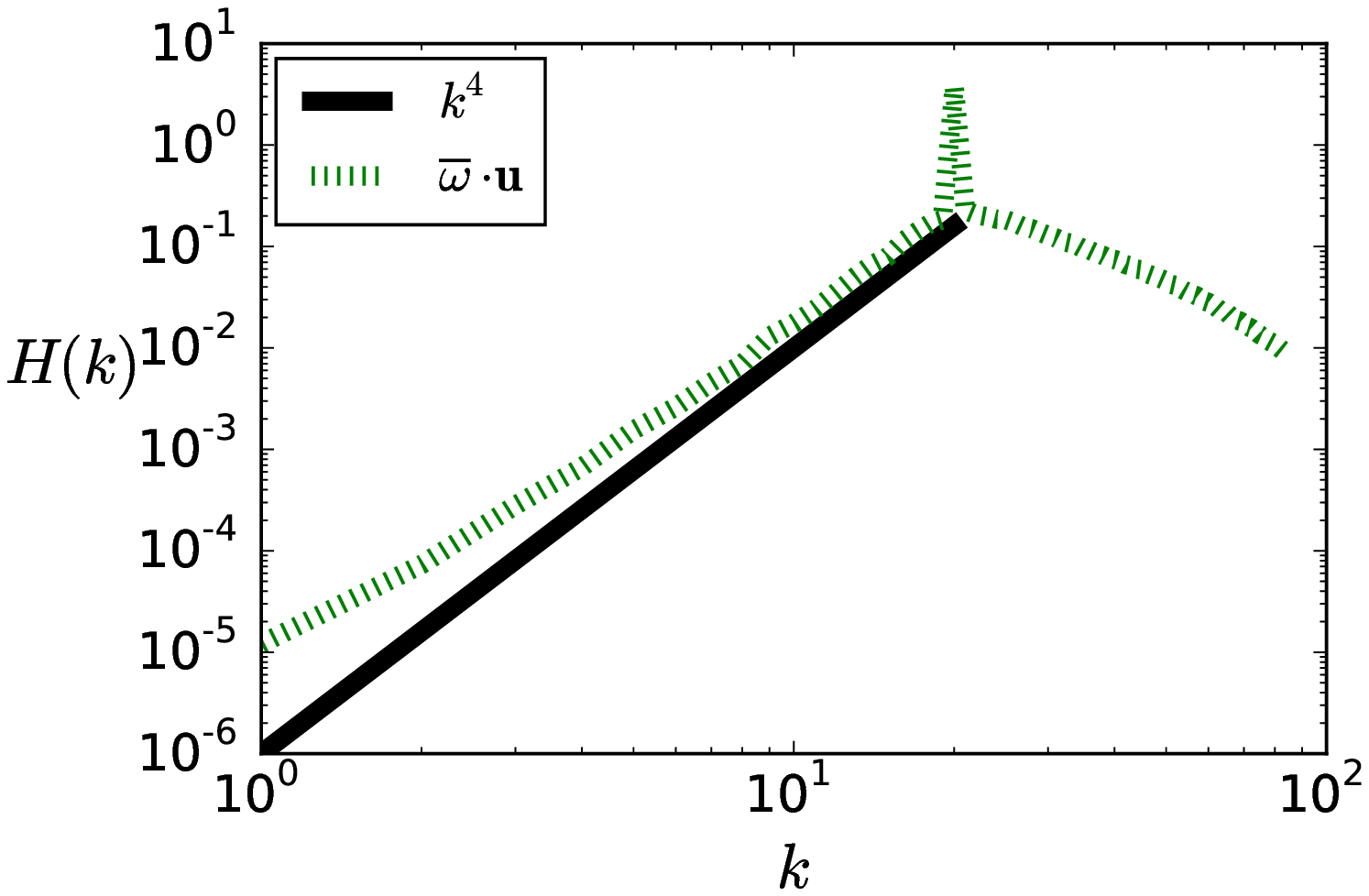}
    \begin{tikzpicture}
        \node[anchor=south west,inner sep=0] (image) at (0,0) 
        { \includegraphics[height=5.8cm, trim= 0 0 0 0, clip=true]{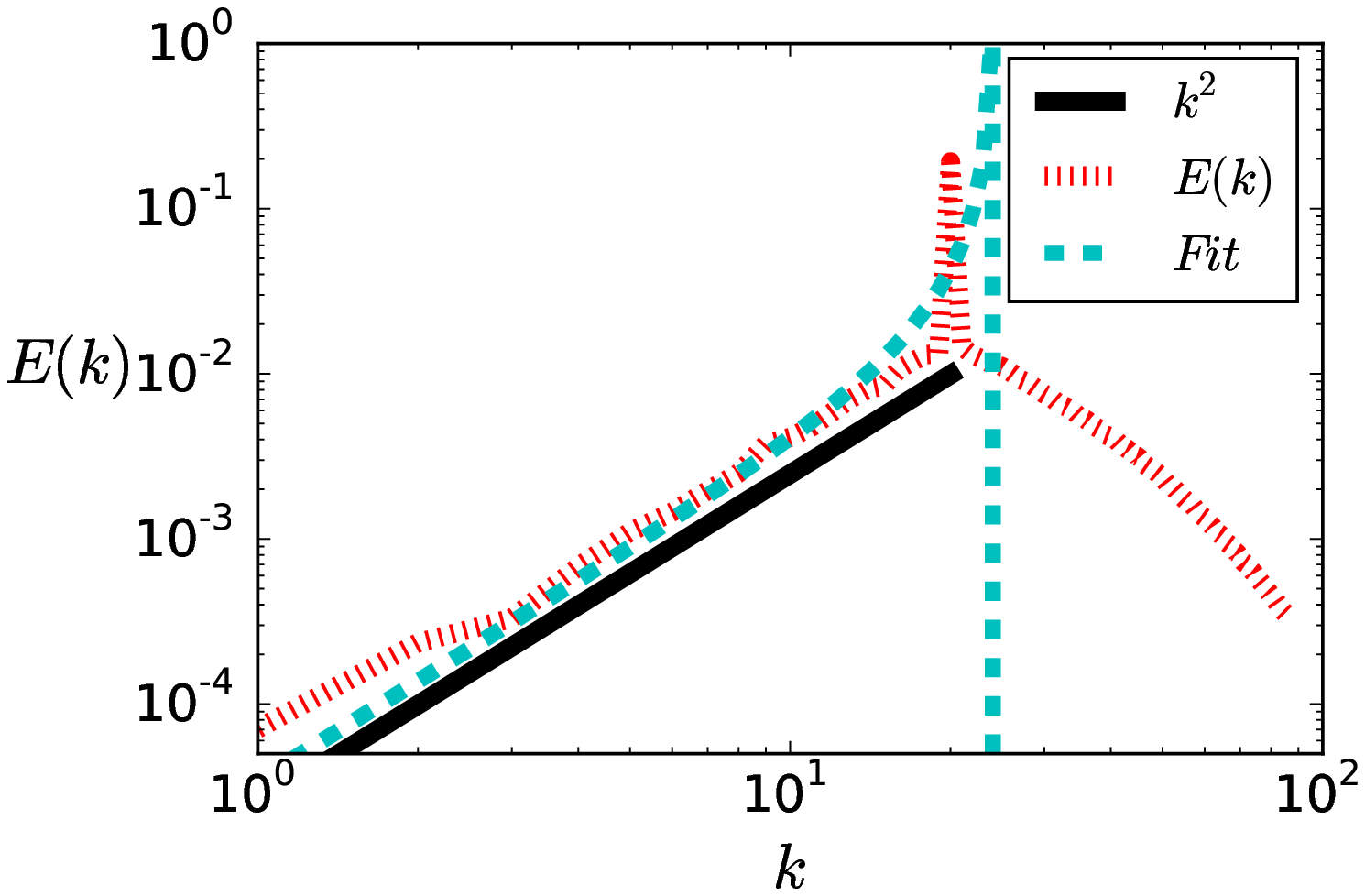}};
        \begin{scope}[x={(image.south east)},y={(image.north west)}]
            \node[anchor=south west,inner sep=0] (image) at (0.2,0.55){
			 \includegraphics[width=3.45cm,trim= 5 10 5 4, clip=true]{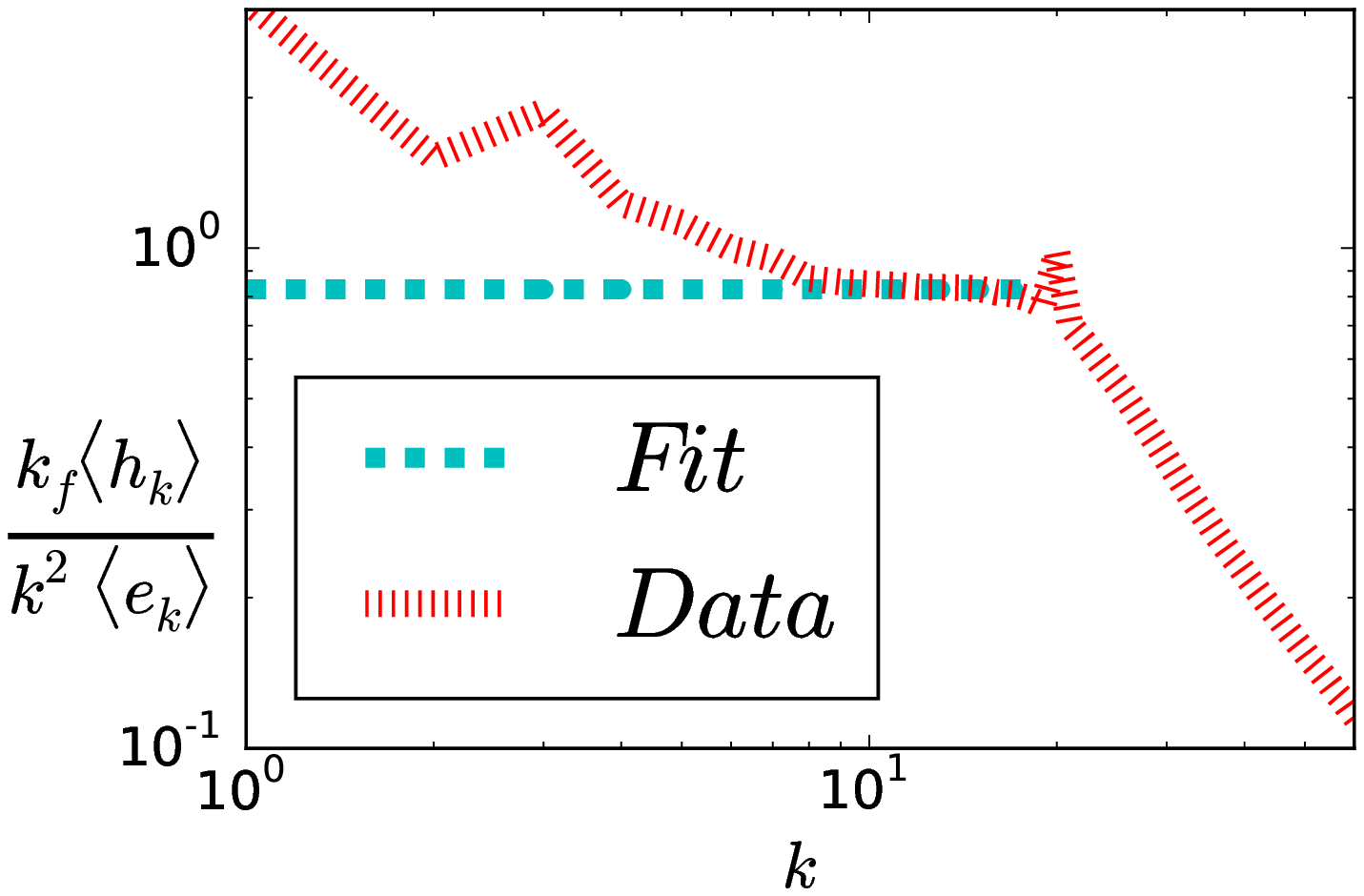} };
        \end{scope}
    \end{tikzpicture}
  \caption[Navier-Stokes: energy and helicty spectrum]
  {Spectrum of general-periodic DNS of the Navier-Stokes equation 
  with an \ABC{} forcing.
  Left) Helicity spectrum in dotted line and the absolute equilibrium power law in full line.
  Right) Energy spectrum in dotted line with absolute equilibrium power law
  in full line and absolute equilibrium fit in dashed line. The insert represents 
  $ \kf{} \langle h_{\kvec} \rangle / (k^2 \langle e_{\kvec} \rangle )$ and the asymptotic value
  used to make the absolute equilibrium fit.
  }
  \label{fig:fig13}
\end{figure}

The left panel of fig.~\ref{fig:fig13} represents the helicity spectrum of a solution 
of the Navier-Stokes equation with an \ABC{} forcing. In the scales slightly larger
than the forcing scale, the helicity spectrum is in good agreement with the 
$k^4$-power law of the absolute equilibrium prediction. But in the largest scale, the
helicity spectrum has a deviation from the $k^4$-power law.

To compare absolute equilibrium solutions of the \TEE{} and the large scale modes
of solutions of the \NSE{}, we introduce an analogue of the Kraichnan number for the \NSE{}. 
The equivalent of the maximal wavenumber \km{} in the truncated Euler problem is assumed 
to be the forcing wavenumber \kf{} in the case of the \NSE{}. Two expressions can be used 
to compute the local Kraichnan number: either 
$ \kf{} \langle h_{\kvec} \rangle / (k^2 \langle e_{\kvec} \rangle )$ 
coming from eq.~\eqref{eq:KrSt} or 
$	\kf{}
	\left( \langle e^{+}_{\kvec} \rangle - \langle e^{-}_{\kvec} \rangle \right) /
	\left( k \left(\langle e^{+}_{\kvec} \rangle + \langle e^{-}_{\kvec} \rangle \right) \right)$
coming from eq.~\eqref{eq:CrHe}.
Both expressions are equivalent and give the same numeric results presented in the insert 
of the right panel of fig.~\ref{fig:fig13}. The local Kraichnan number is not independent of 
the wavenumber and has an important peak in the large scales. As the wavenumber reaches 
the forcing wavenumber, the local local Kraichnan number goes to a constant value equal 
to $0.83 \,$. The result is consistent with the $\vert \Kr{} \vert \leq 1$ bound of absolute 
equilibrium solutions.
This asymptotic value is then used to plot the absolute equilibrium fit in the right 
panel of fig.~\ref{fig:fig13}. In the large scales, the data from the DNS is slightly above the fit,
while the data is below the fit near the forcing wavenumber. The difference between the fit and 
the data near the forcing wavenumber can be related to the presence of the forcing. The difference 
found in the large scales could be related to finite size effects such as three-mode interaction with
the forcing like in \cite{cameron_large-scale_2016}.

\begin{figure}[!ht]
  \centering
  \includegraphics[height=\fheight , trim= 0 0 0 0, clip=true]{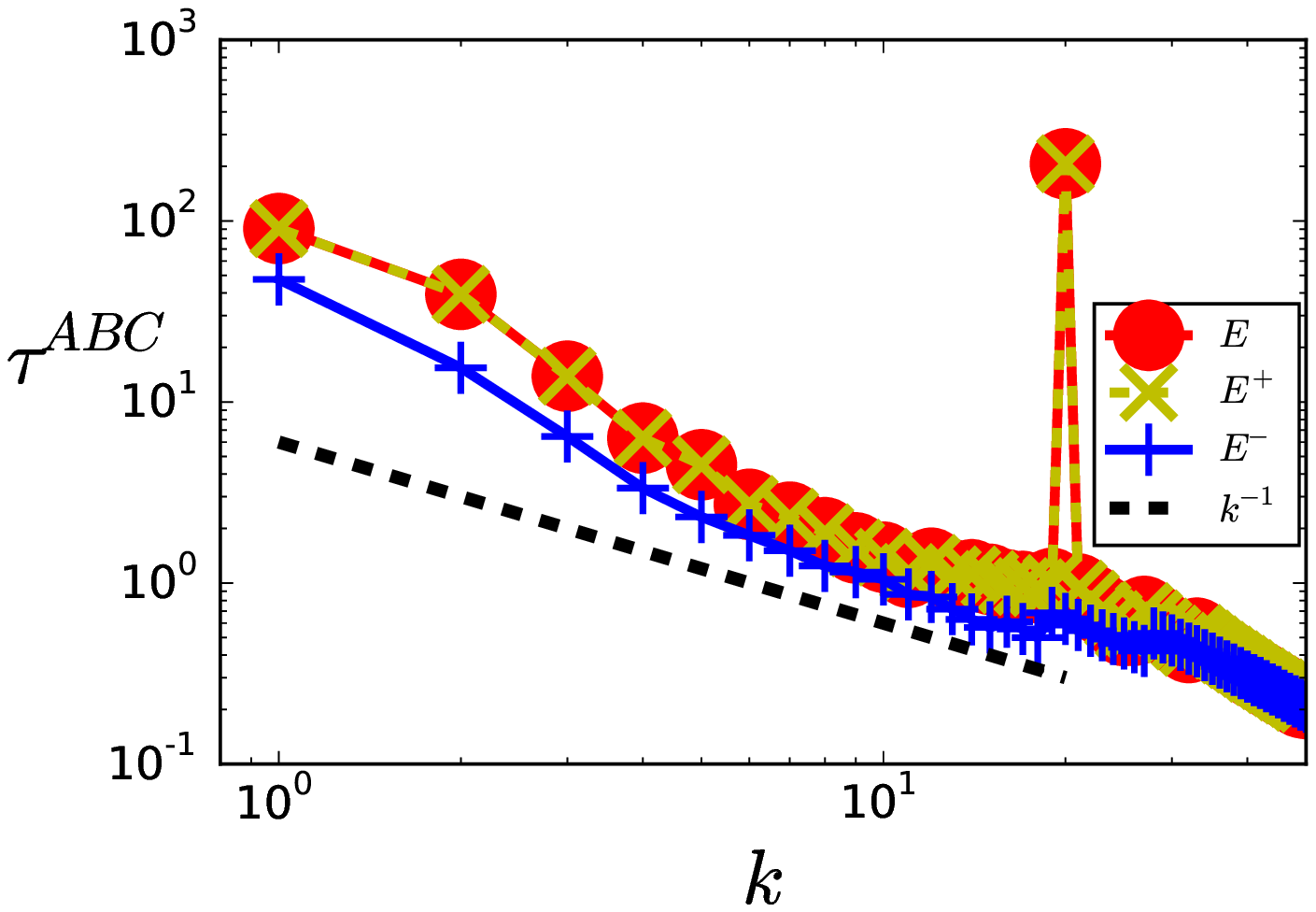}
  \includegraphics[height=\fheight , trim= 0 0 0 0, clip=true]{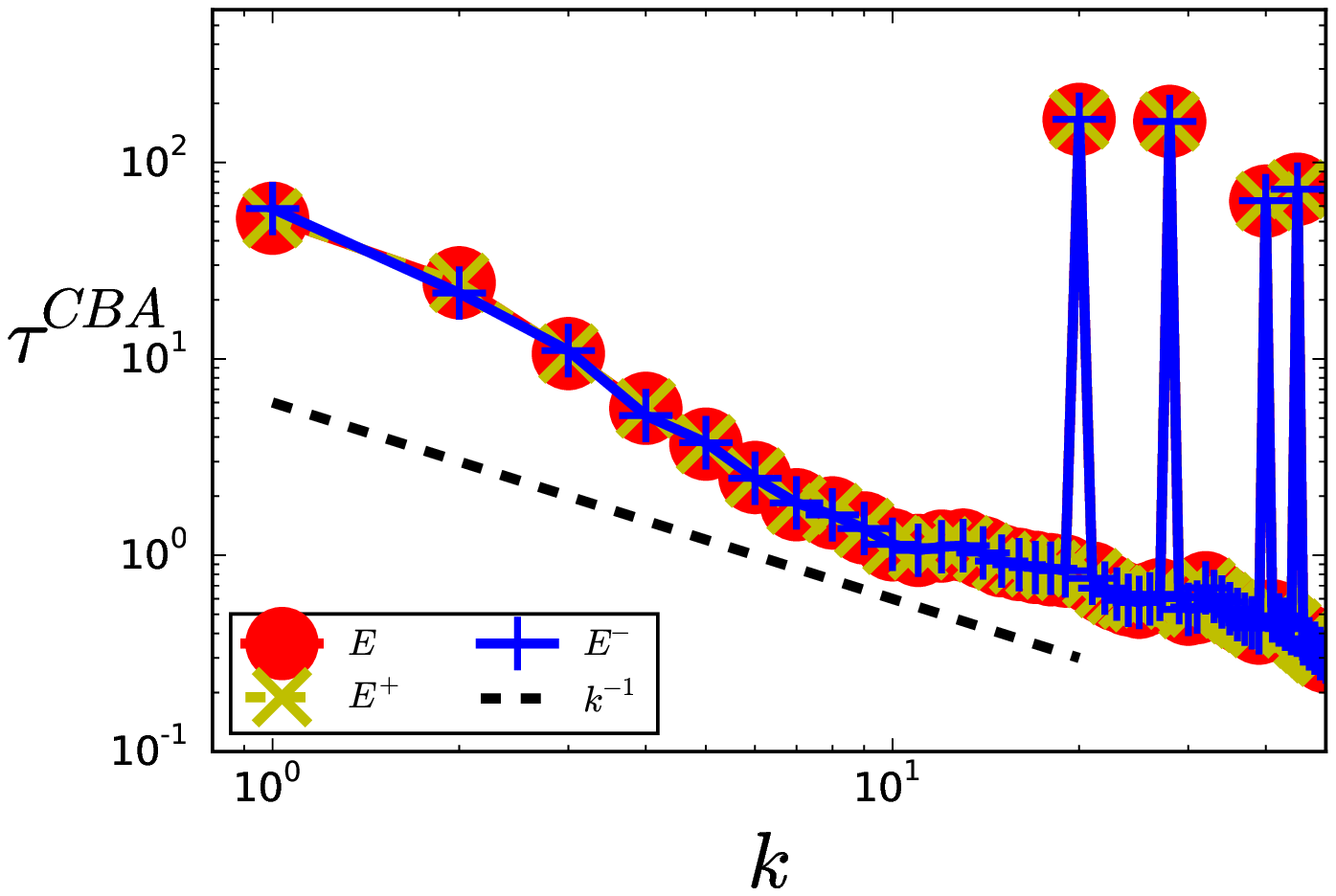}
  \caption[Navier-Stokes: correlation time]
  {Correlation time of general-periodic DNS of the 
  Navier-Stokes equation. 
  Left) \ABC{} forcing. 
  Right) \CBA{} forcing.}
  \label{fig:fig14}
\end{figure}

\begin{figure}[!ht]
  \centering
  \includegraphics[height=\fheight , trim= 0 0 0 0, clip=true]{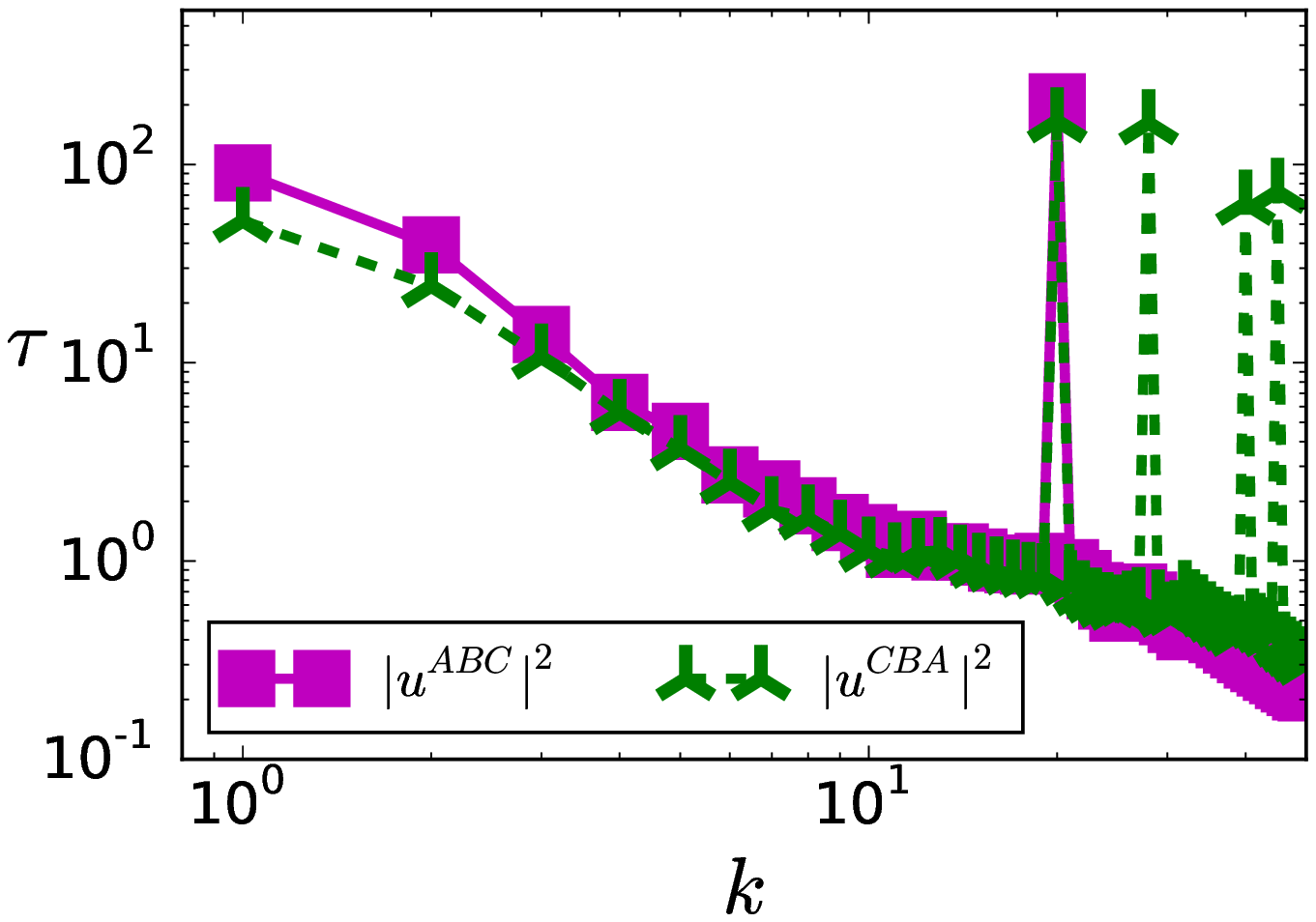}
  \includegraphics[height=\fheight , trim= 0 0 0 0, clip=true]{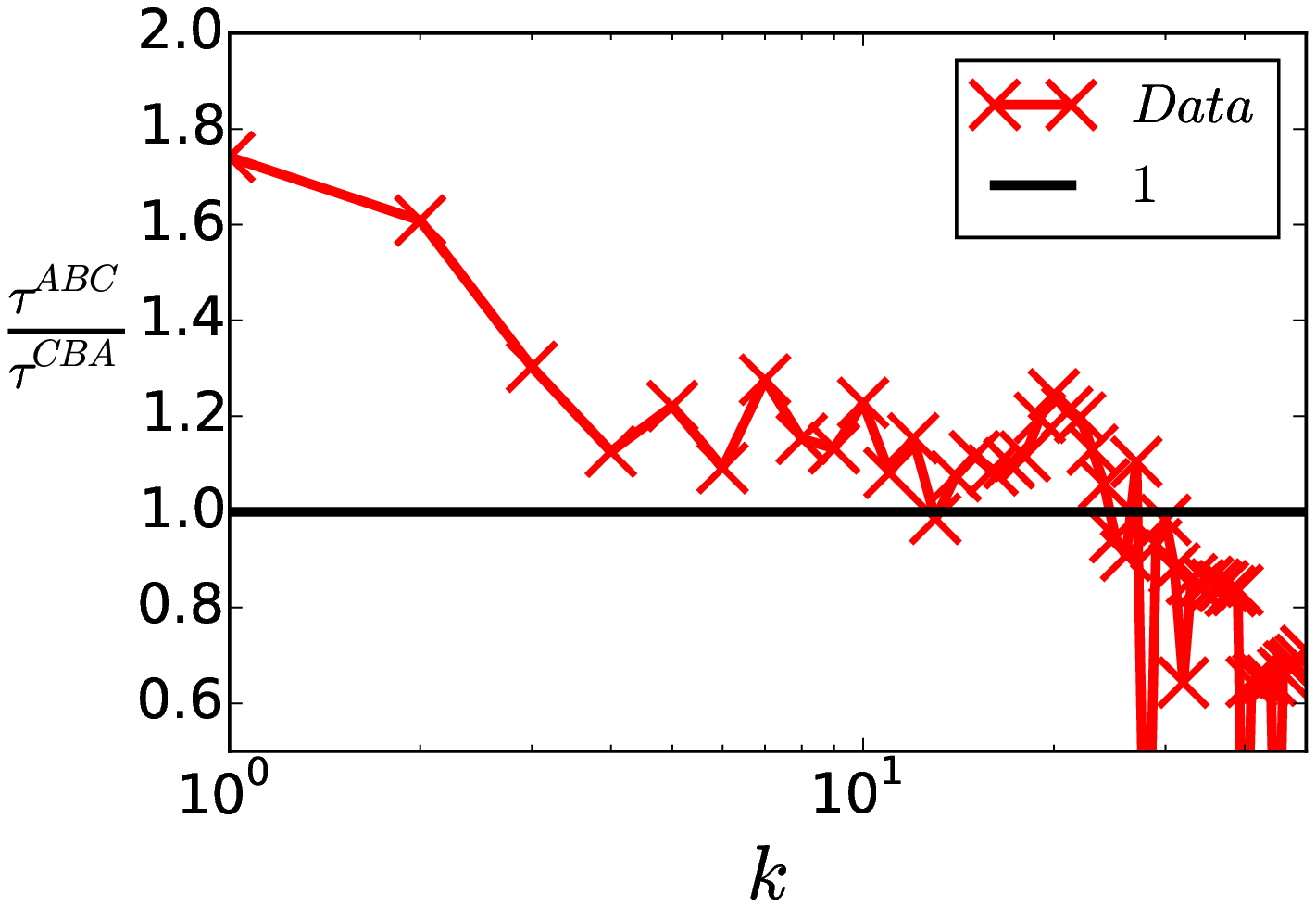}
  \caption[Navier-Stokes: impact of helicity on correlation time]
  {Comparison of the correlation time of $[0 ; 2\pi]^{3}$-periodic DNS 
  solutions of the Navier-Stokes equation forced with \ABC{} and 
  \CBA{} forcings.
  Left) Energy spectrum.
  Right) Energy ratio.}
  \label{fig:fig15}
\end{figure}

The correlation time of DNS of the \NSE{} with an \ABC{} forcing increases in the large scales 
as shown in the left panel of fig.~\ref{fig:fig14}. In the inertial domain, the correlation time 
decreases as the wavenumber increases. These observations are consistent with the results 
of helical DNS of the \TEE{} and the \TG{} symmetric DNS of the the \NSE{}.
However, the correlation time has some elements which do not appear in the helical DNS of the \TEE{}
presented in fig.~\ref{fig:fig8}. The correlation time of positive helical components of the velocity 
is always greater than its negative counterpart, especially in the large scales. 
This separation between the correlation time of the different helical components is not
observed in the right panel of fig.~\ref{fig:fig14} representing DNS of the \NSE{} with a 
\CBA{}. Contrary to the \ABC{} forced flow, the correlation times of \CBA{} forced flows
exhibit a set of peaks in the inertial domain corresponding to harmonics of the forcing.
For the \ABC{} forcing and \CBA{} forcing, the correlation time in the large scales 
deviates from the $k^{-1}$-power law observed in DNS of \TEE{}. However, for similar 
scale separations, the results form \TG{} DNS also deviated from the $k^{-1}$-power law
in the left panel of fig.~\ref{fig:fig11}. 
Fig.~\ref{fig:fig15} compares the correlation times of the \ABC{} forcing and \CBA{} forcing. 
The correlation time of the two flows seems very close when observing 
the energy spectrum presented in the left panel. However, the ratio presented in the right
panel shows that the correlation times of the helical \ABC{} forcing are larger than their 
non-helical \CBA{} counterparts. The differences between the two correlation times reaches 
nearly twenty percents for all the wavenumbers smaller than the forcing wavenumber
and is much more important in the two shells with smallest $k$.

\section{Conclusion}
\label{sec:ccl}

In this study, we examined analytically and numerically the temporal dynamics of absolute 
equilibrium solutions of the truncated Euler equation and compared them to the large scale 
modes of solutions of the Navier-Stokes equation.
We calculated the Eulerian parabolic auto-correlation time of velocity modes of absolute
equilibrium solutions of the \TEE{} under the assumption that 
the velocity modes follow a Boltzmann-Gibbs distribution.
For non-helical or slightly helical flow, $\vert \Kr{} \vert < 0.8 $,
the correlation time decreases inversely proportionally to the wavenumber and the square 
root of energy $\tau\propto 1/(k E^{1/2})$. On the other hand, when the flow is highly helical 
$\vert \Kr{} \vert > 0.9$, the correlation time depends on the helicity and is inversely proportional 
to the square root of the wave number  $\tau\propto 1/(k^{1/2} H^{1/2})$. This new power law 
behavior holds for wavenumbers that satisfy $ (1 - \Kr{}) \ln (1 - \Kr{}) \ll k/ \km{} \ll 1$,
while the $k^{-1}$ is recovered at even smaller wavenumbers. 
%
These results and the assumptions on which they were based 
were tested against the results of DNS of the \TEE{} showing excellent agreement:
the PDFs of the modes of velocity displayed the characteristics of Gaussian variables,
their energy spectrum followed the laws predicted by the absolute equilibrium theory,
and the two predicted power laws for the correlation time for helical and non-helical flows
were recovered. We have thus demonstrated that,in the long time limit, the \TEE{} 
thermalizes to an absolute equilibrium state whose statistics and time-correlation can 
be derived from a Boltzmann-Gibbs distribution. 

The results are less clear for the large scales of the forced \NSE{}. 
Simulations using the non-helical \TG{} geometry, that allowed us to reach larger scale separation, 
displayed considerable agreement with the absolute equilibrium. 
The velocity modes at large scales followed Gaussian distributions with standard deviations 
consistent with equipartition; the correlation time followed a power law
compatible with the $k^{-1}$ scaling for almost two decades. 
This is in agreement with the absolute equilibrium at large scales.
The agreement however was less strong with the results of DNS of general-periodic flows.
The large scale spectra were compatible with the absolute equilibrium predictions for a limited range.
The largest scales deviated significantly, being more energetic and more helical than predicted. 
Note that this deviation was also observed in \cite{dallas_statistical_2015}. 
Furthermore,
while the correlation time of the helical flows was shown to decrease with helicity
as in thermalization theory the measured correlation time deviates significantly from the predicted 
$k^{-1}$ and $k^{-1/2}$ power laws. 
We note however that a similar deviation is also observed in non-helical \TG{} symmetric DNS at the 
smallest boxes used (see fig.~\ref{fig:fig10}), so these deviations could be due to insufficient scale separation.

The reason for which the largest scales of Navier-Stokes solutions deviate from absolute 
equilibrium still remains an open question. As discussed above, a possible cause for this 
deviation could be related to the range in the large scales or the range of Reynolds number. 
For the moment, however, we can not exclude the possibility that large scales 
lack universality with respect to forcing or that large scales instabilities spoil the 
absolute equilibrium properties. These questions are left for future studies.

\acknowledgments 

This work was granted access to the HPC resources of MesoPSL financed by the Region 
Ile de France and the project Equip@Meso (reference ANR-10-EQPX-29-01) of 
the programme Investissements d'Avenir supervised by the Agence Nationale pour 
la Recherche and the HPC resources of GENCI-TGCC-CURIE \& GENCI-CINES-JADE 
(Project No. x20162a7620) where the present numerical simulations have been performed.
The authors are grateful to Pablo \textsc{Mininni} and Patricio \textsc{Clark di Leoni}
for their useful discussion on spatio-temporal spectra.

\section{Appendix: Correlation time -- parabolic hypothesis}      
\label{sec:apx:ct}

The derivation of the correlation time can be done using a projection operator 
on incompressible flows, as in ref. \cite{cichowlas_effective_2005}. However, this method 
is not able to assess the properties of the helical components of the velocity. In order to 
quantify these properties, the standard framework is the Craya-Herring helical basis 
\cite{craya_contribution_1957,herring_approach_1974}. Within this decomposition, 
the \TEE{} is expressed as
\begin{align}
	\left ( \partial_t u^{s_{\kvec}}_{\kvec} \right)^{*}
	= \sum_{ \substack{\kvec+\pvec+\qvec=0  \\ 
	s_{\pvec} \, ,\, s_{\qvec}} }
	( s_{\pvec}p - s_{\qvec}q )
	\left( - \frac{1}{4} \hvec^{s_{\kvec}}_{\kvec} \cdot 
	\hvec^{s_{\pvec}}_{\pvec} 
	\times
	\hvec^{s_{\qvec}}_{\qvec} 
	\right)
	u^{s_{\pvec}}_{\pvec} u^{s_{\qvec}}_{\qvec} 
	= \sum_{ \substack{\kvec+\pvec+\qvec=0  \\ 
	s_{\pvec} \, ,\, s_{\qvec}} }
	C^{s_{\kvec} s_{\pvec} s_{\qvec}}_{\kvec \pvec \qvec}
	u^{s_{\pvec}}_{\pvec} u^{s_{\qvec}}_{\qvec} \,,
\end{align}
where $s_{\kvec}$ denotes the sign of the helical component at mode $\kvec$, 
$u^{s_{\kvec}}_{\kvec}$ denotes the helical component of the velocity of 
sign $s_{\kvec}$ at mode $\kvec$ and $\hvec^{s_{\kvec}}_{\kvec}$
denotes the complex unitary helical vector of the Craya-Herring basis satisfying: 
$\roT \hvec^{s_{\kvec}}_{\kvec} = s_{\kvec} k \hvec^{s_{\kvec}}_{\kvec} $.
The Craya-Herring tensor 
$C^{s_{\kvec} s_{\pvec} s_{\qvec}}_{\kvec \pvec \qvec}$ 
is symmetric on its last two variables:
$C^{s_{\kvec} s_{\pvec} s_{\qvec}}_{\kvec \pvec \qvec}
= C^{s_{\kvec} s_{\qvec} s_{\pvec}}_{\kvec \qvec \pvec}$, 
and also satisfies :
$C^{s_{\kvec} s_{\pvec} s_{\pvec}}_{\kvec \pvec \pvec} = 0$.
With the Craya-Herring helical decomposition, the average correlation of the temporal 
derivative of the velocity can be derived with the assumption that all helical components 
are independent Gaussian variables. The derivation leads to
\begin{align}
	\langle \vert \partial_t \bar{u}^{s_{\kvec}}_{\kvec}\vert^2 \rangle
	& = 
	\left \langle
	\sum_{ \substack{\kvec+\pvec_{1}+\qvec_{1}=0  \\ 
	s_{\pvec_{1}} \, ,\, s_{\qvec_{1}}} }
	\sum_{ \substack{\kvec+\pvec_{2} +\qvec_{2}=0  \\ 
	s_{\pvec_{2}} \, ,\, s_{\qvec_{2}}} }
	C^{s_{\kvec} s_{\pvec_{1}} s_{\qvec_{1}}}_{\kvec \pvec_{1} \qvec_{1}}
	u^{s_{\pvec_{1}}}_{\pvec_{1}} u^{s_{\qvec_{1}}}_{\qvec_{1}}
	\left [	
	C^{s_{\kvec} s_{\pvec_{2}} s_{\qvec_{2}}}_{\kvec \pvec_{2} \qvec_{2}}
	u^{s_{\pvec_{2}}}_{\qvec_{2}} u^{s_{\pvec_{2}}}_{\qvec_{2}}
	\right ]^{*} 
	\right \rangle	 	 	
	\\ & =\sum_{s_{1} \, ,\, s_{2} } \mathcal{S}^{s_{1}  s_{2}} 
	\quad \text{where} \quad
	\mathcal{S}^{s_{1}  s_{2}} = 
	\sum_{ \substack{\kvec+\pvec+\qvec=0  \\ 
	s_{\pvec}=s_{1} \, ,\, s_{\qvec}=s_{2}} }
	2 \left \vert 
	C^{s_{\kvec} s_{\pvec} s_{\qvec}}_{\kvec \pvec \qvec}
	\right \vert^2
	\left \langle \vert \uvec^{s_{\pvec}}_{\pvec} \vert^2\right \rangle
	\left \langle \vert \uvec^{s_{\qvec}}_{\qvec} \vert^2 \right \rangle \, .
	\label{eq:ch:evol}
\end{align}
where $\mathcal{S}^{s_{1}  s_{2}} $ corresponds to the sum of the triadic interact of 
sign $s_{1}$ and $s_{2}$.

\begin{figure}[!htb]
	\begin{center}
		\begin{tikzpicture}
		\begin{scope}
		\clip (2,2) circle (3cm);
		\fill[black!30] (4,2) circle (3cm);
		\end{scope}
		\draw[line width=.5mm,blue,dotted]  (2,2) circle (3cm);
		\draw[line width=.5mm,blue,dotted]  (4,2) circle (3cm);
		\draw[line width=.3mm,black] (-1,2) -- (7,2) node[above=2mm,left=0mm] {$\km{}+k$} ;
		\node[above=2mm,right=-1mm] at (-1,2) {$-\!\km{}$};
		\node[below=4.5mm,left=-1mm] at (1,2) {$k\!-\!\km{}$};
		\node[below=4.5mm,right=-1mm] at (5,2) {$\km{}$};
		\draw[line width=.3mm,black] (2,2) circle (0.1cm) node[below=2mm] {$O$} ;
		\draw[line width=.5mm,black] (4,2) circle (0.1cm) node[below=2mm] {$k$};
		\draw[line width=.5mm,green	] (2,2) -- (-0.1213203436,-0.121213203436) ;
		\draw[line width=.5mm,green	] (4,2) -- (6.1213203436,-0.121213203436);
		\draw[line width=.5mm,red,dashed] (2,2) -- (4,2);
		\draw[line width=.5mm,yellow,dashed] (3,2) node[below=1.5mm,black] {$\frac{1}{2}k$} -- 
		(3,4.8284271247461898) ;
		\draw[line width=.4mm,->] (3.2,3) -- (3.8,3) node[anchor=north east] {$\parallel$};
		\draw[line width=.4mm,->] (3.2,2.99) -- (3.2,3.6) node[anchor=north west] {$\perp$};
		\end{tikzpicture}
	\end{center}
	\caption[Diagram of the integration domain]
	{Diagram of a cut of the integration domain. The dark surface 
	corresponds to the integration domain. The dotted lines correspond to 
	the limit of the circles of radius \km{} and of center $0$ or $k$. The thick 
	full line corresponds to radii of the previously described circles. The dark 
	dashed line corresponds to the distance between the center of the two 
	circles. The bright dashed line corresponds to the maximal length possible
	for $q_{\perp}$.}
	\label{fig:apx1}
\end{figure}
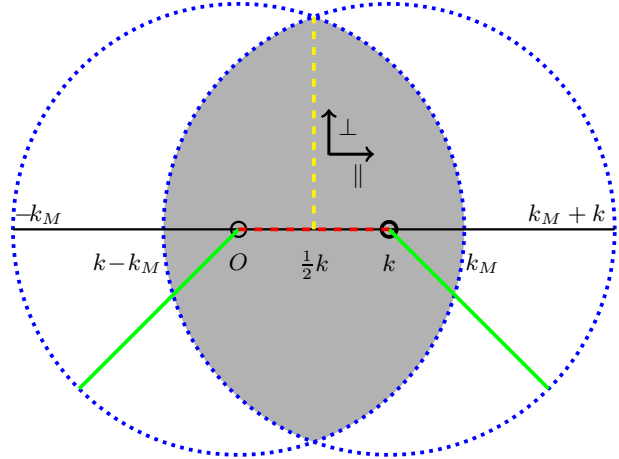

In eq.~\eqref{eq:ch:evol}, the truncation condition, $ \uvec( k > \km{}) = 0 $, has not 
been applied to the velocity fields. The velocity appears in the equation within the 
expression of the average energy with two indices $\pvec$ and $\qvec$, therefore the 
summation must be done at $p \leq \km{}$ and $q \leq \km{}$. The domain prescribed 
by these conditions corresponds to the intersection of two spheres of radii \km{} and 
centers $\pvec$ and $\qvec$.
The triadic condition, $\kvec + \pvec + \qvec = 0$, also implies that the summation over 
$\pvec$ and $\qvec$ can be done over $\qvec$ at fixed $\pvec = \kvec - \qvec $. The 
domain of summation is represented in fig.~\ref{fig:apx1}. This domain is invariant by 
rotation along the axis defined by $\kvec$ in fig.~\ref{fig:apx1}. The centers of the two 
spheres are also located on this direction. The coefficients of the sum in 
eq.~\eqref{eq:ch:evol} are unaffected by the rotation. The summation can thus be 
performed with the variables $q_{\perp}$ and $q_{\parallel}$ where $q_{\perp}$ is the 
projection of the wavevector along the plane orthogonal to $\kvec$ and $q_{\parallel}$ is 
the projection of the wavevector along the axis of rotation.
Taking this new coordinate system, the derivation can be simplified by converting the discrete 
sum into an integral using the equivalence
\begin{align}
	\sum_{ \kvec+\pvec+\qvec=0} 
	\quad \Longleftrightarrow \quad
	\int^{1-\frac{1}{2}m}_{-(1-\frac{1}{2}m)} d q_{\parallel}
	\int^{1 - \left(\vert q_{\parallel}\vert + \frac{1}{2}m \right)^2}_{0} \pi d q^{2}_{\perp}
	\quad \text{where} \quad
	m = \frac{k}{\km{}} \, .
\end{align}

\begin{figure}[!ht]
	\includegraphics[width=5.8cm, trim= 0 0 0 0, clip=true]{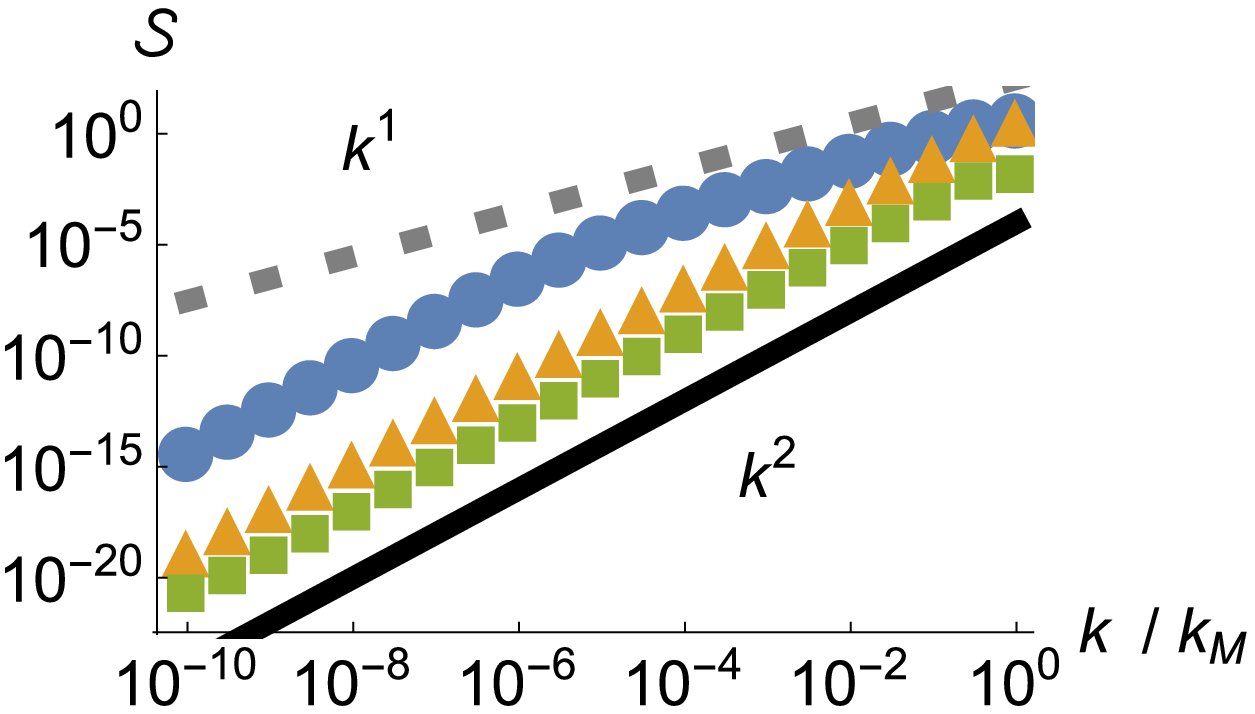}
	\includegraphics[width=5.8cm, trim= 0 0 0 0, clip=true]{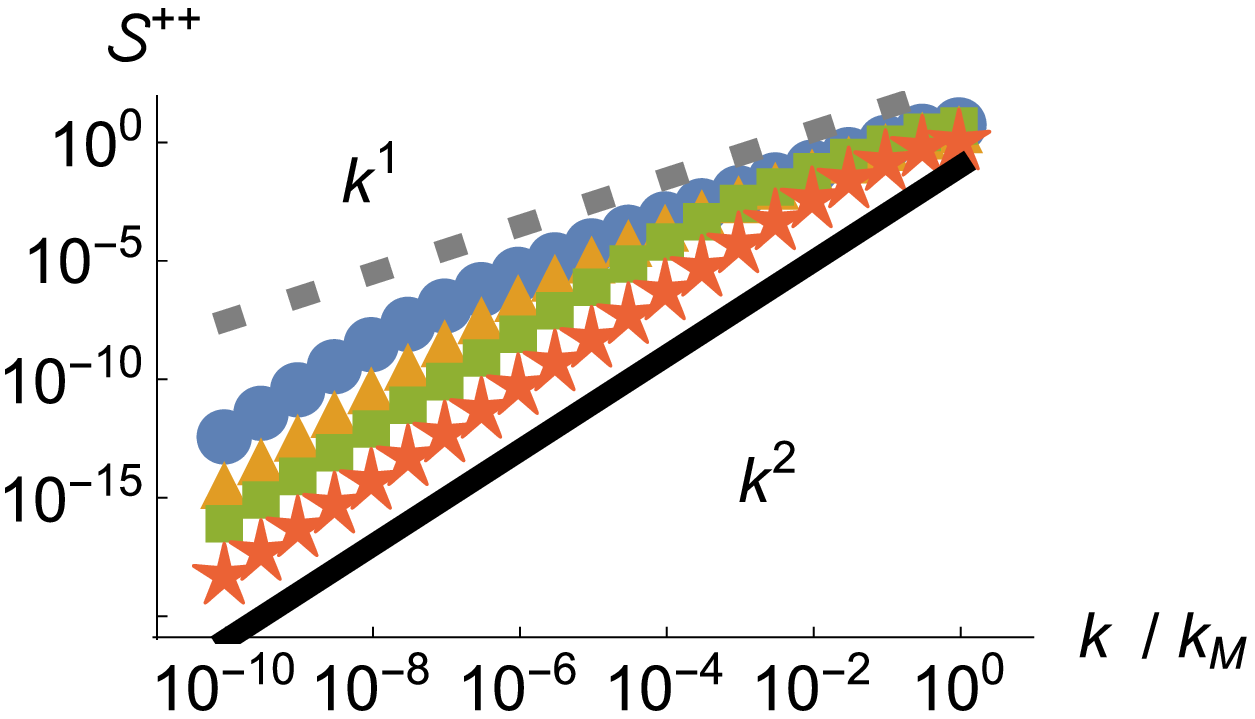}
	\includegraphics[width=5.8cm, trim= 0 0 0 0, clip=true]{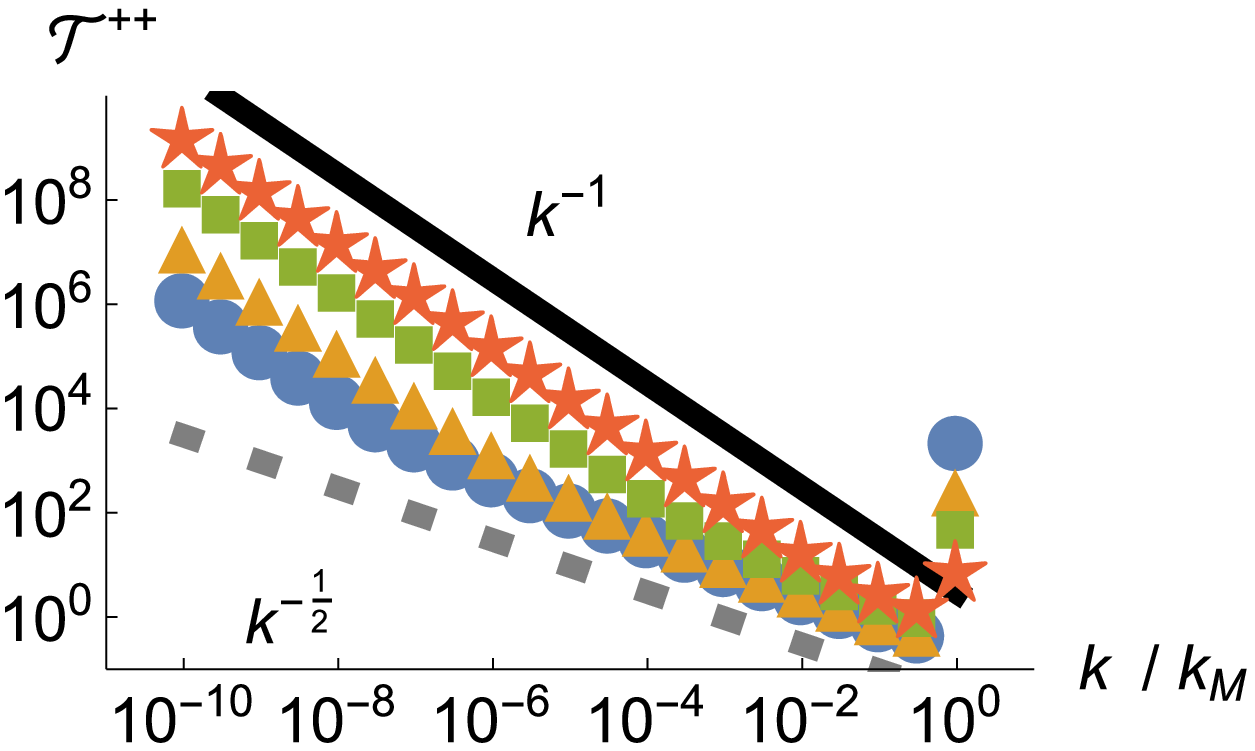}
	\caption[Triadic interaction]
	{Triadic interaction as a function of the wavenumber and associated
	correlation time.
	Left) At $\Kr{} = 1- 10^{-6}$, $\mathcal{S}^{- -}$ plotted with squares, 
	$\mathcal{S}^{+-}$ plotted with triangle,
	$\mathcal{S}^{+ +}$ plotted with discs, 
	$k^{1}$ plotted with a dashed line, 
	$k^{2}$ plotted with a full line. 
	Center) $\mathcal{S}^{+ +}$ plotted at $\Kr{} = 1- 10^{-8}$ with discs, 
	at $\Kr{} = 1- 10^{-6}$ with triangles, at $\Kr{} = 1- 10^{-2}$ with squares,
	at $\Kr{} = 1- 10^{-2}$ with stars, $k^{1}$ with a dashed line, 
	$k^{2}$ with a full line. 
	Right) $\mathcal{T}^{+ +}$ plotted at $\Kr{} = 1- 10^{-8}$ with discs, 
	at $\Kr{} = 1- 10^{-6}$ with triangles, at $\Kr{} = 1- 10^{-2}$ with squares,
	at $\Kr{} = 1- 10^{-2}$ with stars, $k^{-\frac{1}{2}}$ with a dashed line, 
	$k^{-1}$ with a full line. }
	\label{fig:apx3}
\end{figure}

This integral can be computed exactly when $\Kr{}=0$ for $m \ll 1$,
the parabolic correlation time is then expressed as
\begin{align}
	\tau^{s_{\kvec}}_{\kvec} = 
	\sqrt{\frac{ \left \langle \vert u^{s_{\qvec}}_{\qvec}\vert^2 \right \rangle}{
	\langle \vert \partial_t \bar{u}^{s_{\kvec}}_{\kvec}\vert^2 \rangle}}
	= \sqrt{\frac{ 45 \alpha}{ 112}} \frac{1}{k } 
	= \sqrt{\frac{ 15\pi C_{N} }{ 14}} \frac{1}{k \sqrt{E_{tot}}} \,.
\end{align}
Since $E_{tot} \propto \alpha^{-1}$, the correlation time follows the energy-based scaling-law.

When $\Kr \neq 0$, the final integral can be computed using Mathematica 
for the different helical triadic interactions as shown in fig.~\ref{fig:apx3} for 
highly helical flows. Instead of computing all the terms corresponding to all different 
possible triads in the sum, the graph shows the computation triad with different helical 
signs. The left plot of fig.~\ref{fig:apx3} shows three possible sums: 
$\mathcal{S}^{+ +} $ plotted with discs, $\mathcal{S}^{+ -} $ plotted with triangles 
and $\mathcal{S}^{- -} $ plotted with squares. The $\mathcal{S}^{- +} $ sum has 
exactly the same values as the $\mathcal{S}^{+ -} $ sum for symmetry reasons. The 
$\mathcal{S}^{+ +} $ sum dominates the other terms and has a $k^{1}$-scaling at 
large wavenumbers consistent with a helicity-based correlation time, and a 
$k^{2}$-scaling at small wavenumbers consistent with an energy-based correlation time. 
All other terms follow a $k^{2}$-scaling.
The center plot of fig.~\ref{fig:apx3} shows the evolution of the $\mathcal{S}^{+ +} $
for different Kraichnan numbers. As the Kraichnan number goes to one, the domain
where the sum follows a $k^{2}$-scaling widens. The right plot of fig.~\ref{fig:apx3} 
shows the evolution of the correlation time, $\mathcal{T}^{+ +} $, built using 
the sum $\mathcal{S}^{+ +} $. Its evolution is consistent with the evolution of the 
correlation time of the full velocity field presented in the left panel of fig.~\ref{fig:fig1}.

When the Kraichnan number goes to one, the {\it plus-plus} triadic interaction
dominates the non-linear interaction in the \TEE{}. Whether $m \ll \epsilon \ll 1$ or 
$\epsilon \ll m \ll 1$, the correlation time follows the asymptotic expression
\begin{align}
	\mathcal{S}^{+ + } = 2 \pi k^2 \alpha^{-2} \frac{- 2 \ln \epsilon }{ m - A \epsilon \ln \epsilon } 
	\quad \text{with} \quad 
	A = \frac{15}{8}
	\quad \text{thus} \quad 
	\mathcal{T}^{+ + }(\kvec, s_{\kvec} ) = 
	\sqrt{\frac{\langle \vert \uvec^{s_{\kvec}}_{\kvec}\vert^2 \rangle }{\mathcal{S}^{++}}}
	= \sqrt{\frac{ A (1-\Kr{} ) - \frac{k }{ \km{} \ln(1-\Kr{} )} 
	}{ 4 \pi \alpha^{-1} k^2 (1 - s_{\kvec} \Kr{} \frac{k }{ \km{} } ) }} \, .
\end{align} 
In the domain where $ (1-\Kr{} ) \ll \frac{k }{ \km{} } \ll 1$, the correlation time follows
a helicity-based scaling-law.

\section{Appendix: Taylor-Green symmetries}
\label{sec:apx:TGP}
In addition to the condition listed below eq.~\eqref{eq:TG:def}, \TG{} symmetries 
\cite{nore_decaying_1997} impose that
\begin{align}
	u^y(k_x,k_y,k_z) = (-1)^{p+1} u_x(k_y,k_x,k_z)  
	\quad \text{and} \quad
	u_z(k_x,k_y,k_z) = (-1)^{p+1} u_z(k_y,k_x,k_z)  \, ,
\end{align}
where $p$ characterizes the parity of the mode: 
$p=1$ if $k_x$, $k_y$, $k_z$ are all even 
and $p=0$ if $k_x$, $k_y$, $k_z$ are all odd.

Since the flows studied are also incompressible, they also satisfy
$
	\diV \uvec = 0 \iff k_x u_x + k_y u_y + k_z u_z = 0 \, .
$

In a few special cases, modes of \TG{} symmetric flows 
depend on only one independent variable
\begin{itemize}
\item $k_x=k_y$ and $r=0$:
$ \quad
	\uvec^{odd}(k_x,k_x,k_z) = ( \evec_x - \evec_y ) \psi_{0}(k_x,k_z) 
\quad $
where $\psi_{0}$ is a real field.
\item $k_x=k_y$ and $r=1$:
$ \quad
	\uvec^{even}(k_x,k_x,k_z) = 
	(k_z(\evec_x + \evec_y) - (k_x+k_y) \evec_z) \psi_{1}(k_x,k_x,k_z) 
\quad $ 
where $\psi_{1}$ is a real field.
\item $k_x=0$ :
$ \quad
	\uvec(0,k_y,k_z) = (k_z\evec_y - k_y \evec_z) \psi_{2}(k_y,k_z)
\quad $
where $\psi_{2}$ is a real field.
\item $k_y=0$ :
$ \quad
	\uvec(k_x,0,k_z) = (k_z\evec_x - k_y \evec_z) \psi_{3}(k_x,k_z)
\quad $ 
where $\psi_{3}$ is a real field.
\item $k_z=0$ :
$ \quad
	\uvec(k_x,k_y,0) = (k_y\evec_x - k_x\evec_y) \psi_{4}(k_x,k_y) \, ,
\quad $ 
where $\psi_{4}$ is a real field satisfying $\psi_{4}(k_y,k_x) = - \psi_{4}(k_x,k_y)$.
\end{itemize}
The vectors $\evec_{\alpha}$ with $\alpha \in \lbrace x ; y ; z \rbrace$ denote 
the directions of the Cartesian basis.

In the other cases, the \TG{} symmetric modes of flows only depend on two 
independent variables -- $\phi(k_x,k_y,k_z)$ and $\phi(k_y,k_x,k_z)$, where $\phi$ 
is a real field -- and can be written as
\begin{align}
	\uvec(k_x,k_y,k_z) = (k_z\evec_x - k_x \evec_z) \phi(k_x,k_y,k_z) + 
	(-1)^{p+1} (k_z\evec_y - k_y \evec_z) \phi(k_y,k_x,k_z)
	\,.
\end{align}

\section{Appendix: Chi-squared distribution}
\label{sec:apx:chi2}
A Gaussian distribution of average $\mu$ and standard derivation $\sigma$ has a
probability density function defined by
\begin{align}
	G(X \vert \mu, \sigma ) = {\frac {1}{\sqrt {2\sigma ^{2}\pi }}}
	\;e^{-{\frac {(X-\mu )^{2}}{2\sigma ^{2}}}}
	\,.
	\label{eq:gaus}
\end{align}

Chi-squared distributions are defined using $g$ independent Gaussian variables. Let 
$(G_i)_{i \in [\![1 ;  g ]\!]}$ be independent, centered ($\mu=0$), reduced 
($\sigma=1$) Gaussian variables. The sum of their squares, 
$X=\sum _{i=1}^{g}G_{ i }^{2}$, is distributed according to the chi-squared 
distribution with $g$ degrees of freedom denoted as $\chi^{2}_{g} $ and defined by
\begin{align}
	\chi^{2}_{g}(X) = \frac{1}{2^{\frac{g}{2}}\Gamma_{Euler}(\frac{g}{2})} 
	X^{\frac{g}{2}-1} e^{-\frac{X}{2} }
	\,.
	\label{eq:chi2:appx}
\end{align}
where $\Gamma_{Euler}$ denotes Euler's Gamma function.
The power law of the probability density function at small $X$ gives the 
number of degrees of freedom of the system, and the exponential fit at large 
$X$ validates the Gaussian decay of the probability density function.
\begin{align}
	\log \chi^{2}_{g}(X) \underset{X \to 0}{=} 
	\left( \frac{g}{2}-1\right) \log( X )
		\quad \text{thus} \quad
	g \underset{X \to 0}{=} 2 \left( \frac{\log \chi^{2}_{g}(X) }{\log( X )} + 1\right)
	\label{eq:logchi2}
\end{align}

\section{Appendix: Computation of the correlation time}
\label{sec:apx:cort}
In order to produce spatio-temporal spectrum, the velocity field is 
outputted at a regular time interval. These outputs form a dataset of the velocity
field in the $\kvec-t$ space. However, keeping in memory the entire $N^{3}$ DNS, 
$N$ being the resolution, is too demanding in storage memory. To reduce the 
volume of the dataset without loosing the properties of the different modes, only the 
six planes: $k_x = \lbrace 0 ; 1 \rbrace$, $k_y = \lbrace 0 ; 1 \rbrace$ and 
$k_z = \lbrace 0 ; 1 \rbrace$ are outputted for the \TG{} symmetric TYGRE code 
and only the three planes: $k_x=0$, $k_y=0$ and $k_z=0$ are outputted for the 
general-periodic GHOST code. The velocity time series are then multiplied by 
an apodization function \cite{weisstein_apodization_2017} and Fourier-transformed to 
form a dataset in the $\kvec-\omega$ space. The power spectrum $s(\kvec, \omega)$ 
is then computed by taking the modulus square of the velocity and summing over 
the different Cartesian directions. The isotropic power spectrum $S(k, \omega)$ 
is computed by summing the power spectrum over the modes of same 
wavenumber. A binning of spacing of one is used to compute the isotropic power 
spectrum. The closest integer smaller than $k+\frac{1}{2} $ is used to define the bin number. 
The left panel of fig.~\ref{fig:fig4} represents the power spectrum of a truncated Euler DNS
computed using this method.
The correlation function $\Gamma(k,t)$ is then computed using Wiener-Khinchin theorem 
by performing a Fourier-transform of the isotropic power spectrum and normalizing 
the function. The right panel of fig.~\ref{fig:fig4} represents the correlation function of a 
truncated Euler DNS computed using this method.

Finally, the correlation time can be computed by doing a fit of the correlation 
function in a well-resolved domain as shown in the left panel of fig.~\ref{fig:fig5}. 
The time where the correlation function reaches half-height, $\tau_{\frac{1}{2}}$,
can be evaluated numerically. 
The same algorithm can be used to find the correlation function of the positive 
and negative helical modes of the velocity field using eq.~\eqref{eq:CrHe}. 
The steps of the procedure are summed up in the algorithm presented below.
\begin{figure}[!ht]
	\begin{minipage}[c]{0.42\textwidth}
	\rule{\textwidth}{0.4pt}
		\begin{algorithmic}[1]
			\REQUIRE $\uvec(\kvec, n\Delta t)$, $n$, $\Delta t$, 
			\STATE $\uvec(\kvec, \omega) = D\mathcal{F} 
			[ apo( n, \Delta t ) \uvec(\kvec, n \Delta t) ](\omega)$
			\STATE $s(\kvec, \omega) = \sum_{i} 
			\left \vert \uvec_{i}(\kvec, \omega) \right \vert^{2}$
			\STATE $S(k, \omega) = \sum_{\kvec} \indic_k s(\kvec, \omega) $
			\STATE $\gamma( k, t) = D\mathcal{F}^{-1} [ S(k, \omega) ](t)$
			\STATE $\Gamma( k, t) = \gamma( k, t) / \gamma( k, 0) $
			\STATE $\tau(k) = Solve [t, \Gamma( k, t), 1/2] $ 
		\end{algorithmic}
	\rule{\textwidth}{0.4pt}
	\end{minipage}\hfill
	\begin{minipage}[c]{0.5\textwidth}
		\captionsetup{labelformat=empty}
		\caption[Correlation time algorithm]{Algorithm to compute the correlation time. 
		$ D\mathcal{F} $ denotes the discrete Fourier-transform,
		$ D\mathcal{F}^{-1} $ denotes the discrete inverse Fourier-transform,
		$\indic$ denotes the characteristic function satisfying 
		$\indic ( bool ) = 1$ if $(k-\frac{1}{2} < \vert \kvec \vert \leq k+\frac{1}{2} ) $ and $0$ otherwise, 
		$apo( n, \Delta t )$ denotes an apodization function and
		$Solve[t, \Gamma( k, t), 1/2]$ denotes a function that finds the 
		smallest positive $t$ satisfying $\Gamma( k, t)= 1/2$.} 
		\label{fig:appx3}
	\end{minipage}
\end{figure}

\bibliography{tauK,tauKadd}

\end{document}